\documentclass[10pt,a4paper]{article}

\usepackage[a4paper,margin=2.5cm]{geometry}

\usepackage[T1]{fontenc}
\usepackage{lmodern}
\usepackage{microtype}

\usepackage{amsmath,amssymb,amsfonts,mathtools,bm}
\usepackage{siunitx}

\DeclareMathOperator{\erf}{erf}

\usepackage{graphicx}
\usepackage{subcaption}
\usepackage{tikz}
\usetikzlibrary{arrows.meta}

\usepackage{array}
\usepackage{booktabs}
\usepackage{enumitem}
\usepackage{float}
\usepackage{setspace}

\usepackage{xcolor}
\usepackage{soul}
\usepackage{lipsum}

\usepackage[hidelinks]{hyperref}

\tikzset{
    box/.style={
        draw,
        rounded corners,
        align=center,
        minimum width=3.5cm,
        minimum height=1.1cm,
        font=\small
    },
    boxA/.style={box, fill=blue!5},
    boxB/.style={box, fill=green!5},
    boxC/.style={box, fill=orange!5},
    boxD/.style={box, fill=purple!5},
    arrow/.style={
        -Latex,
        thick
    }
}

\newcommand{\rev}[1]{%
    \ifmmode\else\leavevmode\fi
    {\color{blue}#1}%
}

\begin{document}

\title{\textbf{A multi-physics Eulerian framework for long-term contrail evolution (MultiCon): Single plume analysis}}

\author{
Amin Jafarimoghaddam\thanks{Corresponding author: ajafarim@pa.uc3m.es}
\and
Manuel Soler\thanks{masolera@ing.uc3m.es}\\
Department of Aerospace Engineering, UC3M, Madrid, Spain
}
\maketitle
	\textbf{ABSTRACT}\\
	\begingroup
Persistent contrails may evolve into contrail cirrus, whose long-term evolution strongly influences their radiative forcing. We present MultiCon, a reduced-order, multi-physics Eulerian framework for simulating the long-term evolution of individual aircraft contrail plumes while retaining essential microphysical processes at low computational cost. The model resolves vertical plume evolution through a coupled PDE system in the \((z,t)\) domain, while horizontal spread is tracked using the method of characteristics and a quasi-Gaussian kernel. The current implementation adopts a locally monodisperse approximation. The framework couples the transport of ice-particle number and mass, as well as water-vapor transport, with an additional equation governing particle-shape evolution. It accounts for two processes that have received comparatively little attention in contrail modeling: ice-crystal habit evolution and turbulence-modulated bulk settling. Habit-dependent growth and sublimation are represented through an approximate inherent growth factor (IGF) parameterization and a diffusion-limited formulation incorporating latent-heat effects. Furthermore, the model incorporates a reduced-order closure for the effective bulk settling velocity, using a Burgers-type model to describe the transition from turbulence-reduced settling towards the terminal velocity of individual particles. We evaluate plume evolution under stationary and time-dependent atmospheric conditions and compare MultiCon with CoCiP and APCEMM for a common reference case and a parametric sweep. Interestingly, across the parameter sweep, MultiCon and APCEMM exhibit similar distributions for most plume diagnostics. In terms of computational cost, CoCiP and MultiCon both operate in the range of tenths of a second to a few seconds per plume simulation, with MultiCon typically requiring approximately 2–3 times more computational time than CoCiP while remaining in the same order of magnitude. By comparison, for comparable temporal and vertical resolutions, APCEMM requires runtimes on the order of minutes, potentially approaching an hour or more depending on the particle-size bin resolution. Finally, owing to MultiCon's combination of spatially resolved quasi-2D cross-sectional dynamics, selected microphysical processes, and computational efficiency, the solver provides a basis for future global-scale contrail assessments.

	\endgroup 
		
\section{Introduction}
Contrails, visible ice‐crystal clouds generated by aircraft engine exhaust, are now considered to be one of the leading anthropogenic contributors to radiative forcing, with a warming effect at least comparable to that of {aviation-related} CO{$_2$} emissions, albeit with a high level of uncertainty \cite{1}. After rapid nucleation process and initial growth phase, contrails may persist for hours, undergoing complex interactions between ambient winds, turbulent mixing, and microphysical growth processes before dissipating or transitioning into cirrus clouds. Capturing these evolution stages is essential for accurate estimation of contrail‐induced climate impacts and for the development of mitigation strategies. Contrail modeling approaches typically fall into the following categories: (i) high-resolution simulations of individual contrails; (ii) Lagrangian plume models that follow contrail segments with prescribed microphysics; and (iii) parameterized schemes in global models \cite{2}. 
	
Although contrail dynamics have been the subject of extensive research, no existing numerical framework has yet demonstrated sufficient accuracy to reproduce contrail evolution over climatological timescales, nor to rigorously quantify the associated radiative forcing (\cite{3}). Most existing models lack experimental validation and often rely on oversimplified macro- and micro-scale physics. Combined with limitations in the computational framework, this has led to inconsistent results across current contrail models (\cite{4}).

% \subsection{Physical Stages of Contrail Evolution}
% Two primary modeling frameworks can be employed to study the formation and persistence of aircraft contrails. In the Lagrangian framework, particles are followed along their trajectories and the governing dynamics are applied along each path. By contrast, the Eulerian approach describes the evolution of particle‐size (or mass) distributions via the Population Balance Equation (PBE) (\cite{38}), often cast as a coupled system of advection–diffusion equations (ADE) with suitable closure assumptions (e.g.\ \cite{37}).  In this study, we focus on the \textbf{long‐term evolution} regime (after the jet‐induced wake has decayed) and introduce a novel Eulerian formulation that conserves both particle number and mass by retaining the zeroth and first moments of the PBE.

 Contrail development progresses through three to four overlapping regimes (\cite{2}):

A. \text{Jet/vortex‐regime (vortex roll-up) and nucleation (seconds):} Immediately downstream of the engine exit, extremely high supersaturations drive rapid homogeneous or heterogeneous nucleation of very small ice crystals.  Hydrodynamic shear and coherent vortex structures dominate their dispersion and early collision/coalescence behavior.

B. \text{Intermediate vortex wake descent/break up (seconds to minutes):} As the strong jet vortices break down, crystals remain small and are mixed by decaying turbulent eddies.  Temperature and humidity start to relax toward ambient values, but settling effects are still minor compared to turbulent mixing.

C. \text{Long‐term diffusion and habit development (tens of minutes to hours):} Once the jet‐induced turbulence has dissipated, only the ambient wind and residual turbulence remain.  In this regime, two competing inertial‐turbulence mechanisms, loitering (enhanced drag from small‐eddy sampling slows settling) and preferential sweeping (partial decoupling drives larger crystals into downward‐moving regions), affect the net particle descent; while ice particles can grow to tens of micrometers featuring different habits. Notably, the net radiative forcing attributable to contrails is overwhelmingly determined during this long-term, contrail-cirrus stage.

 In high-fidelity contrail modeling, both Reynolds-Averaged Navier–Stokes (RANS) and Large Eddy Simulation (LES) frameworks are used to resolve the jet and near-wake stages of an individual contrail plume, as well as its long-term evolution. For example, Unterstrasser et al. have conducted key LES studies on the transition from young contrails to persistent contrail cirrus, including parametric analyses of contrail evolution (\cite{Simon1,Simon2}) and comparisons with natural cirrus, focusing on local-scale interactions (\cite{Simon3,Simon4}). Their work links small-scale contrail microphysics to larger-scale cirrus dynamics, emphasizing factors that control contrail lifetime and climate impact (interested readers are referred to \cite{40,41,42,43,newpaper3} for detailed high-fidelity studies of contrail formation and early plume evolution). These high-fidelity contrail models are mostly designed to simulate the temporal evolution of an individual contrail plume under idealized or synthetic environmental conditions. These models, however, are computationally very intensive and therefore impractical for global-scale analyses that must represent many flights across extensive geographical domains.  

To overcome this limitation, several low- and intermediate-fidelity models and parameterization schemes have been developed in the literature to reduce computational expense while preserving some physical processes governing contrail formation and evolution. Two notable examples are the Contrail Cirrus Prediction model (CoCiP) \cite{7} and the Aircraft Plume Chemistry, Emissions, and Microphysics Model (APCEMM) \cite{8}, both widely discussed in the aviation community.

\subsection{Review of Low- and Intermediate-Fidelity Contrail Models and Research Gaps}
CoCiP (\cite{7}) {is a zero-dimensional model in the spatially resolved sense (or one-dimensional when time is included), which represents the contrail through a Lagrangian Gaussian-plume formulation. Rather than resolving spatially distributed fields, it tracks bulk plume properties, primarily ice mass and particle number, while parameterizing the plume cross-section using a Gaussian distribution. Particle size is diagnosed from the corresponding bulk quantities. Consequently, CoCiP provides a computationally efficient representation suitable for large-scale and global applications. In our computational estimates, its runtime is approximately one tenth of a second to a few seconds for a single-plume simulation.}

In contrast, APCEMM (\cite{8}) provides an intermediate-fidelity representation by resolving the two-dimensional plume cross-section together with the particle-size distribution. Its mature-plume formulation employs a two-dimensional rectilinear grid and a sectional representation of the ice-particle population. In the configuration considered by \cite{4}, the particle-size distribution is represented using 38 bins per computational cell. This additional spatial and microphysical resolution allows APCEMM to capture the nonuniform distribution and differential settling of ice particles across the plume (\cite{12,lateapp}). {However, it also entails a considerably greater computational burden. In the configurations considered in this work, APCEMM requires runtimes ranging from minutes to hours for a single-plume simulation, depending on the spatial and particle-size resolutions and temporal discretization in the $(y,z,m,t)$ domains. Consequently, its computational cost can become restrictive for large ensembles of contrails. Regarding its numerical implementation, although APCEMM resolves spatially distributed fields on a two-dimensional grid, its computational framework relies on the sequential splitting treatment of chemical kinetics and turbulent transport rather than a single, directly coupled PDE formulation.}

{From a numerical-analysis perspective, operator splitting introduces an approximation error when the individual operators do not commute, since their sequential application does not necessarily reproduce the evolution of the combined system }(\cite{splittingmethod}). {The magnitude of this error depends on the degree of coupling between processes, the temporal resolution, and the numerical integration scheme. This becomes particularly relevant for strongly coupled nonlinear systems, motivating the investigation of alternative formulations that retain the principal physical processes within a reduced, directly coupled PDE system.}

Nonetheless, and besides the limitations described above, neither CoCiP nor APCEMM incorporate the following processes that are perceived relevant in the long-term contrail evolution: A) micro-physical habit development \footnote{{Although CoCiP does not evolve ice-crystal habit during deposition and sublimation, it prescribes a habit mixture based on the volume-equivalent spherical radius for radiative-forcing calculations.}} and B) subgrid-scale settling velocity of ice particles.
  
When it comes to habit development, it is well known that persistent contrails and contrail cirrus are dominated by faceted, non-spherical ice crystals that closely resemble those in natural cirrus clouds. Both in situ observations and remote sensing studies consistently demonstrate that contrail ice particles deviate from spherical shapes. For instance, \cite{23} report that contrails and contrail-cirrus clouds are composed almost exclusively of nonspherical ice crystals. Polarimetric lidar measurements further support this by revealing strong light depolarization signals indicative of non-spherical particles. Similarly, CALIPSO satellite retrievals confirm that non-spherical ice dominates persistent contrails~\cite{24}. \cite{25} find that depolarization ratios measured in contrails align well with theoretical predictions for mixtures of randomly oriented, nonspherical ice habits. As contrails age and spread, larger faceted crystals, including plates, columns, and rosettes, become increasingly prevalent. Complex crystal shapes such as irregular bullet rosettes have also been observed. Concurrently, the effective radii of ice crystals increase with contrail age. Recent reviews indicate that, in older contrails (age $> 120$~s), most ice crystals have effective radii ranging from 10 to 150~$\mu$m~\cite{26,27}. Field campaigns have also documented crystals reaching several hundred microns in size~\cite{25}. Specifically, once contrail ice crystals exceed a critical size (typically around $5-10\mu$m diameter), they experience microphysical behaviors analogous to those in natural cirrus clouds. Empirical evidence from in situ sampling at approximately $-61^\circ$C shows that contrail ice crystals with effective radii larger than about 10~$\mu$m predominantly exhibit habits characteristic of natural cirrus: roughly 75\% hexagonal plates, 20\% columns, and a smaller fraction of triangular plates. These habits emerge irrespective of the initial nucleation mechanism. Therefore, mature contrail cirrus transition into the same microphysical regime as natural cirrus, where habit classification based on temperature and supersaturation is appropriate~\cite{23}. However, the dominant ice crystal habits and their variation with contrail age and crystal size remain poorly understood \cite{28}.

The bulk settling velocity of ice particles is also a relevant microphysical process in long-term contrail evolution. The terminal velocity of an isolated, spherical particle (e.g.\ as estimated from Stokes' law) is often used in long-term contrail models such as in COCIP and APCEMM, but it does not capture collective, multiphase effects that determine the effective downward speed of an ensemble of crystals, particularly for contrail models that neglect explicit fluid–solid coupling. The \emph{bulk} settling velocity denotes the ensemble-averaged settling behaviour of particles embedded in a carrier fluid and thus inherently reflects particle–fluid and particle–particle interactions as well as turbulence modulation. Multiphase simulations and laboratory experiments routinely document substantial departures of ensemble settling from single-particle terminal velocities \cite{33,34,35,36}, driven by processes such as \emph{loitering}, and \emph{preferential sweeping}.

\subsection{Rationale for the MultiCon Model}

{The MultiCon model is a reduced-order Eulerian (PDE-based) framework developed to simulate the long-term evolution of aircraft contrails. Its primary objective is to minimize computational cost formulate while retaining the essential physical mechanisms governing contrail evolution through reduced-dimensional PDE system.
}

{A general Eulerian population-balance formulation for a single contrail plume involves three spatial dimensions, together with the internal coordinates describing the ice-particle population: particle mass (polydispersity) and particle geometry (habit and shape). Consequently, in a mathematical sense, the particle distribution function depends on five independent coordinates, namely, $(x,y,z,m,\chi)$, in addition to time. Directly resolving this full-dimensional problem entails a substantial computational burden, as the number of degrees of freedom increases with the discretization of each spatial and internal coordinate. This becomes particularly restrictive for global contrail simulations, which may involve millions of individual plumes. Therefore, computationally intensive single-plume models may become impractical for exhaustive fleet-scale applications. The central motivation behind MultiCon is thus to develop a low-dimensional Eulerian PDE system that retains the essential physics of the full-dimensional problem while incorporating the effects of unresolved dimensions through appropriate mathematical model-reduction techniques.
}

{Therefore, MultiCon aims to establish an intermediate-fidelity framework that combines the computational efficiency required for global-scale applications with a spatially resolved representation of the essential contrail dynamics. Specifically, MultiCon retains the vertical coordinate ($z$) as its principal explicitly resolved transport dimension, resulting in a reduced $(z,t)$ PDE system. This choice is particularly important because vertical particle redistribution is strongly influenced by settling dynamics, whose velocity depends nonlinearly on particle mass, geometry, and ambient conditions. The remaining dimensions and physical processes are incorporated through appropriate reduced-order representations, including modeling of horizontal spreading (via the method of characteristics and quasi-Gaussian kernel) and particle-internal properties. In this manner, MultiCon seeks to retain the essential nonlinear physics of long-term contrail evolution without explicitly resolving the complete multidimensional problem. 
}
{Furthermore, MultiCon incorporates ice-crystal habit dynamics through an additional advection–reaction equation governing the evolution of the particle shape index. This formulation employs an approximate inherent growth factor (IGF) parameterization, coupled with a diffusion-limited, habit-dependent growth and sublimation model incorporating latent-heat effects. Another feature is its low-order closure for the effective bulk settling velocity, which accounts for the competing effects of subgrid-scale turbulence-induced loitering and preferential sweeping. These mechanisms are captured through a Burgers-type model describing the transition from reduced settling velocities under turbulent entrainment towards the particle terminal velocity.
}

% \hl{Finally, for horizontal plume evolution ($x-y$ or equivalently $lat-lon$ coordinates), MultiCon offers two alternative approaches: Eulerian advection–diffusion and Lagrangian tracking. While the former can capture horizontal advection–diffusion coupling and interactions between neighbouring plumes, the latter is more suitable for isolated single-plume assessments. Importantly, the choice between these approaches becomes relevant when extending the framework to aircraft-level simulations involving continuous emissions. Importantly, this choice does not affect the microphysical treatment in single-plume analysis, which remains resolved through the quasi-two-dimensional Eulerian PDE system}

\section{General Mathematical Framework for MultiCon}
\subsection{Modeling Aspects and Assumptions of the Current MultiCon Version}

The development of the present MultiCon formulation relies on the following modeling assumptions and dimensional-reduction procedures:

First, we assume the dominance of ambient wind-driven advection during the long-term evolution of the contrail. Transient aerodynamic effects, including aircraft-induced jets and wake vortices, are regarded as short-term phenomena and are not explicitly resolved during the long-term simulation phase. Their influence is instead incorporated into the initial conditions of the mature plume. Subsequent particle transport is governed primarily by ambient winds, and settling, considering turbulent effects.

{Second, a dimensional-separability approximation is introduced to reduce the governing system. Specifically, following the derivation of the main population-balance and micro-physical PDE system, a separation-of-variables ansatz is applied to obtain the reduced $(z,t)$ PDE system adopted in MultiCon. This allows the principal vertical dynamics to be resolved separately from horizontal plume spreading, without explicitly discretizing the complete spatial domain.
}

Third, ice-particle geometry is assumed to be spheroidal \cite{igf1,igf2}, which allows complex ice-crystal habits, including hexagonal morphologies, to be approximated by equivalent spheroids. Their influence on particle growth, sublimation, and settling is retained through an appropriate shape index.

{Fourth, each computational cell is characterized by a representative ice-particle mass, number concentration, mass concentration, and shape index. Accordingly, the particle population is assumed to be locally monodisperse, with all particles in a given cell represented by a single characteristic mass and geometry. In the population-balance formulation, this corresponds to representing the distribution in the internal coordinates using Dirac delta distributions centered at the local representative mass and shape index. Although the general framework can accommodate polydispersity, the current implementation adopts this locally monodisperse approximation. 
}

{Fifth, lateral plume evolution is approximated using the method of characteristics together with a quasi-Gaussian kernel in a bulk representation. Specifically, the resolved vertical evolution defines the characteristic trajectories, while the unresolved lateral distribution is reconstructed through a quasi-Gaussian representation of the plume cross-section. This avoids introducing an additional explicitly resolved spatial dimension. Notably, the characteristic formulation can also accommodate altitude- and time-dependent wind-shear profiles.
}

{Sixth, due to the dimensional-separability approximation, the nonlinear particle-number loss associated with aggregation cannot be directly incorporated as a fully coupled local source/sink term in the reduced $(z,t)$ PDE system. Instead, the number-concentration field is treated conservatively during transport, while its overall normalization is updated at each time step to account for the reduced number of available ice particles.}

	\subsection{Governing Monodisperse Equations for Bulk Motion}  	
	Based on the above assumptions, we begin modeling the evolution of persistent contrails by the following system of coupled advection--diffusion equations (see Appendix \ref{apB}):  
	\begin{equation}\label{eq1}
		\begin{split}
			\frac{\partial c_N}{\partial t} +   \nabla\cdot\big(\mathbf{v}_p\, c_N\big) &= \nabla\cdot \bigl({{K}}\,\nabla c_N\bigr) + S_{c_N},\\[1mm]
			\frac{\partial c_M}{\partial t} +  \nabla \cdot \big(\mathbf{v}_p\, c_M \big)&= \nabla\cdot \bigl({{K}}\,\nabla c_M\bigr) + c_N\,\rho_\text{dep}\,f_\mathcal{V} + S_{c_M}.
		\end{split}
	\end{equation}  
	In the equations above, $t$ is time, $\mathbf{x}:=(x,y,z)$ is the system coordinate in a Cartesian framework, $c_N:=c_N(\mathbf{x},t)$ and $c_M:=c_M(\mathbf{x},t)$ represent the ice-particle number and mass concentrations, respectively. The bulk velocity of the ice particles is given by $\mathbf{v}_p:=\mathbf{v}_p(\mathbf{x},t)$, and the diffusion coefficient matrix is denoted by ${{K}}$. 
    Moreover, $S_{c_N}:=S_{c_N}(\mathbf{x},t)$ and $S_{c_M}:=S_{c_M}(\mathbf{x},t)$ represent the sink terms for the ice-particle number and mass concentrations, respectively \footnote{As discussed earlier, in the present study the sink term is not modeled directly inside the PDEs (i.e., \(S_c=0\)).}. Finally, $f_\mathcal{V}:=f_\mathcal{V}(a,\phi,\mathbf{x},t)$ represents the volumetric growth rate of individual particles, $\rho_\text{dep}$ denotes the effective deposition density, where $a$ is the ice particle's equatorial radius and $\phi$ is the ice-particle shape index, which will be elaborated in the next section.
	
	\subsection{Governing Equations for Individual Ice Particles' Microphysics}  
    Modeling persistent contrails also requires microphysical equations that describe the evolution of individual ice particles' mass and shape. Accurate parameterization of contrail single-scattering properties, and thus reliable simulation of contrail and contrail-cirrus radiative forcing, requires explicit representation of the crystal nonsphericity (\cite{23}). Employing the habit dynamic framework by \cite{igf1}, and \cite{igf2}, the coupled ODEs describing the evolution of ice crystal mass and shape can be written as: 
	\begin{equation}\label{eq2}
		\begin{split}
			\frac{dm}{dt} = \rho_\text{dep}\,f_\mathcal{V}(a,\phi,\mathbf{x},t),\qquad 
			\frac{d\phi}{d\mathcal{V}} = \frac{\Gamma^*(T,s_i)-1}{\Gamma^*(T,s_i)+2}\,\frac{\phi}{\mathcal{V}},\qquad
			\mathcal{V} = \frac{4\pi}{3}\,a^3\,\phi.
		\end{split}
	\end{equation} 
	In the above, $m$ is the individual ice particle mass, $\mathcal{V}$ corresponds to the volume of an individual ice particle, while $\phi$ represents the individual ice-particle shape index, defined as $\phi := \frac{c}{a}$, where $a$ is the equatorial radius and $c$ is the transverse radius; $\phi > 1$ indicates columnar (prolate) crystals and $\phi < 1$ denotes plate-like (oblate) crystals. $T:=T(\mathbf{x},t)$ is the spatiotemporal temperature field, and $s_i:=s_i(\mathbf{x,t})$ represents the spatiotemporal background/ambient ice supersaturation field. In addition, $\Gamma^*$ is known as Inherent Growth Factor (IGF), and ($f_\mathcal{V}:=f_\mathcal{V}(a,\phi,\mathbf{x},t)$) denotes the volumetric growth rate of individual ice particles (see Appendix \ref{apE}, and \ref{apF})\footnote{Notably, parameterizations of ice-vapor growth exert a first-order control on cold-cloud behavior \cite{10,20}. While laboratory and theoretical advances have yielded useful, approximate growth models, the detailed physical processes governing ice-crystal development remain incompletely characterized. In addition, the ice crystal quantities measured at different temperatures cannot be directly incorporated into the capacitance model, a formulation that nonetheless underlies many atmospheric ice-growth simulations (see e.g., \cite{21,13}). This inconsistency highlights a persistent gap between process-oriented laboratory observations and the bulk parameterization employed in large-scale atmospheric models.}. 
	
%	Individual ice-particle volumetric growth rate ($f_\mathcal{V}:=f_\mathcal{V}(a,\phi,\mathbf{x},t)$) is given by:  
%	\begin{equation}
%		\begin{split}
%			f_\mathcal{V}(\mathbf{x},a,\phi,t) &= \frac{4\pi\,C\,s_i}{\rho_{dep}\left(\frac{R_vT}{e_iD_v} + \frac{L_s}{T\,k_a}\Bigl(\frac{L_s}{R_vT}-1\Bigr)\right)},
%		\end{split}
%	\end{equation}  
%	where $T$ is the temperature, $C:=C(a,\phi)$ is the crystal capacitance, $s_i:=s_i(T)$ is the ice supersaturation, $R_v$ is the water vapor gas constant, $L_s:=L_s(T)$ is the latent heat of sublimation, $e_i:=e_i(T)$ is the saturation vapor pressure with respect to ice, $D_v:=D_v(T,P)$ is the water vapor diffusivity in air with $P=P(T)$ being the pressure, and $k_a$ is the thermal conductivity of air. 
%	
%	Moreover, $\Gamma(T,s_i)=\frac{\alpha_c(T,s_i)}{\alpha_a(T,s_i)}$ determines the distribution of mass deposition onto the crystal faces based on the associated growth ratios (i.e., the inherent growth ratio or habit); see \cite{16} for details. 
%	
%	Based on the available data in \cite{17,18}, we approximate $\Gamma(T,s_i)$ using a neural network function for better accuracy, expressed as $\Gamma(T,s_i) \approx \Gamma^{\mathcal{N}}(T,s_i)$. 

    For completeness, we note that laboratory evidence suggests the aspect ratio of ice crystals remains constant during sublimation. This is attributed to the uniformity of vapor density along the crystal surface, which results in shape preservation throughout the sublimation process \cite{13,14,15}.

\subsection{{Ice supersaturation} \(s_i(\mathbf{x},t)\)}
The local ice supersaturation is defined in terms of the water-vapor mixing ratio as $s_i(\mathbf{x},t)=\frac{q_v(\mathbf{x},t)}{q_s(T,p)}-1$, with $q_s(T,p)=\frac{\epsilon e_{si}(T)}{p-e_{si}(T)},
\epsilon=\frac{R_d}{R_v},$ where \(q_v\) is the water-vapor mixing ratio, \(q_s\) is its saturation value over ice, \(e_{si}(T)\) is the saturation vapor pressure over ice, \(p\) is the ambient pressure, and \(R_d\) and \(R_v\) are the specific gas constants of dry air and water vapor, respectively.

Using the conservative water-vapor balance and the air-continuity equation, the governing evolution equation for ice supersaturation can be expressed as (see Appendix G for the complete derivation):
\begin{equation}
    \frac{D_a s_i}{Dt}
=
-\frac{S_{\rm dep}}{\rho_{\rm air}q_s}
-\frac{1+s_i}{q_s}\frac{D_a q_s}{Dt}
+\frac{1}{\rho_{\rm air}q_s}
\nabla\cdot\left[
\rho_{\rm air}K_v\nabla\big(q_s(1+s_i)\big)
\right].
\end{equation}

Here, \(D_a/Dt=\partial/\partial t+\mathbf{u}\cdot\nabla\) denotes the material derivative following the carrier air, \(\rho_{\rm air}\) is the air density, and \(K_v\) is the effective molecular–turbulent diffusivity tensor of water vapor. Furthermore, \(S_{\rm dep}=c_N\rho_{\rm dep}f_V\) represents the volumetric rate of water-vapor depletion due to ice deposition, where \(\rho_{\rm dep}\) is the effective deposition density and \(f_V\) is the volumetric growth rate of individual ice particles.

In the present MultiCon implementation, explicit water-vapor diffusion is neglected in the supersaturation budget, and consequently \(K_v=0\) \footnote{ $K_v$ can be incorporated through an effective diffusivity, accounting for molecular and turbulent diffusivities: $D_{v,m}$, and $\frac{\nu_{t}}{Sc_{t,v}}$, where \(D_{v,m}\) is the molecular diffusivity of water vapour in air, \(\nu_{t}\) is the turbulent eddy viscosity, and \(Sc_{t,v}\) is the turbulent Schmidt number. The more challenging part is estimating turbulent Schmidt number, since it is flow-dependent. Specifically, the corresponding diffusion contribution to the supersaturation equation, obtained using \(q_v=q_s(1+s_i)\), and for locally uniform \(q_s\), density and diffusivity is \(K_{v}\partial^2s_i/\partial z^2\). Nonetheless, the present MultiCon version neglects this explicit diffusive contribution, and sets \(K_{v}=0\) in the supersaturation equation.}. 
    
	\subsection{General Equations for Single-Plume Evolution}
     As discussed, we can obtain an ADE accounting for the evolution the mass field $m(\mathbf{x},t)$ directly (see the derivation details in Appendix \ref{apB})\footnote{It is noteworthy that the equation governing $m(\mathbf{x},t)$ can also be derived directly from the quotient rule $c_M := m\,c_N$ by defining the net inlet flux into the control volume as $\mathbf{J}_{in}:=m(\mathbf{x},t)c_N(\mathbf{x},t)$, and the net outlet flux expressed as $\mathbf{J}_{out}=\mathbf{J}_{in}+\frac{\partial}{\partial \mathbf{x}_i}\mathbf{J}_{in}$. The derivation is accomplished on using the conservation of the number concentration as the closure equation.}. However, it is not possible to directly formulate an ADE for particle shape evolution in the same sense as for mass concentration. Although a particle entering a control volume transports both its mass and specific shape, only mass is a conserved scalar obeying continuity laws. In contrast, particle shape evolves through micro-physical processes that fundamentally differ from diffusion. Therefore, in the Eulerian framework for particle shape evolution, only an advection term is incorporated to ensure the presence of the shape index $\phi(\mathbf{x},t)$ in all regions where particles are present, followed by a term describing its micro-physical evolution that determines the rate at which particle shapes change. Therefore, the Eulerian equations governing the long-term evolution of contrails are presented as:
	\begin{equation}\label{eq1_1}
		\begin{split} 
			&\frac{\partial c_N}{\partial t} +  \nabla \cdot\big(\mathbf{v}_p c_N\big) = \nabla\cdot \bigl({{K}}\,\nabla c_N\bigr),\\
			&\frac{\partial c_M}{\partial t}
    + \nabla\cdot\left(\mathbf{v}_p c_M\right)
    =
    \nabla\cdot\left({{K}}\nabla c_M\right)
    + \rho_\text{dep}\,
    f_\mathcal{V}(\bar{m},\bar{\phi},\mathbf{x},t)c_N,\\
			&\frac{\partial \phi}{\partial t} + \mathbf{v}_p \cdot \nabla \phi = \frac{\Gamma^*(T,s_i)-1}{\Gamma^*(T,s_i)+2}\,\frac{\phi}{\mathcal{V}}\,f_\mathcal{V},\quad \mathcal{V} = \frac{4\pi}{3}\,a^3\,\phi,\\
            &  \frac{D_a s_i}{Dt}
=-\frac{S_{\rm dep}}{\rho_{\rm air}q_s}
 -\frac{1+s_i}{q_s}\frac{D_a q_s}{Dt}
 +\frac{1}{\rho_{\rm air}q_s}
  \nabla\!\cdot\!\left[
    \rho_{\rm air}\mathbf{K}_v
    \nabla\!\left(q_s(1+s_i)\right)
  \right].
		\end{split}
	\end{equation}

    In the following section, we present the theory and formulas for the remaining building block of the Eulerian model, which includes the ice particle velocity $\mathbf{v}_p$.

\section{Ice Particle Velocity \( \mathbf{v}_p \) and Bulk Settling Velocity \(v_s\)} 
The particle velocity field entering the governing equations comprises the local ambient wind velocity and the effective slip velocity of the particle phase relative to the carrier fluid. Formally, the particle slip velocity is defined as $\mathbf{v}_{slp}=\mathbf{v}_p-\mathbf{w}$, where $\mathbf{w}$ denotes the local carrier-air velocity. {For the long-term contrail evolution considered here, the mean particle–air slip is approximated as predominantly vertical, as gravitational settling constitutes its leading-order contribution. Accordingly, the total ice-particle bulk velocity field is defined as:}
\begin{equation}\label{eq1s}
    \mathbf{v}_p = \mathbf{v}_{{slp}} +
    \underbrace{(w_x,w_y,w_z)^\top}_{\scriptsize\shortstack{\text{background velocity} \\ \text{(wind components)}}}
    \approx (0,0,v_s)^\top +
    \underbrace{(w_x,w_y,w_z)^\top}_{\text{background velocity}}
    = (w_x,w_y,w_z+v_s)^\top.
\end{equation}

where $\mathbf{v}_p:=\mathbf{v}_p(\mathbf{x},m,\phi,t)$ is the particle bulk velocity, $\mathbf{v}_{slp}$ represents the bulk slip velocity of the particle phase relative to the local carrier air, while $\mathbf{w}:=(w_x,w_y,w_z)$ denote the ambient wind components and $v_s$ the effective bulk settling velocity.{ In particular, altitude-dependent variations in the horizontal wind components, including vertical wind shear, are retained in the carrier-air velocity field and are accounted for separately through the characteristic-based cross-sectional formulation presented in Sec. 4.1.}

{In the present implementation, unresolved vertical air motions are represented through a bounded, spatially stochastic velocity field sampled from a Gaussian random distribution}, \(w_z(z,t)\), {with an imposed magnitude limit of} \(0.05\,\mathrm{m\,s^{-1}}\). {Accordingly, the vertical particle velocity is given by} \(v_{p,z}=w_z+v_s\), where \(v_s\) {denotes the signed bulk settling velocity.}

Unlike the terminal velocity of an isolated particle, the bulk settling velocity accounts for the collective behavior of ice particles subject to turbulent mixing. In particular, it incorporates the competing effects of loitering, which reduces the effective settling rate through turbulent entrainment, and preferential sweeping, which can enhance downward transport through inertial interactions with turbulent eddies. In MultiCon, these mechanisms are represented through a reduced-order Burgers-type equation for the bulk settling velocity:
\begin{equation}
\begin{split}
&\frac{\partial v_s}{\partial t}
+v_s\frac{\partial v_s}{\partial z}
=\nu_{t,\mathrm{ef}}\frac{\partial^2v_s}{\partial z^2},\\
&v_s(z_{\mathrm{ref}},t=0)=0,\qquad
v_s(z\ll z_{\mathrm{ref}},t=0)=v_{\mathrm{ter}},
\qquad
\lim_{t\to\infty}v_s(z,t)=v_{\mathrm{ter}}.    
\end{split}
\end{equation}

Here, \(\nu_{t,\mathrm{ef}}\) is the effective turbulent eddy viscosity, \(z_{\mathrm{ref}}\) denotes the reference height, and \(v_{\mathrm{ter}}\) is the terminal settling velocity of an individual ice particle. The physical assumptions underlying this closure, together with the reduction of the particle-phase momentum equation, the estimation of the turbulent relaxation length, the corresponding closed-form solution, and the terminal-velocity parameterization are detailed in Appendix \ref{apB}; whereas the performance of the bulk-settling-velocity formulation is presented as a preliminary assessment in Appendix \ref{apC}.

	{\section{{MultiCon Reduced Model and Solution Methodology for Single-Plume Evolution}}\label{sec5}
    Although the governing equations (Eqs. \ref{eq1_1}) are inherently three‐dimensional, requiring a fully resolved 3D mesh and often forcing a trade‐off between resolution and domain size, the system admits a separable structure under physically justified assumptions. Exploiting this separability permits high‐resolution solutions over extended and sensitive vertical domain without resorting to coarse‐grained approximations and high computational times.
		
	We initialize the contrail by releasing ice particles into a three‐dimensional region immediately following the earlier stages that the contrail has been through (roughly, in the order of below 1 minute after the plume birth), prescribing initial fields as:
	\begin{equation}
		c_N(\mathbf{x},0)=c_{N,0}, 
		\quad
		c_M(\mathbf{x},0)=c_{M,0},
		\quad
		\phi(\mathbf{x},0)=\phi_0.
	\end{equation}
    For single-plume analysis, we assume $	T=T(z,t)$, and, $s_i=s_i(z,t)$, yielding  $f_\mathcal{V}=f_\mathcal{V}(z,t)$ and $m=m(z,t)$.  Therefore, we adopt the following separable ansatz:
	\begin{equation}
		c_N(x,y,z,t) \;=\; F(x,y,t)\,g(z,t).
		\label{eq:separable_ansatz}
	\end{equation}
	% with a closure for the turbulent diffusivities $\widetilde{\mathcal D}_{xx}$, $\widetilde{\mathcal D}_{yy}$, and $\widetilde{\mathcal D}_{zz}$.
Here, $x$ denotes the contrail direction, $y$ the horizontal direction normal to the contrail axis, and $z$ the vertical direction.

  Enforcing vanishing fluxes at \(x,y\to\pm\infty\) and \(z\to\pm\infty\) together with the normalization
	\(\displaystyle\int_{-\infty}^{+\infty}g(z,t)\,\mathrm{d}z=1\), and the system decouples into:}
    {
	\begin{equation}\label{eq:decoupled_system}
		\begin{split} 
&\left\{
\begin{aligned}
\frac{\partial F}{\partial t}
+ w_x(x,y,t)\frac{\partial F}{\partial x}
+ w_y(x,y,t)\frac{\partial F}{\partial y}
&=
{ K}_{x}
\frac{\partial^2F}{\partial x^2}
+
{ K}_{y}
\frac{\partial^2F}{\partial y^2}
\end{aligned}
\right\} \rightarrow \textbf{horizontal ADE (Eulerian tracker)}.
			\\[2mm]
			&\left\{
\begin{aligned} &\frac{\partial g}{\partial t}
			+\frac{\partial}{\partial z}\left(v_{p,z}(z,t)g\right)
			=
			{ K}_{z}
			\frac{\partial^2 g}{\partial z^2},
			\\
			&\frac{\partial M}{\partial t}
+\frac{\partial}{\partial z}\left(v_{p,z}(z,t)M\right)
=
K_z\frac{\partial^2 M}{\partial z^2}
+\rho_{\mathrm{dep}}f_{\mathcal{V}}(z,t) g,
			\\
			&\frac{\partial \phi}{\partial t}
			+ v_{p,z}(z,t)\,\frac{\partial \phi}{\partial z}
			=
			\frac{\Gamma^*\bigl(T(z,t),s_i(z,t)\bigr)-1}{\Gamma^*\bigl(T(z,t),s_i(z,t)\bigr)+2}
			\;\frac{\phi}{\mathcal{V}}\;f_\mathcal{V}(z,t),\\
			&\frac{\partial s_i}{\partial t}+w_z\frac{\partial s_i}{\partial z}
			= -\frac{1}{q_s}\frac{S_{\rm dep}}{\rho_{\rm air}}
			- \frac{1+s_i}{q_s}\frac{D q_s}{D t}
			+ \frac{s_{\rm env}-s_i}{\tau_{\rm entrain}},\end{aligned}
\right\} \rightarrow \textbf{plume evolution in \(z\)-\(t\) space}.
		\end{split}
	\end{equation}
    }

where \(M=m(z,t)g(z,t)\) represents the mass-weighted
vertical distribution of the ice particles, 
and \(g\) the normalized vertical particle-number
distribution. The volumetric deposition rate is given by
\(S_{\mathrm{dep}}=c_{N}\rho_{\mathrm{dep}}
f_\mathcal{V}\), where \(c_{N}\) is
the three-dimensional particle-number concentration.

{The particle-number
concentration employed in the present implementation is:}
\begin{equation}
c_{N}
=
\frac{n_{\ell}(t)}{B(t)}
f_{\mathrm{cw}}(t)g(z,t),
\qquad
f_{\mathrm{cw}}(t)
=
\operatorname{erf}
\left(
\frac{B(t)}{2\sqrt{2}\sigma_y(t)}
\right),
\end{equation}
where \(n_{\ell}(t)\) represents the line-integrated particle number (which is treated separately
through bulk turbulent and aggregation losses), \(B(t)=\sqrt{8}\sigma_y(t)\) is the effective
crosswind width, \(\sigma_y(t)\) is the crosswind standard
deviation, and \(f_{\mathrm{cw}}\) is a finite-width
Gaussian correction. This correction accounts for the
fraction of the Gaussian crosswind distribution contained
within the effective plume width. 

The supersaturation equation incorporates the local
depletion associated with ice growth and a reduced
relaxation term representing the exchange of water vapor
between the plume and its environment. Here,
\(s_{\mathrm{env}}\) denotes the ambient ice
supersaturation, and \(\tau_{\mathrm{entrain}}\) is the
effective environmental relaxation timescale. 

\textbf{Remark 1:}
In the present single-plume implementation, the vertical
evolution of the particle-number distribution, mass
distribution, particle shape, and ice supersaturation is
solved on a one-dimensional vertical grid. The transport
equations are discretized using implicit upwind
advection schemes and solved through a tridiagonal
matrix algorithm (TDMA).

\textbf{Remark 2:}
Although the dimensional-separability formulation
admits a horizontal Eulerian advection--diffusion
equation, this equation is not explicitly solved in
the present single-plume implementation. Instead,
the horizontal plume extent is represented through
a quasi-Gaussian crosswind approximation, while the
effect of vertical wind shear on the cross-sectional
geometry is incorporated through a one-dimensional
ensemble of characteristic trajectories. This
approach retains the principal effects of horizontal
spreading and shear-induced cross-sectional distortion
without requiring an additional explicitly resolved
spatial dimension.

\begin{figure}[H]
	\centering
	\includegraphics[width=0.9\textwidth]{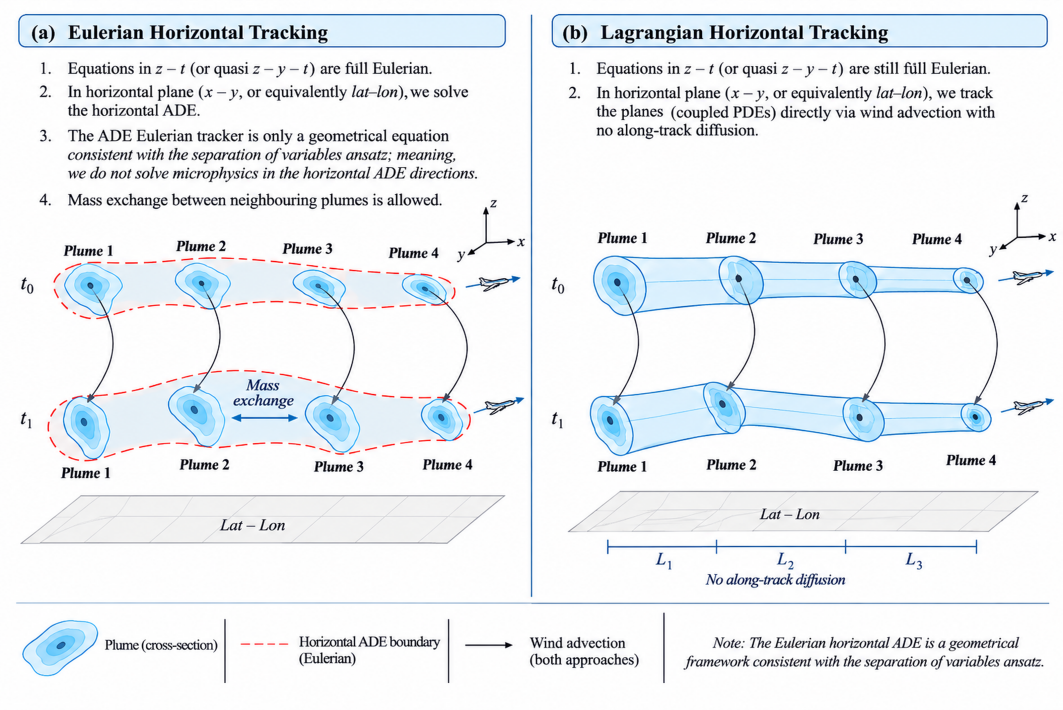}
     \caption{MultiCon solves the vertical plume evolution in the same Eulerian way in both modes and differs only in how mass is exchanged between multiple co-existing plumes. This is applicable to the case of single-flight simulations: it is outside the scope of the present single-plume study. Schematic of the two horizontal tracking modes, with four plume
cross-sections along a flight track shown at two times $t_0$ and $t_1$ above the latitude--longitude plane; $x$ is the
flight direction, $y$ the crosswind direction and $z$ the vertical. (a) Eulerian horizontal tracking: a horizontal
advection--diffusion equation (ADE) is solved over the region enclosed by the dashed orange line; it is a geometrical
tracker consistent with the separation-of-variables ansatz, no microphysics is solved in the horizontal directions, and
neighbouring plumes can exchange mass. (b) Lagrangian horizontal tracking: each cross-section is advected
with the wind, the segments of lengths $L_1$ to $L_3$ between them are not diffused along the track, and plumes do not
exchange mass.}
	\label{trackingmodes}
\end{figure}

\textbf{Remark 3:}
The horizontal Eulerian tracker
 remains a possible alternative for
single-flight simulations involving multiple
coexisting plume segments. In such applications,
it can represent horizontal advection--diffusion
coupling and mass exchange between neighboring
plumes, which are not explicitly resolved in the
present isolated single-plume calculation (see Fig. \ref{trackingmodes}). 
% In other words, the Eulerian tracker extends the geometrical resolution of a contrail model, and not its microphysical resolution; meaning that the microphysics still only appear as reaction terms in z-t PDEs. 

\textbf{Initialization:} The initial time \(t=0\) corresponds to the handoff
from the wake-vortex regime to the long-term plume
evolution considered here. Consequently, the solver
requires initial estimates of the vertical
particle-number distribution \(g(z,0)\), the
individual particle mass \(m(z,0)\), the particle
shape index \(\phi(z,0)\), the line-integrated
particle number \(n_{\ell}(0)\), and the initial
crosswind standard deviation \(\sigma_y(0)\).
Assuming an initially spherical particle population
and a Gaussian vertical distribution, these
initial conditions are specified as follows:

1) The initial representative particle radius
\(r_0\) determines the individual particle mass, $m_{i,0}=\frac{4}{3}\pi r_0^3\rho_{\mathrm{ice}}.$

2) The initial vertical standard deviation
\(\sigma_{z,0}\) determines the normalized
Gaussian particle-number distribution, $g(z,0)=
\frac{1}{\sqrt{2\pi}\sigma_{z,0}}
\exp\left[
-\frac{(z-z_r)^2}{2\sigma_{z,0}^2}
\right],$ where \(z_r\) denotes the initial reference height
of the plume.

3) The initial line-integrated particle number
\(n_{\ell,0}\), together with the initial crosswind
standard deviation \(\sigma_{y,0}\), determines
the initial nominal particle-number concentration
through the cross-sectional reconstruction described
above. The initial particle shape is prescribed as
\(\phi(z,0)=1\), corresponding to spherical crystals. These initial quantities can be obtained from
existing early-stage contrail parameterizations,
such as CoCiP (\cite{7}), or prescribed
according to the particular simulation case.

We develop the characteristic-based cross-sectional
representation and its associated covariance
description in Section 4.1. The resulting effective
plume dimensions and area are subsequently coupled
to the entrainment timescale and the
geometrical contribution to the bulk particle-number
loss. The latter is formulated separately in Section 4.2.

% ---- previous version of this figure (kept for reference) ----
% ---- TikZ version ----
% \begin{figure}[H]
%  \centering
%  \includegraphics[width=0.9\textwidth]{Figures/Paper/fig01_tracking_modes}
%  \caption{MultiCon solves the vertical plume evolution in the same Eulerian way in both modes and differs only in how plume
% cross-sections are carried in the horizontal plane. Schematic of the two horizontal tracking modes, with four plume
% cross-sections along a flight track shown at two times $t_0$ and $t_1$ above the latitude--longitude plane; $x$ is the
% flight direction, $y$ the crosswind direction and $z$ the vertical. (a) Eulerian horizontal tracking: a horizontal
% advection--diffusion equation (ADE) is solved over the region enclosed by the dashed orange line; it is a geometrical
% tracker consistent with the separation-of-variables ansatz, no microphysics is solved in the horizontal directions, and
% neighbouring plumes can exchange mass (double arrow). (b) Lagrangian horizontal tracking: each cross-section is advected
% with the wind, the segments of lengths $L_1$ to $L_3$ between them are not diffused along the track, and plumes do not
% exchange mass. Blue contours: plume cross-section; curved black arrows: wind advection from $t_0$ to $t_1$. Not to scale.}
%  \label{trackingmodes}
% \end{figure}

\subsection{{Cross-sectional geometry and wind-shear coupling}}
Let us retain the normal
wind shear can be represented by:
\begin{equation}
      u_y(z,t)=u_y(z_r,t)+S_y(z-z_r),
  \qquad
  S_y=\frac{\partial u_y}{\partial z}.
\end{equation}
The height-independent component translates the cross-section without
changing its shape. Streamwise shear $S_x=\partial u_x/\partial z$ is not
included because its effect is along the unresolved $x$ direction and
would require along-track concentration gradients and interactions
between neighbouring plume segments, which are outside the present
single-plume formulation.

The shear-induced lateral displacement is estimated from the
characteristic system:
\begin{equation}
      \frac{dZ_q}{dt}=\mathbf{v}_{p,z}(Z_q,t),
  \qquad
  \frac{dY_q}{dt}=S_y\,[Z_q(t)-z_r],
  \qquad
  Z_q(0)=z_q,\quad Y_q(0)=0,
\end{equation}
where $\mathbf{v}_{p,z}$ is the total vertical particle velocity, and $z_r$ is the initial number-weighted reference height. Mapping the characteristic
displacements onto the current vertical grid gives $y_s(z,t)$. Combining
this displacement with the current vertical number profile defines the
cross-sectional covariance matrix $\boldsymbol{\Sigma}(t)$ described in
Appendix~H. The associated Gaussian-equivalent area and projected
crosswind and vertical dimensions are:
\begin{equation}\label{eq12}
      A(t)=2\pi\sqrt{\det\boldsymbol{\Sigma}(t)},
  \qquad
  B(t)=\sqrt{8\sigma_{yy}(t)},
  \qquad
  D(t)=\sqrt{8\sigma_{zz}(t)}.
\end{equation}
The ellipse serves solely as a moment-based geometrical descriptor. For visualization purposes however, a two-dimensional Gaussian-equivalent cross-sectional representation is reconstructed, centered at the characteristic displacement and parameterized by the crosswind variance obtained from the same spread–shear decomposition. 

\textbf{Water Vapor Mixing Closure}: The full-dimensional water-vapour balance in Sec. 2.4 accounts for vapour depletion by ice deposition and, in its general form, spatially resolved transport and exchange with the surrounding atmosphere. Following the dimensional-separability approximation, however, the cross-sectional vapour fluxes are no longer explicitly resolved by the reduced \((z,t)\) formulation. In the present implementation, explicit water-vapour diffusion is also neglected. To retain the leading-order effect of the unresolved plume–environment exchange, we introduce a reduced mixing tendency (residual):
\begin{equation}
    \mathcal R_{\mathrm{mix}}^{\mathrm{red}}
=
\frac{s_{i,\mathrm{env}}-s_i}{\tau_{\mathrm{entrain}}},
\qquad
s_{i,\mathrm{env}}=\frac{q_{v,\mathrm{env}}}{q_s}-1,
\end{equation}
where \(s_{i,\mathrm{env}}\) is the ambient ice supersaturation sampled at the plume-element location and \(\tau_{\mathrm{entrain}}\) is the effective entrainment timescale. The residual represents the net tendency of the unresolved ambient vapour exchange to relax the plume supersaturation toward its environmental value. 

To establish its position in the governing supersaturation equation, we introduce a general mixing operator \(\mathcal E_{\mathrm{mix}}[q_v]\), representing the corresponding unresolved vapour-mixing contribution. The reduced water-vapour balance can then be expressed in material form as:
\begin{equation}
    \frac{D_a q_v}{Dt}
=
-\frac{S_{\mathrm{dep}}}{\rho_{\mathrm{air}}}
+\mathcal E_{\mathrm{mix}}[q_v],
\end{equation}
where \(D_a/Dt\) denotes the material derivative following the carrier air. Differentiating the identity \(q_v=q_s(1+s_i)\) along the air motion, one arrives at:
\begin{equation}
    \frac{D_a s_i}{Dt}
=
-\frac{S_{\mathrm{dep}}}{\rho_{\mathrm{air}}q_s}
-\frac{1+s_i}{q_s}\frac{D_a q_s}{Dt}
+\frac{\mathcal E_{\mathrm{mix}}[q_v]}{q_s}.
\end{equation}

Equation (18) shows that the unresolved vapour-mixing contribution enters the supersaturation balance as an additive source/sink term, normalized by the local saturation mixing ratio \(q_s\). A convenient bulk closure for this contribution is the first-order relaxation model:
\begin{equation}\label{eq16}
    \mathcal E_{\mathrm{mix}}[q_v]
=
\frac{q_{v,\mathrm{env}}-q_v}{\tau_{\mathrm{entrain}}},
\end{equation}

where \(q_{v,\mathrm{env}}\) is the ambient water-vapour mixing ratio and \(\tau_{\mathrm{entrain}}\) characterizes the rate of exchange between the plume and its surroundings. On using \(q_v=q_s(1+s_i)\), we get:
\begin{equation}
    \frac{D_a s_i}{Dt}
=
-\frac{S_{\mathrm{dep}}}{\rho_{\mathrm{air}}q_s}
-\frac{1+s_i}{q_s}\frac{D_a q_s}{Dt}
+\frac{1}{\tau_{\mathrm{entrain}}}
\left(\frac{q_{v,\mathrm{env}}}{q_s}-1-s_i\right).
\end{equation}

By defining \(s_{i,\mathrm{env}}=q_{v,\mathrm{env}}/q_s-1\), the final term is precisely the reduced mixing residual, i.e., $  \frac{1}{\tau_{\mathrm{entrain}}}
\left(\frac{q_{v,\mathrm{env}}}{q_s}-1-s_i\right)=  \mathcal R_{\mathrm{mix}}^{\mathrm{red}}
=
\frac{s_{i,\mathrm{env}}-s_i}{\tau_{\mathrm{entrain}}}$. 

\textbf{Entrainment Timescale}: To determine the entrainment timescale, we consider a Lagrangian plume element containing an air mass \(M_a\), which entrains an additional ambient air mass \(\Delta M_a\) during a time interval \(\Delta t\). Assuming instantaneous mixing within the element, conservation of water vapour gives the post-mixing vapour mixing ratio (Schumann, 2012):
\begin{equation}
    q_v(t+\Delta t)
=
\frac{M_aq_v(t)+\Delta M_a q_{v,\mathrm{env}}}
{M_a+\Delta M_a}.
\end{equation}
Taking the differential limit, with \(\Delta M_a/M_a\ll 1\), yields:
\begin{equation}
    \frac{dq_v}{dt}
=
\frac{1}{M_a}\frac{dM_a}{dt}
(q_{v,\mathrm{env}}-q_v),
\end{equation}
which, upon comparison with Eq. (\ref{eq16}), identifies the entrainment timescale as:
\begin{equation}
    \tau_{\mathrm{entrain}}
=
\frac{M_a}{dM_a/dt}.
\end{equation}
We now relate this mass-based definition to the evolving plume geometry. The air mass contained in a plume element can be approximated by \(M_a\simeq\rho_{\mathrm{air}}L_x A(t)\), where \(L_x\) is its streamwise length and \(A(t)\) is its cross-sectional area, defined in Eq. (\ref{eq12}). Assuming approximately constant air density and streamwise length over the entrainment timescale, the relative air-mass growth rate becomes:
\begin{equation}
    \frac{1}{M_a}\frac{dM_a}{dt}
\simeq
\frac{1}{A(t)}\frac{dA}{dt}=\frac{1}{\tau_{entrain}}.
\end{equation}

% Finally, the saturation mixing ratio \(q_s\) is diagnosed from the local temperature and pressure fields, rather than treated as an independent prognostic variable. In particular, according to Eq. (3),
% \begin{equation}
%     q_s(z,t)
% =
% \frac{\epsilon e_{si}(T(z,t))}
% {p(z,t)-e_{si}(T(z,t))},
% \qquad
% \epsilon=\frac{R_d}{R_v},
% \end{equation}
% where \(e_{si}(T)\) is the saturation vapour pressure over ice. Its material derivative in Eq. (21) is therefore evaluated through the local temperature and pressure variations:
% \begin{equation}
%     \frac{D_a q_s}{Dt}
% =
% \frac{\epsilon}{[p-e_{si}(T)]^2}
% \left[
% p\frac{de_{si}}{dT}\frac{D_aT}{Dt}
% -
% e_{si}(T)\frac{D_ap}{Dt}
% \right].
% \end{equation}
% Thus, the saturation mixing ratio and its evolution are consistently determined by the prescribed thermodynamic fields \(T(z,t)\) and \(p(z,t)\), completing the reduced supersaturation closure used in the \((z,t)\) plume-evolution system.
\subsection{{Bulk Ice-number Loss and Differential-settling Aggregation}}

The line-integrated ice number \(n_\ell\) is subject to
two bulk loss mechanisms: turbulence-induced particle
loss and differential-settling aggregation. Their
combined contribution is expressed as (\cite{7}):
\begin{equation}
\frac{dn_\ell}{dt}
=
-\underbrace{
E_T\left[
\frac{ K_y}{\max(B(t),D(t))^2}
+
\frac{K_z}{D_{\mathrm{eff}}(t)^2}
\right]n_\ell
}_{\text{Turbulent loss}}
-
\underbrace{
E_A\frac{K_{\mathrm{grav}}}{A(t)}
n_\ell^2
}_{\text{Aggregation loss}},
\qquad
D_{\mathrm{eff}}=\frac{A(t)}{B(t)}.
\end{equation}

Here, \(E_T\) and \(E_A\) are the turbulent and
aggregation efficiency coefficients, respectively, and \(D_{\mathrm{eff}}\) represents the
equivalent vertical depth associated with the
covariance-based cross-sectional area.

For the aggregation contribution, the starting point
is the differential-settling collision kernel between
two particles: $K_{ij}
=
\pi(r_i+r_j)^2
\left|V_i-V_j\right|,$ where \(r_i\) and \(V_i=|v_{s,i}|\) denote the radius and settling-speed magnitude of particle \(i\),
respectively. Rather than explicitly evaluating
particle-pair interactions, the present implementation
estimates the collision kernel from the active vertical
computational cells at the preceding time step.
Specifically, the particle radius in each active cell
is reconstructed from its representative mass and
deposition density as $r_i=
\left(
\frac{3m_i}{4\pi\rho_{\mathrm{dep},i}}
\right)^{1/3},
r_m=\frac{r_{\max}+r_{\min}}{2}.$ Here, \(r_m\) is an approximate representative collision
radius obtained as the midpoint between the minimum
and maximum particle radii across the active cells.
Similarly, the root-mean-square relative settling speed
is approximated by: $\Delta v_{\mathrm{rms}}
=
\sqrt{2\,\operatorname{Var}(V_i)},
K_{\mathrm{grav}}
=
\pi(2r_m)^2\Delta v_{\mathrm{rms}}.$ The factor of two follows from the relation
\(\mathbb{E}[(V_i-V_j)^2]=2\operatorname{Var}(V)\)
for two independent samples drawn from the same
velocity distribution. Accordingly, the resulting
collision kernel represents an RMS-based approximation
of the differential-settling effect, rather than an
explicit summation over particle pairs.

\section{{Results and discussion}}
\label{sec:results}

{The results are presented in five parts.} First, the evolution of a single plume is examined under stationary and time-dependent atmospheric forcing. The sensitivity of the {time-integrated optical depth and total extinction} to selected closure parameters is then assessed. {The final parts compare MultiCon with CoCiP and APCEMM for a common case, extend that comparison to a parametric sweep of the inputs, and quantify the computational cost of the three approaches.}

\subsection{Plume evolution and optical depth}
\label{subsec:plume_evolution}

Two idealized atmospheric cases are considered. In both cases, the environmental temperature and ice-supersaturation fields are prescribed on the same ten-layer vertical partition. The layers are indexed from top to bottom and defined relative to the initial plume reference altitude:
\begin{equation}
\begin{aligned}
 \mathcal{L}_1 &: z>200\,\mathrm{m},&
 \mathcal{L}_2 &: 0<z\leq 200\,\mathrm{m},\\
 \mathcal{L}_3 &: -200<z\leq 0\,\mathrm{m},&
 \mathcal{L}_4 &: -400<z\leq -200\,\mathrm{m},\\
 \mathcal{L}_5 &: -600<z\leq -400\,\mathrm{m},&
 \mathcal{L}_6 &: -800<z\leq -600\,\mathrm{m},\\
 \mathcal{L}_7 &: -1000<z\leq -800\,\mathrm{m},&
 \mathcal{L}_8 &: -1200<z\leq -1000\,\mathrm{m},\\
 \mathcal{L}_9 &: -1400\leq z\leq -1200\,\mathrm{m},&
 \mathcal{L}_{10} &: z<-1400\,\mathrm{m}.
\end{aligned}
\label{eq:environment_layers}
\end{equation}
The first and last layers are truncated by the computational domain, $-2500\leq z\leq 500\,\mathrm{m}$. Both $T$ and $s_{i,\mathrm{env}}$ are spatially uniform within each layer; the same altitude-to-layer mapping is used for both fields. For time-dependent forcing, the prescribed layer values are linearly interpolated in time between successive input profiles, supplied at 600 s intervals. {Thus, layers 1--4 correspond to $z>-400\,\mathrm{m}$ and layers 5--10 to $z\leq-400\,\mathrm{m}$; similarly, layers 1--5 and 6--10 are separated at $z=-600\,\mathrm{m}$.}

{In the stationary case, the temperature is $218\,\mathrm{K}$ in all ten layers, while the environmental ice supersaturation is $s_{i,\mathrm{env}}=0.10$ in layers 1--4 and $-0.05$ in layers 5--10.} {In the time-dependent case, the temperature is vertically uniform and prescribed as $T(t)=218+10\cos(2\pi t/T_f)\,\mathrm{K}$, with $T_f=15\,\mathrm{h}$, while the environmental ice supersaturation is $0.05+0.05\sin(2\pi t/T_f)$ in layers 1--5 and $-0.10+0.05\sin(2\pi t/T_f)$ in layers 6--10.} {The initial supersaturation profiles correspond to $t=0$; the time-dependent temperature likewise starts at $228\,\mathrm{K}$ in all layers, with $218\,\mathrm{K}$ denoting its temporal mean.} {The stationary case is also the baseline of the three-model comparison of \mbox{Sect.~\ref{subsec:model_comparison}}.}

Both simulations use an initial particle radius of $r_0=1\,\mu\mathrm{m}$, a line-integrated particle number of $n_{\ell,0}=10^{12}\,\mathrm{m}^{-1}$, initial vertical and crosswind standard deviations of $\sigma_{z,0}=50\,\mathrm{m}$ and $\sigma_{y,0}=30\,\mathrm{m}$, respectively, and an initial shape factor of unity. The normal crosswind shear used in the long-term calculation is $S_y=5\times10^{-4}\,\mathrm{s}^{-1}$. The vertical domain is $-2500\leq z\leq500\,\mathrm{m}$, discretized using 800 points; the integration period is 15 h with a 120 s time step \footnote{{The discretization sensitivity analysis showed that a vertical grid spacing of approximately 4 m or finer yields a relative error below 1\% in the 10-hour mean plume center-of-mass altitude, compared with the high-resolution reference solution. Temporal sensitivity is negligible for time steps below approximately 1 min and remains relatively weak for time steps below 3 min.}}. Here, $z$ denotes altitude relative to the initial plume reference level.

\begin{figure}[p]
 \centering
 \includegraphics[width=0.70\textwidth]{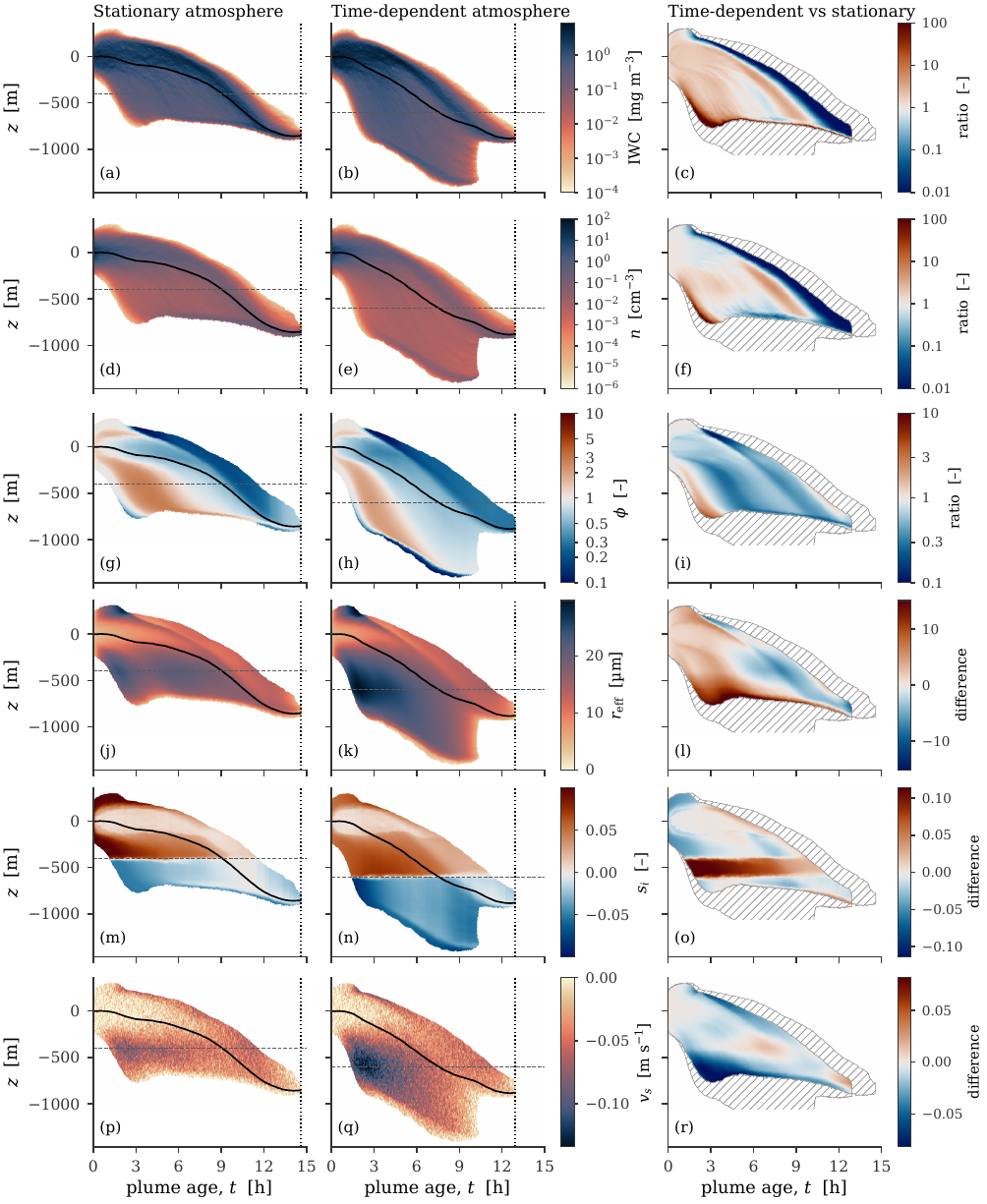}
 \caption{{The time-dependent plume carries more ice than the stationary one in its upper part during the first hours and less
afterwards, and it ends earlier (12.9\,h against 14.6\,h). Altitude--time evolution of six MultiCon diagnostics for the
stationary atmosphere (left column; $T=218\,\mathrm{K}$, $s_{i,\mathrm{env}}=0.10$ above $z=-400\,\mathrm{m}$ and $-0.05$
below), the time-dependent atmosphere (middle column; $T=218+10\cos(2\pi t/T_f)\,\mathrm{K}$,
$s_{i,\mathrm{env}}=0.05+0.05\sin(2\pi t/T_f)$ above $z=-600\,\mathrm{m}$ and $-0.10+0.05\sin(2\pi t/T_f)$ below, with
$T_f=15\,\mathrm{h}$) and their comparison (right column); initial plume and shear as given in the text, identical in the
two cases. $z$ is altitude relative to the initial plume level. Rows: (a--c) ice water content, (d--f) ice number
concentration, (g--i) habit index, (j--l) effective radius, (m--o) ice supersaturation, (p--r) bulk settling velocity.
The left and middle panels of a row share one colour scale. Right column: ratio time-dependent/stationary on a
logarithmic scale for the first three rows, difference time-dependent minus stationary (in the units of the row) for
the last three; red means larger in the time-dependent case; it is drawn only where both plumes exist, and hatching
marks where only one of them does. Solid black line: centre of ice mass; dashed grey line: altitude at which the
prescribed supersaturation changes sign; dotted vertical line: plume end. White: IWC below
$10^{-4}\,\mathrm{mg\,m^{-3}}$ or after the plume end.}}
 \label{fig:plume_fields}
\end{figure}

{\mbox{Figure~\ref{fig:plume_fields}} shows} the six diagnostic fields used to characterize the plume evolution. The fields are plotted on common time and vertical coordinates, with identical colour limits between the two cases wherever practicable. Logarithmic colour scales are used only for quantities spanning several orders of magnitude. {The six plotted variables are the ice water content ($\mathrm{mg\,m^{-3}}$), the ice-crystal number concentration ($\mathrm{cm^{-3}}$), the habit index $\phi$ (dimensionless; below one for plate-like and above one for column-like crystals), the effective radius ($\mu\mathrm{m}$), the ice supersaturation (dimensionless), and the bulk settling velocity ($\mathrm{m\,s^{-1}}$, negative downwards).} {In both cases the plume sinks slowly while it remains in supersaturated air: the centre of ice mass leaves the supersaturated layers after 9.1\,h in the stationary case ($z=-400\,\mathrm{m}$) and after 7.6\,h in the time-dependent case ($z=-600\,\mathrm{m}$), and the plume then sublimates in the subsaturated air below, its centre reaching $z=-855\,\mathrm{m}$ and $z=-879\,\mathrm{m}$, respectively.} {The plume ends after 14.6\,h in the stationary case and after 12.9\,h in the time-dependent case, the plume end being defined as the first time after its maximum at which the vertically integrated ice mass falls below $10^{-3}$ of that maximum.} Together, these fields describe the height-resolved plume response; the optical-depth {and total-extinction histories} below provide complementary column-integrated measures.

The optical-depth diagnostic is evaluated as
\begin{equation}
 \tau_c(t)=\frac{n_{\ell}(t)f_{\mathrm{cw}}}{B(t)}
 \int Q_{\mathrm{ext}}(z,t)A_{\mathrm{proj}}(z,t)g(z,t)\,\mathrm{d}z,
 \label{eq:tau_multicon}
\end{equation}
where $Q_{\mathrm{ext}}$ is the extinction efficiency, $A_{\mathrm{proj}}$ is the orientation-averaged projected particle area, $g$ is the normalized vertical number distribution, $B=\sqrt{8}\,\sigma_y$ is the effective crosswind width, and $f_{\mathrm{cw}}=\operatorname{erf}[B/(2\sqrt{2}\,\sigma_y)]$ is the finite-width correction. The extinction calculation uses a wavelength of $550\,\mathrm{nm}$ and a refractive index of 1.31. {The optical depth $\tau_c$ is thus the vertical optical depth of the plume averaged over its effective width: it measures how opaque the contrail is.}

{How much radiation the contrail as a whole intercepts depends, in addition, on how wide it is.} {This is quantified by the total extinction per unit length of contrail, introduced by \mbox{\cite{4}}.} {Let $\beta_{\mathrm{ext}}=n\,Q_{\mathrm{ext}}A_{\mathrm{proj}}$ denote the extinction coefficient ($\mathrm{m^{-1}}$), with $n$ the ice-crystal number concentration.} {The vertical optical depth at the crosswind position $y$ is:}
\begin{equation}
 {\tau_y(y,t)=\int_{-\infty}^{\infty}\beta_{\mathrm{ext}}(y,z,t)\,\mathrm{d}z ,}
 \label{eq:tau_y}
\end{equation}
{and, by the Beer--Lambert law, a vertically incident beam crossing the plume at that position is attenuated by the fraction $1-\exp(-\tau_y)$.} {Integrating this fraction across the plume gives the total extinction:}
\begin{equation}
 {E(t)=\int_{-\infty}^{\infty}\left[1-\exp\left(-\tau_y(y,t)\right)\right]\,\mathrm{d}y ,}
 \label{eq:integrated_extinction}
\end{equation}
{which has units of length: it is the width of a completely opaque strip that would remove the same amount of radiation per unit length of contrail.} {For an optically thin plume, $E\simeq\int\tau_y\,\mathrm{d}y$, and with \mbox{Eq.~(\ref{eq:tau_multicon})}}
\begin{equation}
 {E(t)\simeq n_{\ell}(t)\int Q_{\mathrm{ext}}(z,t)A_{\mathrm{proj}}(z,t)g(z,t)\,\mathrm{d}z=\frac{\tau_c(t)\,B(t)}{f_{\mathrm{cw}}} ,}
 \label{eq:extinction_thin}
\end{equation}
{that is, the total extinction is the product of the mean optical depth and the plume width.} {The two diagnostics are therefore complementary: as a plume spreads, $\tau_c$ decreases through dilution, whereas $E$ can keep increasing for as long as the plume widens faster than it thins.} {\mbox{Equation~(\ref{eq:extinction_thin})} is the form evaluated in this section, so that $E$ is obtained in the same way for all the models compared below.} 

% \hl{It overestimates \mbox{Eq.~(\ref{eq:integrated_extinction})} where the plume is optically thick: for a Gaussian crosswind profile of $\tau_y$, the time integrals reported below for the two cases of this subsection would be 3\,\% and 4\,\% lower.}

The time-integrated diagnostics {considered in this section are}
\begin{equation}
 J_{\tau}=\int_0^{T} \tau_c(t)\,\mathrm{d}t,\qquad
 {\widehat{E}=\int_0^{T} E(t)\,\mathrm{d}t,}\qquad T=15\,\mathrm{h},
 \label{eq:tau_time_integral}
\end{equation}
{expressed in hours and in metre-hours, respectively; both integrands are zero after the plume end.}

\begin{figure}[htbp]
 \centering
 \includegraphics[width=\textwidth]{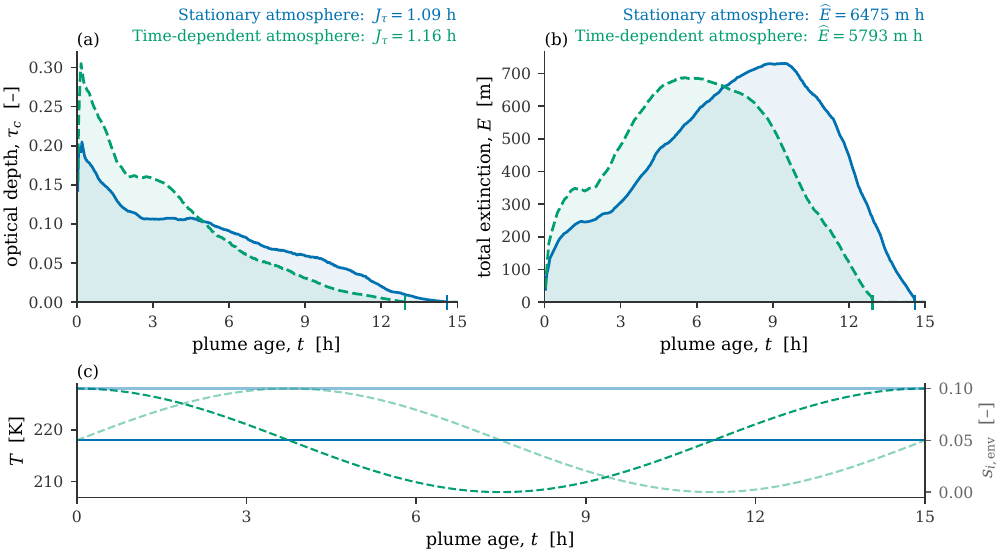}
 \caption{{The two diagnostics rank the cases in opposite order: the time-dependent atmosphere has the larger
time-integrated optical depth but the smaller time-integrated total extinction. (a) MultiCon optical depth $\tau_c$
(Eq.~(\ref{eq:tau_multicon}), 550\,nm) and (b) total extinction $E$ (Eq.~(\ref{eq:extinction_thin})) against plume age,
for the stationary atmosphere (solid blue) and the time-dependent atmosphere (dashed green); forcing as in
Fig.~\ref{fig:plume_fields}. Shaded areas are the 15\,h integrals $J_{\tau}$ and $\widehat{E}$, whose values are written
above the panels. Maxima: $\tau_c=0.20$ and $E=731\,\mathrm{m}$ (stationary), $\tau_c=0.31$ and $E=687\,\mathrm{m}$
(time-dependent). Both quantities are zero after the plume end, marked by a tick on the time axis (14.6 and 12.9\,h).
(c) Prescribed forcing of the layers in which the plume is initialized: temperature (dark lines, left axis) and
environmental ice supersaturation (light lines, right axis), with the colours and line styles of (a) and (b).}}
 \label{fig:tau_cases}
\end{figure}

{\mbox{Figure~\ref{fig:tau_cases}} compares the histories of the two diagnostics.} {The optical depth peaks about ten minutes after initialization, at 0.20 in the stationary case and at 0.31 in the time-dependent case, and falls below 0.1 after about five hours in both.} {The corresponding values of $J_{\tau}$ are 1.09\,h and 1.16\,h.} {The total extinction evolves differently: it increases while the plume widens faster than it thins, reaches a maximum of 731\,m at 9.5\,h in the stationary case and of 687\,m at 5.4\,h in the time-dependent case, and then decreases to zero at the plume end.} {Its time integral is $\widehat{E}=6475\,\mathrm{m\,h}$ in the stationary case and $5793\,\mathrm{m\,h}$ in the time-dependent case.} {The two diagnostics therefore rank the cases in opposite order: the time-dependent plume is optically thicker during the first five hours, which gives it the larger $J_{\tau}$ (by 6\,\%), but it ends 1.7\,h earlier and has the smaller $\widehat{E}$ (by 11\,\%).} {This illustrates that $J_{\tau}$ is not a substitute for $\widehat{E}$.}

The IWC-weighted vertical centroid is calculated from the first vertical moment divided by the vertically integrated IWC. {Its trajectory is reported up to the plume end defined above.} Once the remaining IWC becomes negligible, the centroid is no longer a robust measure of plume position and is therefore omitted. The altitude--time fields are independently inspected using the common IWC plotting threshold of $10^{-4}\,\mathrm{mg\,m^{-3}}$; structures below this threshold are not included in the displayed plume boundary.

\subsection{Sensitivity to closure parameters}
\label{subsec:sensitivity}

The sensitivity analysis considers $s_{\mathrm{crit}}$ (i.e., the correction function used in habit dynamics for very low supersaturation levels), $K_y$, $K_z$, $E_T$, and $E_A$. {It is carried out for the stationary case of \mbox{Sect.~\ref{subsec:plume_evolution}} with a full factorial design: each parameter takes three values and all $3^5=243$ combinations are simulated.} {The values are $s_{\mathrm{crit}}\in\{0.02,\,0.03,\,0.04\}$, $K_y\in\{10,\,25,\,40\}~\mathrm{m^2\,s^{-1}}$, $K_z\in\{0.05,\,0.10,\,0.20\}~\mathrm{m^2\,s^{-1}}$, and $E_T,E_A\in\{0.5,\,0.75,\,1.0\}$.} {They span the interval considered for each parameter and include the nominal values used in all the other simulations of this paper, $s_{\mathrm{crit}}=0.02$, $K_y=25~\mathrm{m^2\,s^{-1}}$, $K_z=0.10~\mathrm{m^2\,s^{-1}}$, and $E_T=E_A=0.5$, so that the nominal simulation is one member of the ensemble.} {The response variables are $J_{\tau}$ and $\widehat{E}$ from \mbox{Eq.~(\ref{eq:tau_time_integral})}.}

\begin{figure}[htbp]
 \centering
 \includegraphics[width=\textwidth]{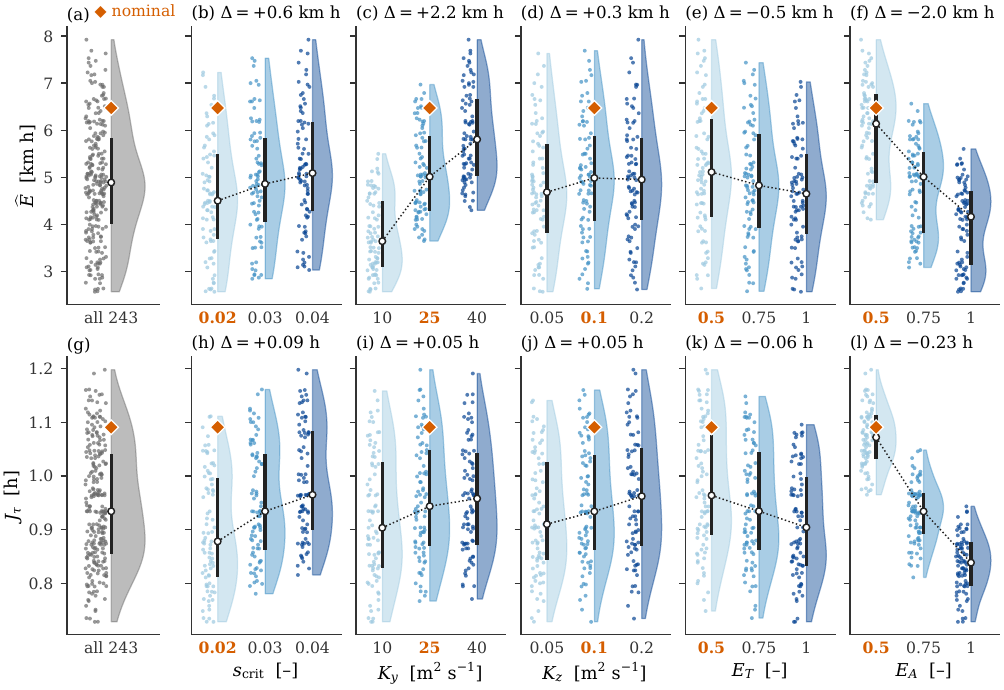}
 \caption{{The crosswind diffusivity $K_y$ and the aggregation efficiency $E_A$ control the time-integrated total
extinction, whereas $E_A$ is the main control of the time-integrated optical depth. Stationary case of
Sect.~\ref{subsec:plume_evolution} ($T=218\,\mathrm{K}$, $s_{i,\mathrm{env}}=0.10$ above $z=-400\,\mathrm{m}$ and $-0.05$
below). Full factorial design: each of the five closure parameters takes the three values on the abscissa of its panel,
and all $3^5=243$ combinations are simulated. Top row (a--f): time-integrated total extinction $\widehat{E}$; bottom row
(g--l): time-integrated optical depth $J_{\tau}$; both over 15\,h. (a, g) All 243 simulations. (b--f, h--l) The same
simulations grouped by the value of one parameter, 81 per value (light to dark blue), the other four parameters taking
all their combinations; $\Delta$ is the difference between the medians at the highest and the lowest value. In each
panel: half violin, kernel density; dots, the individual simulations; white circle and black bar, median and interquartile range; dotted line, medians joined. Orange diamond and
orange tick label: the nominal simulation and the nominal parameter values ($s_{\mathrm{crit}}=0.02$,
$K_y=25\,\mathrm{m^2\,s^{-1}}$, $K_z=0.1\,\mathrm{m^2\,s^{-1}}$, $E_T=E_A=0.5$), for which $\widehat{E}=6.47\,\mathrm{km\,h}$
and $J_{\tau}=1.09\,\mathrm{h}$.}}
 \label{fig:sensitivity}
\end{figure}

{\mbox{Figure~\ref{fig:sensitivity}} shows, for each parameter, the distribution of the two responses over the 81 simulations that share each of its three values, the other four parameters taking all their combinations.} {A parameter is influential when this distribution shifts from one value to the next.} {The time-integrated total extinction is controlled by $K_y$ and $E_A$: between the lowest and the highest value of the parameter its median changes by $+2.2\,\mathrm{km\,h}$ for $K_y$ and by $-2.0\,\mathrm{km\,h}$ for $E_A$, against $0.6\,\mathrm{km\,h}$ or less for $s_{\mathrm{crit}}$, $K_z$, and $E_T$.} {The time-integrated optical depth responds mainly to $E_A$ ($-0.23\,\mathrm{h}$, against $+0.09\,\mathrm{h}$ for $s_{\mathrm{crit}}$ and $0.06\,\mathrm{h}$ or less for the other three).} {The different response of the two diagnostics to $K_y$ follows from \mbox{Eq.~(\ref{eq:extinction_thin})}: a larger crosswind diffusivity widens the plume, which increases $E$ but not the mean optical depth.} {Over the whole ensemble $\widehat{E}$ ranges from 2.6 to $7.9\,\mathrm{km\,h}$ and $J_{\tau}$ from 0.73 to 1.20\,h.} {The nominal simulation ($\widehat{E}=6.47\,\mathrm{km\,h}$, $J_{\tau}=1.09\,\mathrm{h}$) lies above about 87\,\% of the ensemble in both diagnostics, because the nominal values of $s_{\mathrm{crit}}$, $E_T$, and $E_A$ are the lowest of their intervals.} Having established the response of MultiCon to its closure parameters, the following subsection compares its predictions with two established contrail models under a common atmospheric and initial-plume configuration.

\subsection{Comparison with CoCiP and APCEMM}
\label{subsec:model_comparison}

\begin{figure}[t]
 \centering
 \includegraphics[width=0.95\textwidth]{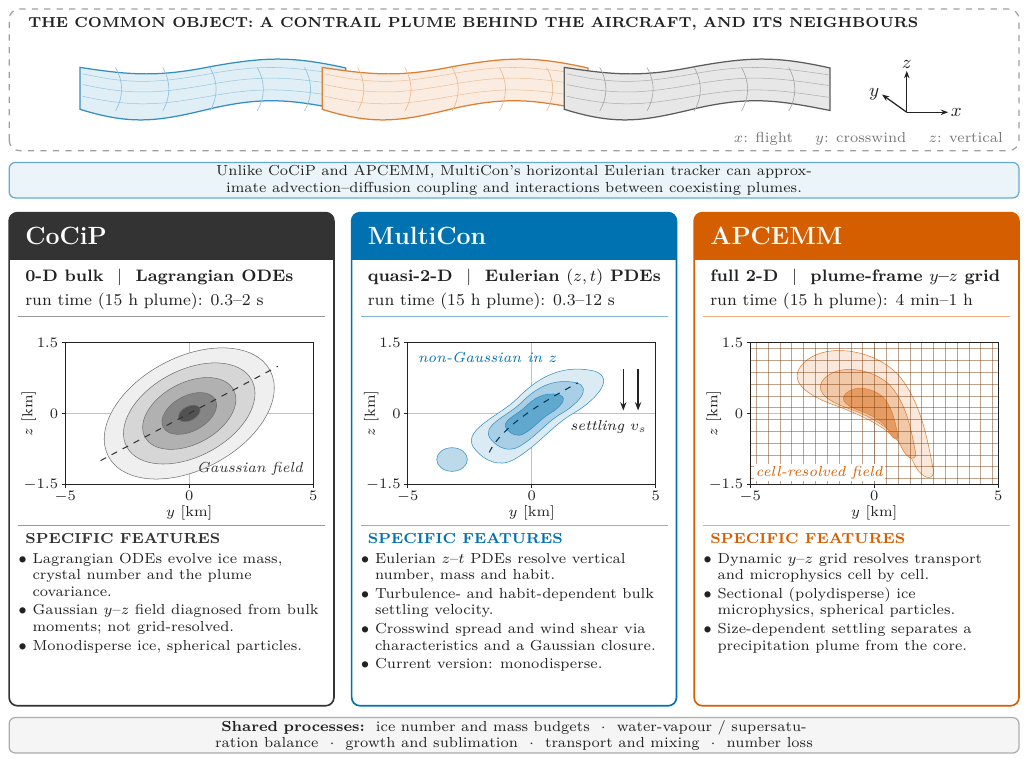}
 \caption{The three models represent the same contrail plume with increasing spatial and microphysical resolution, from a bulk
Gaussian (CoCiP) through a vertically resolved profile (MultiCon) to a two-dimensional grid (APCEMM).}
 \label{fig:model_comparison}
\end{figure}

To provide context for the quantitative comparison,  Fig.~\ref{fig:model_comparison} summarizes the principal  characteristics of CoCiP, MultiCon, and APCEMM. These differences are important when 
interpreting the subsequent comparisons, since the  models do not resolve the plume evolution at the  same spatial or microphysical level.

% ---- previous version of this figure (kept for reference) ----
% \begin{figure*}[t]
%     \centering
%     \includegraphics[width=0.95\textwidth]
%     {contrail_model_schematic_final}
%     \caption{Schematic comparison of the CoCiP, 
%     MultiCon, and APCEMM modelling frameworks. 
%     The figure summarizes their respective plume 
%     representations, resolved dimensions, 
%     microphysical treatments, and the principal 
%     physical processes represented in the three models.}
%     \label{fig:model_comparison_old}
% \end{figure*}
% ---- TikZ version ----

The three-model comparison uses a stationary environment at $218\,\mathrm{K}$ in all ten layers, with $s_{i,\mathrm{env}}=0.10$ in layers 1--4 and $-0.05$ in layers 5--10; the initial ice supersaturation is 0.10. The same layer mapping defined by Eq.~(\ref{eq:environment_layers}) is used for both environmental fields, so the supersaturation transition is located at $z=-400\,\mathrm{m}$ relative to the initial plume reference level. The initial plume is specified by $r_0=1\,\mu\mathrm{m}$, $n_{\ell,0}=10^{12}\,\mathrm{m}^{-1}$, $\sigma_{z,0}=50\,\mathrm{m}$, and $\sigma_{y,0}=30\,\mathrm{m}$, with an initial shape factor of unity, zero initial cross-covariance, and normal crosswind shear $S_y=5\times10^{-4}\,\mathrm{s}^{-1}$. All models are evaluated over the same 15 h interval and at matched output times. Model-specific initialization quantities required by APCEMM, including the mapping of the monodisperse initial condition to its sectional size grid and any additional aerosol or chemical inputs, are documented in Table~\ref{tab:initial_conditions}. {The three models start from this prescribed plume, so that their own representations of the jet and vortex phases are not used.} {APCEMM is run with 38 size bins spaced geometrically between radii of 0.05 and $80\,\mu\mathrm{m}$, a time step of 300\,s, the vertical grid spacing of MultiCon (3.75\,m) and the diffusivities of MultiCon ($K_y=25~\mathrm{m^2\,s^{-1}}$, $K_z=0.1~\mathrm{m^2\,s^{-1}}$).} {CoCiP is run with its implementation in the pycontrails library (version 0.64), with a time step of 120\,s and its own diffusivity closures.}

\begin{table}[htbp]
 \centering
 \caption{Common comparison conditions and model-specific {implementation. In the three models the plume is initialized
 directly with the common values, so that the jet and vortex phases are not simulated.}}
 \label{tab:initial_conditions}
 \begin{tabular}{p{0.20\linewidth}p{0.25\linewidth}p{0.47\linewidth}}
  \hline
  Quantity & Common value & Model-specific implementation \\
  \hline
  Temperature & $218\,\mathrm{K}${, uniform and constant} & {MultiCon: prescribed in the ten layers. CoCiP and APCEMM: horizontally uniform,
  isothermal meteorological fields (234\,hPa at the plume level)} \\
  Environmental $s_i$ & $0.10$ above $-400\,\mathrm{m}$; $-0.05$ below & {Same boundary in the three models; given to CoCiP and APCEMM as the
  specific humidity of the meteorological fields} \\
  {Wind shear} & {$5\times10^{-4}\,\mathrm{s^{-1}}$, normal to the contrail} & {CoCiP and APCEMM: wind increasing linearly with altitude, normal to the
  flight track; CoCiP applies its subgrid-scale enhancement to this shear} \\
  Initial radius & $1\,\mu\mathrm{m}$ & {CoCiP: volume-mean radius. APCEMM: crystals distributed between the two bins adjacent
  to $1\,\mu\mathrm{m}$, conserving number and ice mass} \\
  Initial line number & $10^{12}\,\mathrm{m}^{-1}$ & {Prescribed directly in the three models; no vortex-phase losses are applied} \\
  Initial plume widths & $\sigma_{z,0}=50\,\mathrm{m}$; $\sigma_{y,0}=30\,\mathrm{m}$ & {Zero initial cross-covariance. CoCiP: width and depth
  $\sqrt{8}\,\sigma_{y,0}$ and $\sqrt{8}\,\sigma_{z,0}$. APCEMM: Gaussian distribution on its grid} \\
  Particle-size resolution & Monodisperse in MultiCon & {CoCiP: bulk (one volume-mean radius). APCEMM: 38 size bins, radii from 0.05 to
  $80\,\mu\mathrm{m}$} \\
  {Turbulent diffusivities} & {$K_y=25$, $K_z=0.1\,\mathrm{m^2\,s^{-1}}$} & {Prescribed in MultiCon and APCEMM. CoCiP: its own closures; the horizontal
  one is proportional to the wind shear} \\
  {Processes not included} & {Radiative heating; water vapour of the exhaust} & {CoCiP: radiation inputs set to zero. APCEMM: no exhaust water vapour, no
  enhanced diffusion of the young plume, no temperature perturbation} \\
  {Numerical resolution} & {15\,h; output every 2 to 5\,min} & {MultiCon: $\Delta z=3.75\,\mathrm{m}$, $\Delta t=120\,\mathrm{s}$. APCEMM: $\Delta z=3.75\,\mathrm{m}$,
  $\Delta t=300\,\mathrm{s}$. CoCiP: $\Delta t=120\,\mathrm{s}$} \\
  \hline
 \end{tabular}
\end{table}

\begin{figure}[htbp]
 \centering
 \includegraphics[width=\textwidth]{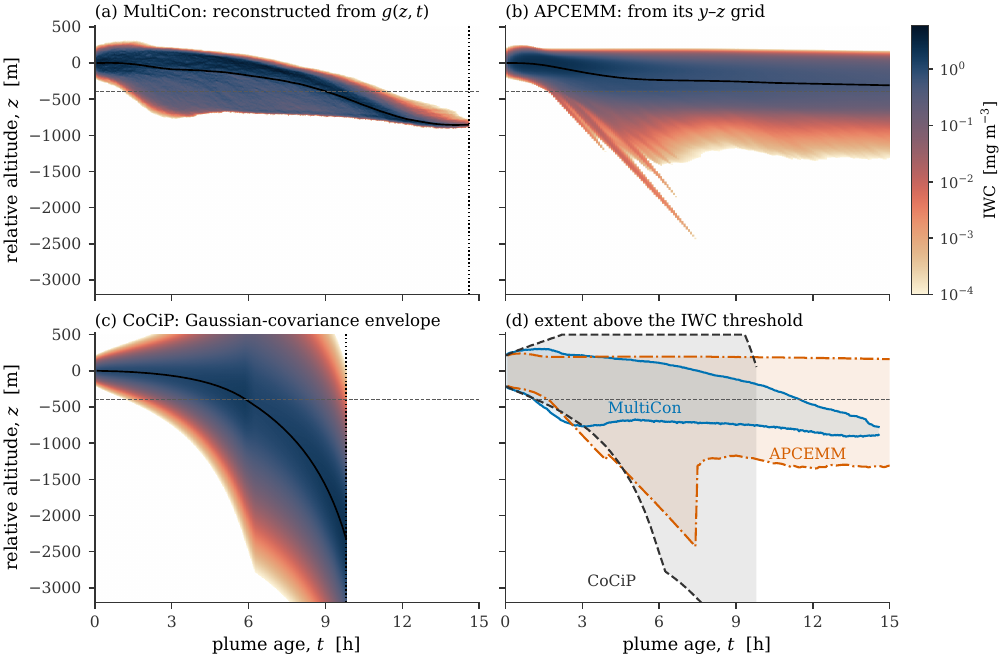}
 \caption{MultiCon and APCEMM keep a plume core near flight level while ice falls below the moist layer; CoCiP displaces a single
Gaussian downwards and ends earlier. Ice water content against plume age and altitude relative to the initial plume
level for the common comparison case (218\,K; $s_{i,\mathrm{env}}=0.10$ above $z=-400\,\mathrm{m}$ and $-0.05$ below;
$r_0=1\,\mu\mathrm{m}$, $n_{\ell,0}=10^{12}\,\mathrm{m^{-1}}$, $\sigma_{z,0}=50\,\mathrm{m}$, $\sigma_{y,0}=30\,\mathrm{m}$;
crosswind shear $5\times10^{-4}\,\mathrm{s^{-1}}$; 15\,h). The plotted quantity is the same for the three models: the
crosswind integral of IWC divided by $\sqrt{8}\,\sigma_y/\operatorname{erf}(1)$, i.e.\ the crosswind mean over the
effective width. (a) MultiCon, from its vertical Eulerian field. (b) APCEMM, reduced from its two-dimensional grid;
APCEMM is run without its temperature perturbation. (c) CoCiP: not a prognostic field but, at each output time, the
vertical profile of its Gaussian given by the covariance matrix. (d) Lowest and highest altitude with IWC above the
threshold: solid blue MultiCon, dash-dotted orange APCEMM, dashed black CoCiP. One logarithmic colour scale and one
threshold, $10^{-4}\,\mathrm{mg\,m^{-3}}$, for all panels; white is below the threshold or after the plume end. Solid black line:
centre of ice mass; dashed grey line: base of the moist layer; dotted vertical line: plume end (MultiCon ends at 14.6\,h,
APCEMM is still present at 15\,h, CoCiP ends at 9.8\,h). APCEMM: 38 size bins, $\Delta t=300\,\mathrm{s}$, MultiCon
vertical spacing; CoCiP: pycontrails implementation.}
 \label{fig:iwc_zt}
\end{figure}

Figure~\ref{fig:iwc_zt} compares the IWC evolution in the relative-altitude--time plane. A common threshold of $10^{-4}\,\mathrm{mg\,m^{-3}}$ is applied to all models. APCEMM provides a grid-resolved cross-sectional field; {its reduction to a vertical profile is the crosswind integral of the IWC divided by the effective width $\sqrt{8}\,\sigma_y/\operatorname{erf}(1)$, i.e. the crosswind mean of the IWC over the effective width.} {The same quantity is plotted for the three models.} MultiCon provides a vertical Eulerian field from which the crosswind distribution is reconstructed. CoCiP does not resolve a vertical concentration field; its plotted extent is therefore reconstructed from the time-dependent Gaussian geometry and covariance matrix. The CoCiP representation is a threshold-derived envelope at each output time and should not be interpreted as a continuous prognostic IWC field.

\begin{figure}[htbp]
 \centering
 \includegraphics[width=\textwidth]{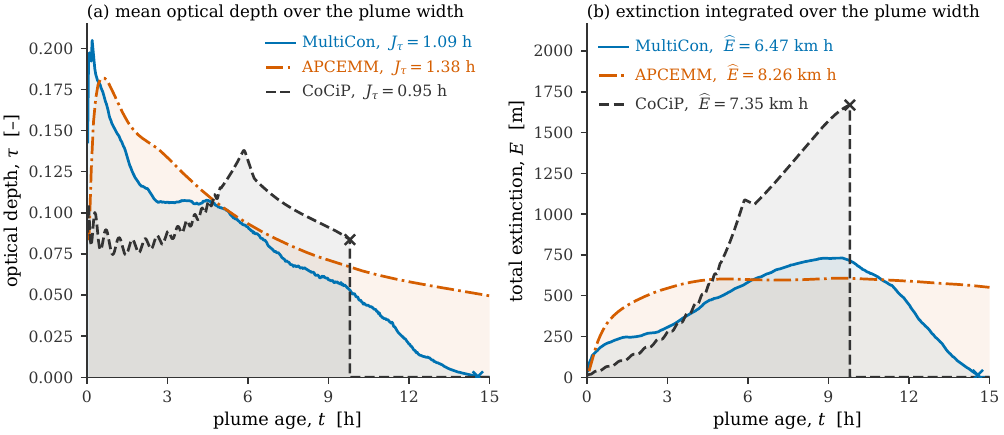}
 \caption{{The optical depth of the three models decays with plume age, but their total extinction does not: it grows
while the plume widens faster than it thins. Time-integrated total extinction ranks the models APCEMM $>$ CoCiP $>$
MultiCon (8261, 7354 and $6475\,\mathrm{m\,h}$). Common comparison case (218\,K, $s_{i,\mathrm{env}}=0.10$ above
$z=-400\,\mathrm{m}$ and $-0.05$ below, shear $5\times10^{-4}\,\mathrm{s^{-1}}$). (a) Optical depth $\tau$, the vertical
optical depth averaged over the effective plume width $B=\sqrt{8}\,\sigma_y$. (b) Total extinction
$E=\tau B/\operatorname{erf}(1)$ (Eq.~(\ref{eq:extinction_thin})), the vertical optical depth integrated across the plume.
The same definitions are used for the three models. Maxima of $E$: 731\,m at 9.5\,h (MultiCon), 608\,m at 9.4\,h
(APCEMM) and 1669\,m at 9.8\,h (CoCiP). Shaded areas are the 15\,h integrals $J_{\tau}$ and $\widehat{E}$ given in the
keys. Solid blue MultiCon, dash-dotted orange APCEMM, dashed black CoCiP. A cross marks the end of a model's contrail
(MultiCon at 14.6\,h; APCEMM does not end within 15\,h; CoCiP at 9.8\,h); after it $\tau$ and $E$ are zero up to 15\,h,
so that all integrals cover the same interval, 2\,min to 15\,h.}}
 \label{fig:tau_models}
\end{figure}

{The optical depth and the total extinction defined in \mbox{Sect.~\ref{subsec:plume_evolution}} are compared in \mbox{Fig.~\ref{fig:tau_models}}.} {They are evaluated in the same way for the three models: $E$ is the extinction coefficient integrated over the plume cross-section (\mbox{Eq.~(\ref{eq:extinction_thin})}), and the optical depth is $\tau=\operatorname{erf}(1)\,E/B$ with $B=\sqrt{8}\,\sigma_y$, which for MultiCon coincides with \mbox{Eq.~(\ref{eq:tau_multicon})}.} {For APCEMM, $E$ and $\sigma_y$ are obtained from its two-dimensional extinction and IWC fields; for CoCiP, from its optical depth and plume width.}

{The optical depth decays with plume age in MultiCon and APCEMM, from maxima of 0.21 and 0.18 reached within the first hour; in CoCiP it stays below 0.14.} {The total extinction behaves differently, because the plumes widen while they thin.} {In MultiCon, $E$ increases to 731\,m at 9.5\,h and then decreases to zero at the plume end (14.6\,h).} {In APCEMM, $E$ reaches about 590\,m after three hours and stays within 8\,\% of that value until the end of the simulation: the widening of the plume compensates the decay of its optical depth, while its ice mass and crystal number remain nearly constant.} {In CoCiP, $E$ increases to 1669\,m until its contrail ends at 9.8\,h.} {The time-integrated total extinction $\widehat{E}$ is 6475, 7354, and $8261\,\mathrm{m\,h}$ for MultiCon, CoCiP, and APCEMM, respectively, and the time-integrated optical depth $J_{\tau}$ is 1.09, 0.95, and 1.38\,h.}

\begin{figure}[p]
 \centering
 \includegraphics[width=\textwidth]{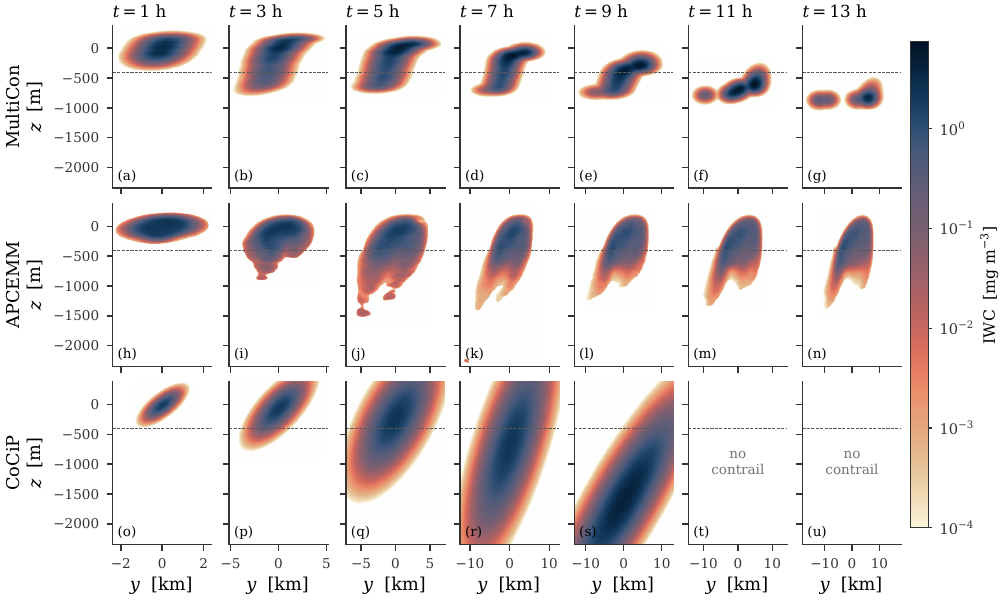}
 \caption{At late ages MultiCon and APCEMM both show a sheared plume core with ice falling below the moist layer, whereas CoCiP
has one tilted Gaussian and no contrail after 9.8\,h. Common comparison case
(218\,K, $s_{i,\mathrm{env}}=0.10$ above $z=-400\,\mathrm{m}$ and $-0.05$ below, shear $5\times10^{-4}\,\mathrm{s^{-1}}$); APCEMM temperature perturbation off. $y$ is the crosswind distance from the plume centre, oriented
so that the upper part of the plume is displaced towards positive $y$; $z$ is altitude relative to the initial plume
level. {Ice water content in the crosswind--vertical plane at plume ages} of 1, 3, 5, 7, 9, 11 and 13\,h
(columns) for MultiCon (a--g), APCEMM (h--n) and CoCiP (o--u). MultiCon: reconstructed from its vertical IWC profile, the
characteristic crosswind displacement of each level and a quasi-Gaussian crosswind spread; where the reconstruction width is not defined at the
requested age the nearest valid output time is used. APCEMM: its native two-dimensional field. CoCiP: the Gaussian
defined by its covariance matrix, clipped at the frame. One logarithmic colour scale and one threshold ($10^{-4}\,\mathrm{mg\,m^{-3}}$,
white below) for all panels; one altitude range and one crosswind range per column. Dashed grey line: base of the moist
layer. `No contrail': the model's contrail has ended.}
 \label{fig:cross_sections}
\end{figure}

{Cross-sectional fields are compared at matched plume ages of 1, 3, 5, 7, 9, 11, and 13 h (\mbox{Fig.~\ref{fig:cross_sections}}), thereby} including the late-time plume structure. The MultiCon distribution is reconstructed from its vertical IWC profile, characteristic lateral displacement, and Gaussian crosswind spread; APCEMM uses its native two-dimensional field, while the CoCiP field is reconstructed from its evolving Gaussian covariance. {MultiCon and APCEMM both show a plume core that stays near the flight level and is sheared in the crosswind direction, with ice falling below the base of the moist layer.} {In APCEMM a fall streak extends to about 2.4\,km below the flight level, whereas the MultiCon plume stays within about 0.9\,km of it.} {CoCiP represents the plume as a single tilted Gaussian, which deepens until it reaches the upper bound that CoCiP imposes on the plume depth (1500\,m, after about six hours) and descends as a whole; its centre is 2.3\,km below the flight level when the contrail ends at 9.8\,h.}

\begin{figure}[htbp]
 \centering
 \includegraphics[width=\textwidth]{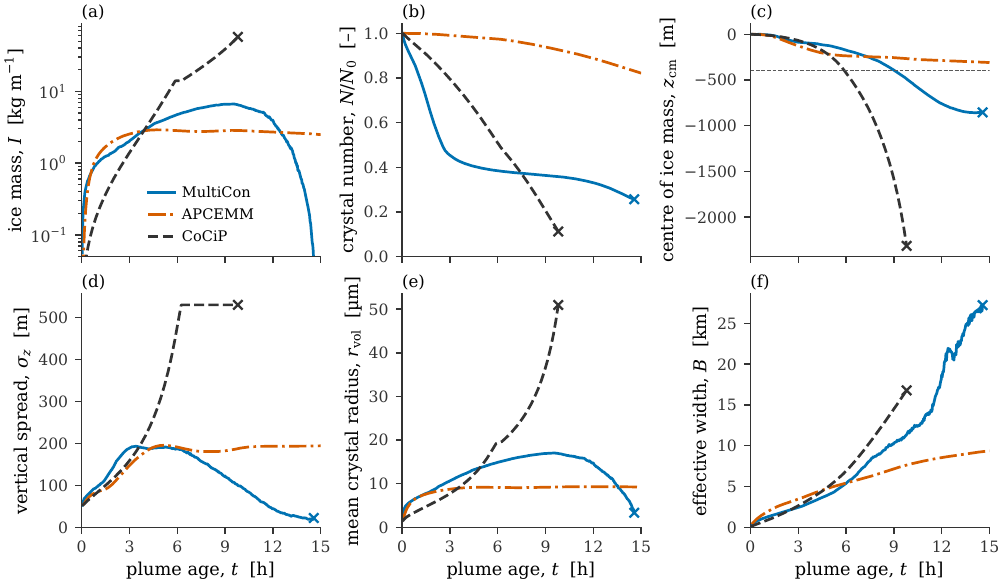}
 \caption{The three models differ most in how much ice the plume holds and how many crystals survive: peak ice mass 6.7 (MultiCon),
2.9 (APCEMM) and $57.2\,\mathrm{kg\,m^{-1}}$ (CoCiP); crystals left at the end 26, 82 and 11\,\%. Integral diagnostics of
the common comparison case (218\,K, $s_{i,\mathrm{env}}=0.10$ above $z=-400\,\mathrm{m}$, shear
$5\times10^{-4}\,\mathrm{s^{-1}}$) against plume age, computed with the same definitions for the three models. (a) Ice
mass per unit flight-path length (logarithmic scale). (b) Crystal number per unit length, relative to its initial
value. (c) IWC-weighted vertical centroid; dashed grey line: base of the moist layer. (d) IWC-weighted vertical standard
deviation. {(e) Mean ice-crystal radius: the volume-mean radius of the whole plume,
$r_{\mathrm{vol}}=[3I/(4\pi\rho_{\mathrm{ice}}N)]^{1/3}$, from the ice mass $I$ and the crystal number $N$ per unit length
of each model ($\rho_{\mathrm{ice}}=917\,\mathrm{kg\,m^{-3}}$).} (f) Effective width $B=\sqrt{8}\,\sigma_y$, from the
IWC-weighted crosswind standard deviation. Solid blue MultiCon, dash-dotted orange APCEMM, dashed black CoCiP. A cross
marks the end of a contrail (column ice below $10^{-3}$ of its peak; CoCiP: its own end of life); nothing is drawn after
it.}
 \label{fig:integral_metrics}
\end{figure}

A small set of integral diagnostics is used to complement the field comparisons: total ice mass per unit flight-path length, total particle number per unit flight-path length, IWC-weighted vertical centroid, IWC-weighted vertical spread, {mean ice-crystal radius, and effective plume width.} {The centroid and spread are reported up to the end of each contrail.} The corresponding histories are shown in Fig.~\ref{fig:integral_metrics}. {The models agree most closely during the first three to four hours in the geometry of the plume: at 3\,h the vertical spread is between 147 and 189\,m, the effective width between 2.4 and 3.4\,km, and the centroid between 68 and 128\,m below the flight level.} {The largest differences occur in the ice budget at later ages: the maximum ice mass is 6.7, 2.9, and $57.2\,\mathrm{kg\,m^{-1}}$ for MultiCon, APCEMM, and CoCiP, the fraction of crystals left at the end of the contrail is 26, 82, and 11\,\%, and the mean crystal radius reaches 17, 9, and $51\,\mu\mathrm{m}$, respectively.} These field-level and integral comparisons provide the basis for the final assessment of the computational cost required by each framework.

\subsection{Parametric sweep}
\label{subsec:parametric_sweep}

{The comparison above concerns a single environment. To examine whether the differences between the three models persist when the inputs change, a parametric sweep of 28 cases is carried out: a baseline and 27 variations, in each of which one input (or one group of related inputs) is changed while the others keep their baseline values. The cases are listed in \mbox{Table~\ref{tab:sweep_cases}}. The sweep baseline differs from the comparison case of \mbox{Sect.~\ref{subsec:model_comparison}}: it has no wind shear, a 400\,m deep ice-supersaturated layer centred on the plume and a 12\,h integration period. The three models start from the same post-vortex plume in every case, and all of them are reduced to the same diagnostics per unit flight-path length, computed with the definitions used above.}

\begin{table}[htbp]
 \centering
 \caption{Cases of the parametric sweep. The baseline is listed first; each variation changes one input (or one group of
 related inputs) with respect to it. $z$ is altitude relative to the plume centre after the vortex phase; ISSR:
 ice-supersaturated layer.}
 \label{tab:sweep_cases}
 \begin{tabular}{p{0.27\linewidth}p{0.66\linewidth}}
  \hline
  \multicolumn{2}{l}{\textit{Baseline}}\\
  Atmosphere & $T_0=222\,\mathrm{K}$ at 10.5\,km (244.7\,hPa), lapse rate $6.5\,\mathrm{K\,km^{-1}}$; ISSR 400\,m deep
  ($-200<z\leq200\,\mathrm{m}$) with $s_{i,\mathrm{env}}=0.10$; $s_{i,\mathrm{env}}=-0.15$ outside it; no wind shear\\
  Aircraft and initial plume & B737-800-like, soot number emission index $\mathrm{EI}_n=10^{14}\,\mathrm{kg^{-1}}$;
  $n_{\ell,0}=3.1\times10^{11}\,\mathrm{m^{-1}}$, $r_0=1.9\,\mu\mathrm{m}$, $\sigma_{y,0}=9.5\,\mathrm{m}$,
  $\sigma_{z,0}=29\,\mathrm{m}$\\
  Diffusivities & $K_z=0.1\,\mathrm{m^2\,s^{-1}}$, $K_y=15\,\mathrm{m^2\,s^{-1}}$ (MultiCon and APCEMM)\\
  Integration period & 12\,h\\
  \hline
  \multicolumn{2}{l}{\textit{Variations}}\\
  Soot number emission index & $\mathrm{EI}_n=10^{13}$, $10^{15}\,\mathrm{kg^{-1}}$\\
  ISSR supersaturation & $s_{i,\mathrm{env}}=0.05$, 0.20, 0.30\\
  ISSR depth & 200, 800, 1600\,m\\
  Subsaturation outside the ISSR & $s_{i,\mathrm{env}}=-0.05$, $-0.30$\\
  Temperature & $T_0=212$, 217\,K\\
  Lapse rate & 0 (isothermal), $3\,\mathrm{K\,km^{-1}}$\\
  Wind shear & 2, 4, $8\times10^{-3}\,\mathrm{s^{-1}}$\\
  Initial plume size & $\sigma_{z,0}=60$, 120\,m; $\sigma_{y,0}=50$, 150\,m; both ($\sigma_{z,0}=120\,\mathrm{m}$,
  $\sigma_{y,0}=150\,\mathrm{m}$)\\
  Initial state & that of Sect.~\ref{subsec:plume_evolution}\\
  Diffusivities & $K_z=0.05$, $0.30\,\mathrm{m^2\,s^{-1}}$; $K_y=5$, $30\,\mathrm{m^2\,s^{-1}}$\\
  \hline
 \end{tabular}
\end{table}

\begin{figure}[p]
 \centering
 \includegraphics[width=0.95\textwidth]{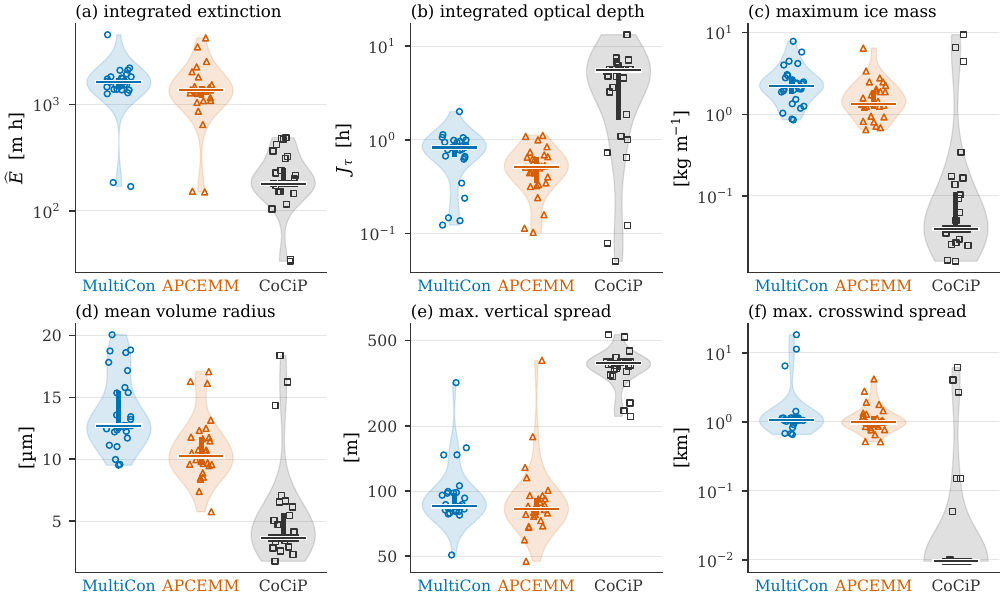}
 \caption{{Over the 28-case sweep MultiCon and APCEMM have similar distributions of every summary diagnostic except crystal
size, whereas CoCiP differs in most of them. Distribution over the 28 cases (one input varied at a time about a
baseline of 222\,K, ice supersaturation 0.10 in a 400\,m deep layer, no shear, 12\,h; same post-vortex plume in the
three models) of (a) the time-integrated total extinction $\widehat{E}$ (medians 1624, 1369 and
$180\,\mathrm{m\,h}$ for MultiCon, APCEMM and CoCiP), (b) the time-integrated optical depth $J_{\tau}$, (c) the maximum
ice mass per unit length, (d) the ice-mass-weighted mean volume radius, (e) the maximum vertical and (f) the maximum
crosswind standard deviation of the plume. In each panel: shaded violin, kernel density; open markers, the individual
cases (circles MultiCon, triangles APCEMM, squares CoCiP; horizontal position jittered for legibility only); thick bar,
interquartile range; horizontal line, median. Logarithmic ordinates in (a), (b), (c), (e) and (f). In (f) the CoCiP
spread stays at its initial value in the 25 cases without wind shear, because its horizontal diffusivity is
proportional to the shear and is zero in those cases; it grows in the three cases with shear.}}
 \label{fig:sweep_all}
\end{figure}

{\mbox{Figure~\ref{fig:sweep_all}} shows the distribution over the 28 cases of six summary diagnostics for each model. MultiCon and APCEMM have similar distributions of most of diagnostics, whereas CoCiP differs from both in most of them.} {The crosswind spread of CoCiP requires a remark: its horizontal diffusivity is proportional to the wind shear, $D_H=0.1\,S\,D^2$ with $D$ the plume depth \mbox{\cite{7}}, and therefore vanishes in the 25 cases of the sweep without shear, in which the CoCiP plume keeps its initial width.} {MultiCon and APCEMM are instead given a constant crosswind diffusivity, and their plumes spread in every case.}

\begin{figure}[p]
 \centering
 \includegraphics[width=0.76\textwidth]{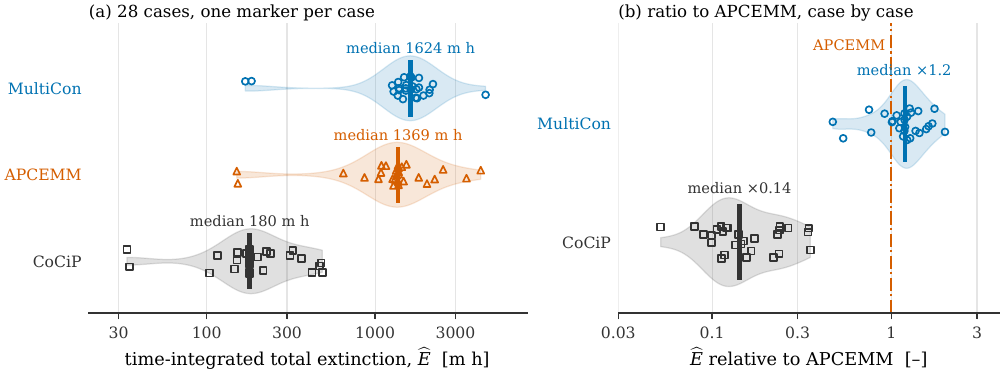}
 \\[1.5ex]
 \includegraphics[width=0.76\textwidth]{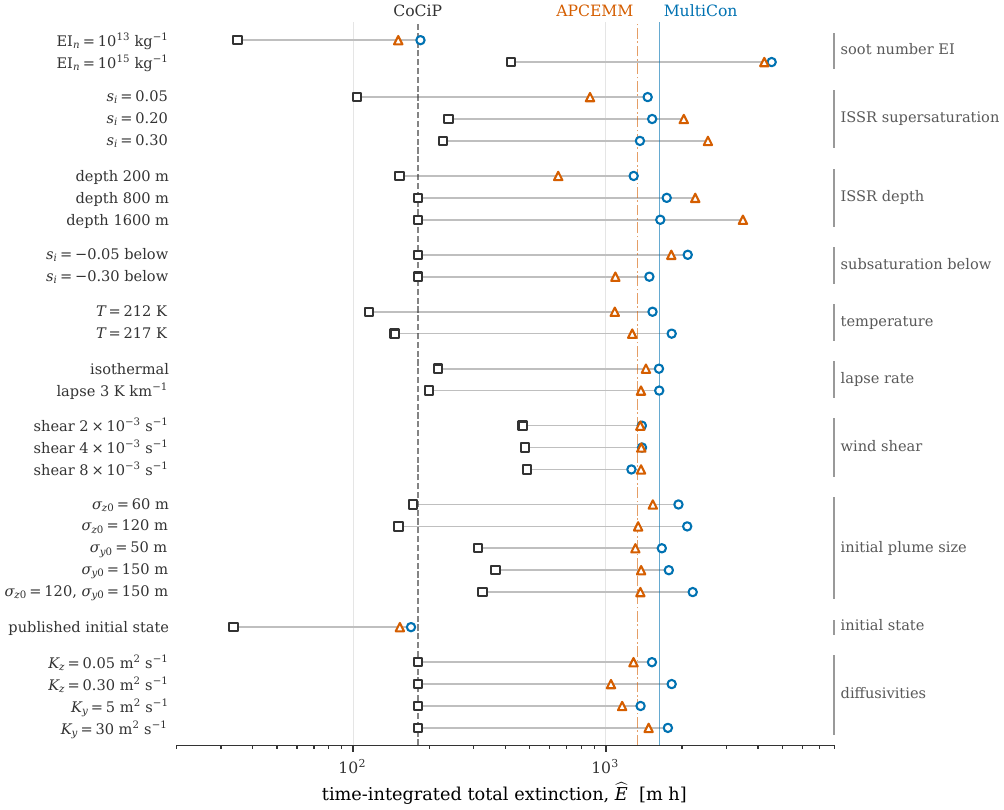}
 \caption{Over the 28-case parametric sweep MultiCon stays within a factor of about 2.1 of APCEMM in time-integrated total
extinction (median ratio 1.19, interquartile range 1.06--1.37; largest difference: ISSR depth 1600\,m), whereas CoCiP
gives a median of 0.14 times APCEMM and lies 3 to 19 times below it. Cases as in Table~\ref{tab:sweep_cases}; the three
models start from the same post-vortex plume; APCEMM temperature perturbation off. \textbf{Top}: (a) $\widehat{E}$, the
integral over the contrail lifetime of the cross-section-integrated extinction, for each model; one open marker per case
(vertical position jittered for legibility only), shaded violin: kernel density of $\log\widehat{E}$, thick bar: median.
(b) The same cases as ratios to APCEMM; the dash-dotted line at 1 is APCEMM. \textbf{Bottom}: $\widehat{E}$ of the three
models for each of the 27 one-at-a-time variations, grouped by the input that is varied (right margin); the grey
segment joins the three models of a case. Vertical lines: the baseline case of each model (solid MultiCon, dash-dotted
APCEMM, dashed CoCiP). Circles MultiCon, triangles APCEMM, squares CoCiP. Logarithmic abscissae. APCEMM is run
single-threaded; CoCiP: pycontrails implementation.}
 \label{fig:sweep_extinction}
\end{figure}

{The time-integrated total extinction is examined in more detail in \mbox{Fig.~\ref{fig:sweep_extinction}}. Over the 28 cases the ratio of the MultiCon value of $\widehat{E}$ to the APCEMM value has a median of 1.19 and an interquartile range of 1.06--1.37, and it stays within a factor of about 2.1 in every case; the largest difference occurs for the 1600\,m deep supersaturated layer. CoCiP gives a median of 0.14 times the APCEMM value and lies 3 to 19 times below APCEMM in the individual cases.}

\subsection{Computational cost}
\label{subsec:computational_cost}

\begin{figure}[htbp]
 \centering
 \includegraphics[width=0.78\textwidth]{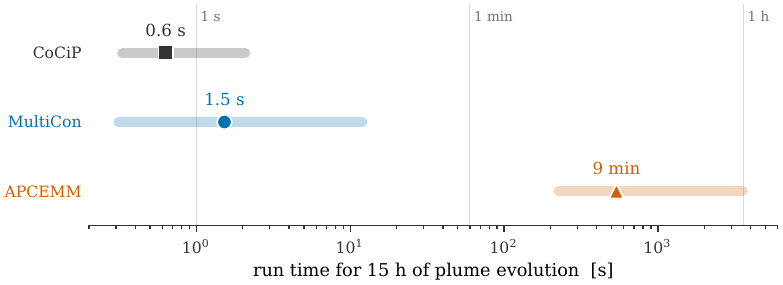}
 \caption{{MultiCon and CoCiP simulate 15\,h of plume evolution in about a second, APCEMM in minutes to an hour. Wall-clock
run time of one simulation of the common comparison case (218\,K, $s_{i,\mathrm{env}}=0.10$ above $z=-400\,\mathrm{m}$,
shear $5\times10^{-4}\,\mathrm{s^{-1}}$), single thread, same processor; logarithmic abscissa. Marker and label: the
set-up used for the comparison of Sect.~\ref{subsec:model_comparison} (Table~\ref{tab:runtime}). Shaded bar: range over
all the tested set-ups (CoCiP: 50 time steps from 30 to 300\,s; MultiCon: the same time steps with vertical spacings
from 2 to 10\,m; APCEMM: 20 configurations with time steps from 30 to 300\,s, vertical spacings from 2.5 to 10\,m and
30 to 80 size bins). Grey lines mark 1\,s, 1\,min and 1\,h.}}
 \label{fig:runtime}
\end{figure}

Wall-clock time is measured on the same hardware using the same timing boundaries for all models; {each simulation covers the 15\,h of the common comparison case of \mbox{Sect.~\ref{subsec:model_comparison}} and runs on a single thread of the same processor (Apple M2 Max).} {CoCiP is evaluated for 50 time steps between 30 and 300\,s;} MultiCon is evaluated at the same time steps while varying the vertical grid spacing from 2 to 10 m {(450 simulations).} {APCEMM is evaluated for 20 configurations, with time steps between 30 and 300\,s, vertical spacings between 2.5 and 10\,m, and 30 to 80 size bins.} {The run times are summarized in \mbox{Table~\ref{tab:runtime}} and \mbox{Fig.~\ref{fig:runtime}}.}

{For the numerical set-up of the comparison, a 15\,h simulation takes 0.6\,s with CoCiP, 1.5\,s with MultiCon, and 9\,min with APCEMM, i.e. MultiCon is about 350 times faster than APCEMM and about 2.5 times slower than CoCiP.} {Over all the set-ups tested, the run time ranges from 0.3 to 2.1\,s for CoCiP, from 0.3 to 12\,s for MultiCon, and from 4 to 59\,min for APCEMM.} 
% {At these rates, 100\,000 plume segments would require about 18\,h (CoCiP), 42\,h (MultiCon), and 1.7 years (APCEMM) of single-thread computation.}

{The three models differ in how much their result depends on the numerical set-up (\mbox{Table~\ref{tab:runtime}}).} {The time-integrated total extinction of CoCiP does not depend on its time step, varying by less than 3\,\% over the tested range.} {That of MultiCon varies between 5.6 and $8.0\,\mathrm{km\,h}$ and, for the set-up of the comparison, is 6\,\% above the value of its most refined set-up.} {That of APCEMM varies most, between 4.9 and $11.1\,\mathrm{km\,h}$; for the set-up of the comparison it is 22\,\% below the value of its most refined configuration ($\Delta t=60\,\mathrm{s}$, $\Delta z=3\,\mathrm{m}$, 56 size bins), which takes 59\,min.} {MultiCon has a restriction on its numerical set-up: for ratios $\Delta t/\Delta z$ above about $80\,\mathrm{s\,m^{-1}}$, i.e. fine grids combined with long time steps, the solution degrades; this affects 37 of the 450 simulations, which are excluded from the values given here, and the set-up of the comparison ($\Delta t/\Delta z=32\,\mathrm{s\,m^{-1}}$) is well within the usable range.} Conclusions are restricted to the hardware, model versions, and single-plume setup considered here.

\begin{table}[htbp]
 \centering
 \caption{{Run time of a 15\,h simulation of the common comparison case (single thread, same processor) and dependence
 of the time-integrated total extinction $\widehat{E}$ on the numerical set-up. First rows: the tested set-up nearest to that
 used in Sect.~\ref{subsec:model_comparison}. Ranges are over all the tested set-ups (50 for CoCiP, 413 for MultiCon,
 20 for APCEMM). Last row: $\widehat{E}$ of the first rows relative to that of the most refined set-up of the same model
 (CoCiP: $\Delta t=30\,\mathrm{s}$; MultiCon: $\Delta t=30\,\mathrm{s}$, $\Delta z=2\,\mathrm{m}$; APCEMM:
 $\Delta t=60\,\mathrm{s}$, $\Delta z=3\,\mathrm{m}$, 56 size bins).}}
 \label{tab:runtime}
 \begin{tabular}{p{0.30\linewidth}p{0.17\linewidth}p{0.19\linewidth}p{0.22\linewidth}}
  \hline
  {Quantity} & {CoCiP} & {MultiCon} & {APCEMM} \\
  \hline
  {Time step $\Delta t$} & {118\,s} & {118\,s} & {300\,s} \\
  {Vertical spacing $\Delta z$} & {--} & {4.0\,m} & {3.75\,m} \\
  {Size bins} & {--} & {--} & {38} \\
  {Run time} & {0.6\,s} & {1.5\,s} & {9\,min} \\
  {Run time relative to APCEMM} & {1/850} & {1/350} & {1} \\
  {Range of run time} & {0.3--2.1\,s} & {0.3--12\,s} & {4--59\,min} \\
  {$\widehat{E}$ [$\mathrm{km\,h}$]} & {7.3} & {6.1} & {8.3} \\
  {Range of $\widehat{E}$ [$\mathrm{km\,h}$]} & {7.3--7.5} & {5.6--8.0} & {4.9--11.1} \\
  {$\widehat{E}$ relative to the most refined set-up} & {$<1\,\%$} & {$+6\,\%$} & {$-22\,\%$} \\
  \hline
 \end{tabular}
\end{table}

\section{Conclusion}      

This study presented MultiCon, a reduced-order, multi-physics Eulerian framework for simulating the long-term evolution of individual aircraft contrail plumes. The proposed formulation combines the transport of ice-particle number and mass with water-vapour supersaturation and an additional equation governing ice-crystal habit evolution. Two physical mechanisms that are insufficiently represented in many existing reduced-order contrail models are incorporated: habit-dependent ice growth and sublimation, through an approximate inherent growth factor (IGF) parameterization and a diffusion-limited formulation accounting for latent-heat effects; and turbulence-modulated bulk settling, represented by a reduced-order Burgers-type closure. By exploiting dimensional separability and combining vertically resolved Eulerian transport with characteristic-based horizontal plume reconstruction, MultiCon provides a computationally efficient framework for capturing essential processes governing mature contrail evolution.

{The numerical simulations demonstrate the sensitivity of contrail evolution to atmospheric variability and microphysical closure parameters. In particular, comparisons between stationary and time-dependent atmospheric conditions show that changes in temperature and ambient ice supersaturation can affect plume lifetime and optical properties differently. The sensitivity analysis further identifies crosswind diffusivity and the aggregation-related closure parameters as important controls on integrated optical diagnostics, highlighting the importance of appropriately characterizing these processes in long-term contrail simulations.
}

{Comparisons with CoCiP and APCEMM reveal both similarities and differences in the predicted plume evolution. Across the 28-case parametric sweep, MultiCon and APCEMM exhibit broadly similar distributions for most investigated plume diagnostics, with the median ratio of their time-integrated total extinction being 1.19 and differences remaining within a factor of approximately 2.1. However, notable discrepancies persist in particle size and ice-mass evolution, particularly during the later stages of plume development. These differences emphasize the importance of spatial resolution and microphysical representation in determining long-term contrail characteristics.
}

{From a computational perspective, the reference 15-hour simulation requires approximately 1.5 s with MultiCon, compared with 0.6 s for CoCiP and 9 min for APCEMM. Thus, MultiCon retains computational efficiency comparable to CoCiP while being approximately 350 times faster than APCEMM under the tested conditions, offering a potentially advantageous balance between physical representation and computational expense.
}

Further developments should address the current locally monodisperse approximation, improve the representation of particle-size distributions and ice-crystal morphology, and calibrate uncertain closure parameters against observational data. Overall, MultiCon provides a promising foundation for investigating the coupled microphysical and dynamical processes governing long-term contrail evolution, systematic model intercomparison, and future large-scale aviation climate assessments.

\textbf{Data Availability Statement}\\
The codes for generating figures presented in this paper are publicly available at the following repository, enabling reproduction of the results reported in this study: \textcolor{black}{\href{https://doi.org/10.5281/zenodo.21534861}{link}}\\

\textbf{Conflict of Interest}\\
The authors declare no conflict of interest of any type within the present submission.\\

\textbf{Acknowledgment}\\
{The authors acknowledge financial support from Project CLIMATION (PID2024-158160OB-C31), funded by the Spanish Ministry of Science and Innovation, and Project E-CONTRAIL II (101287164), granted by SESAR JU. The authors express their sincere gratitude to Dr. Caleb Akhtar Martínez (University of Cambridge) for his generous and insightful guidance on APCEMM and CoCiP, as well as for valuable scientific discussions. The authors also gratefully acknowledge to Abolfazl Simorgh (Universidad Carlos III de Madrid) for his kind support with the PyContrails implementation and analysis of CoCiP outputs.}

%\bibliographystyle{plain} % Use a built-in bibliography style (e.g., plain, IEEE, etc.)
%\bibliography{Ref} % Specify your .bib file

	\section{Appendix}
	\appendix

\section{Monodisperse Moment Equations from the Population Balance Equation}\label{apA}

In this section, we formulate the macroscopic advection--diffusion equations (ADEs) for the number and mass concentrations of ice particles from the Population Balance Equation (PBE).

\subsection{Population Balance Equation (PBE)}

The PBE for the distribution function $f(\mathbf{x},m,\phi,t)$ is given by:
\begin{equation}\label{PBE}
    \frac{\partial f}{\partial t}
    + \nabla \cdot\big(\mathbf{v}_p\,f\big)
    =
    -\frac{\partial}{\partial m}\Bigl(\dot{m}\, f\Bigr)
    -\frac{\partial}{\partial \phi}\Bigl(\dot{\phi}\, f\Bigr)
    + \nabla \cdot\Bigl({{K}}\, \nabla f\Bigr)
    + S_f,
\end{equation}
where $\mathbf{v}_p(\mathbf{x},m,\phi,t)$ is the particle bulk velocity, defined as:
\begin{equation}\label{eq1}
    \mathbf{v}_p = \mathbf{v}_{{slp}} +
    \underbrace{(w_x,w_y,w_z)^\top}_{\scriptsize\shortstack{\text{background velocity} \\ \text{(wind components)}}}
    \approx (0,0,v_s)^\top +
    \underbrace{(w_x,w_y,w_z)^\top}_{\text{background velocity}}
    = (w_x,w_y,w_z+v_s)^\top.
\end{equation}
In Eq. (\ref{eq1}), $\mathbf{v}_{slp}$ denotes the bulk slip velocity of the particle phase within a fluid undergoing turbulent mixing.

In addition, $\dot{m}=\rho_\text{dep}\,f_\mathcal{V}(m,\phi,\mathbf{x},t)$ is the particle growth rate, $\dot{\phi}$ represents the evolution rate of the particle habit coordinate, ${{K}}$ is the diffusion coefficient, and $S_f(\mathbf{x},m,\phi,t)$ represents source/sink terms. Here, $m$ and $\phi$ represent two distinct internal coordinates: particle mass, accounting for polydispersity, and particle habit, respectively. The latter is not part of the polydisperse description but provides an additional coordinate for representing variations in ice-particle geometry.

The number and mass concentrations are defined as the zeroth and first moments of the distribution function over both internal coordinates:
\begin{equation}\label{eq:moments}
    c_N(\mathbf{x},t)
    =
    \int_{\Phi}\int_0^\infty
    f(\mathbf{x},m,\phi,t)\,dm\,d\phi,
    \qquad
    c_M(\mathbf{x},t)
    =
    \int_{\Phi}\int_0^\infty
    m\,f(\mathbf{x},m,\phi,t)\,dm\,d\phi,
\end{equation}
where $\Phi$ denotes the admissible habit space.

Following the joint monodisperse assumption in the internal coordinates, the particle distribution is represented by a single characteristic mass $\bar{m}(\mathbf{x},t)$ and habit coordinate $\bar{\phi}(\mathbf{x},t)$ at each spatial location:
\begin{equation}\label{eq:monodisperse}
    f(\mathbf{x},m,\phi,t)
    =
    c_N(\mathbf{x},t)\,
    \delta\Bigl(m-\bar{m}(\mathbf{x},t)\Bigr)\,
    \delta\Bigl(\phi-\bar{\phi}(\mathbf{x},t)\Bigr).
\end{equation}
Accordingly, the particle population is locally monodisperse in mass and represented by a single habit at each computational location. Substituting Eq. (\ref{eq:monodisperse}) into Eq. (\ref{eq:moments}) yields:
\begin{equation}\label{eq:monodisperse_moments}
    c_N(\mathbf{x},t)=c_N(\mathbf{x},t),
    \qquad
    c_M(\mathbf{x},t)=\bar{m}(\mathbf{x},t)c_N(\mathbf{x},t).
\end{equation}

Under this approximation, the particle growth rate, and bulk velocity are evaluated at the representative internal coordinates:
\begin{equation}
    \dot{m} \approx \rho_\text{dep}\,
    f_\mathcal{V}(\bar{m},\bar{\phi},\mathbf{x},t),\quad 
    \mathbf{v}_p \approx
    \mathbf{v}_p(\mathbf{x},\bar{m},\bar{\phi},t).
\end{equation}

Assuming vanishing particle fluxes at the boundaries of the internal-coordinate domains, the resulting number- and mass-concentration equations are:
\begin{equation}\label{eq:zeroth_moment}
    \frac{\partial c_N}{\partial t}
    + \nabla\cdot\left(\mathbf{v}_p c_N\right)
    =
    \nabla\cdot\left({{K}}\nabla c_N\right)
    + S_{c_N},\quad 
    \frac{\partial c_M}{\partial t}
    + \nabla\cdot\left(\mathbf{v}_p c_M\right)
    =
    \nabla\cdot\left({{K}}\nabla c_M\right)
    + \rho_\text{dep}\,
    f_\mathcal{V}(\bar{m},\bar{\phi},\mathbf{x},t)c_N
    + S_{c_M},
\end{equation}
where the corresponding source terms are defined as:
\begin{equation}
    S_{c_N}(\mathbf{x},t)
    =
    \int_{\Phi}\int_0^\infty
    S_f(\mathbf{x},m,\phi,t)\,dm\,d\phi,
    \qquad
    S_{c_M}(\mathbf{x},t)
    =
    \int_{\Phi}\int_0^\infty
    m\,S_f(\mathbf{x},m,\phi,t)\,dm\,d\phi.
\end{equation}

\section{Derivation of the Particle-Phase Momentum Equation}\label{apB}
In multiphase flow modeling, the Euler--Euler framework treats both the fluid and particulate phases as interpenetrating continua. To better conceptualize this, consider the particle phase settling within a fluid subject to turbulent mixing. Since the mean fluid motion (i.e., wind components) is already incorporated into the definition of $\mathbf{v}_p$, the fluid velocity appearing in the slip velocity $\mathbf{v}_{\text{slp}}$ accounts only for turbulent fluctuations. Under this setup, the full momentum equation for the particle phase takes the following form \cite{multiphase}:
	\begin{equation} \label{eq:full_mom}
		\frac{\partial (\epsilon_p \rho_p \mathbf{v}_{slp})}{\partial t} + \nabla \cdot \bigl( \epsilon_p \rho_p \, \mathbf{v}_{slp} \mathbf{v}_{slp} \bigr) = -\epsilon_p \, \nabla p + \nabla \cdot {\tau}_p + \epsilon_p \rho_p \, \mathbf{g} + {F}_d,
	\end{equation}
	where \(\epsilon_p\) is the particle volume fraction, \(\rho_p\) is the particle density, \(\mathbf{v}_{slp}\) is the particle bulk slip velocity, \(p\) is the pressure due to the fluid phase, \(\tau_p\) is the particle-phase stress tensor (namely deviatoric steress emerging on the exterior of the control volume), \(\mathbf{g}\) is the gravitational acceleration vector, and \({F}_d\) is the net force per unit volume acting on the particles in the control volume. In many closures the net force is modeled as proportional to \(\epsilon_p\) (e.g., arising from the per-particle contribution multiplied by the local particle concentration, i.e., $F_d = \epsilon_p \, f_d$ where $f_d$ is the average net force experienced by individual particles which is inherently different from the drag force experienced by a single settling particle in an unbounded domain). Local mechanical equilibrium \(p_p=p_f\) is assumed, so that the pressure gradient acting on the particle phase is \(-\epsilon_p\,\nabla p\). The stress tensor can be expressed as: $\tau_p = \epsilon_p \, \mu_{t,p}(\nabla\mathbf{v}_{slp}+\nabla\mathbf{v}_{slp}^\top)$.
	
	Expanding the LHS of Eq. (\ref{eq:full_mom}), we write: 
	\begin{equation}
		\begin{split}
			&\frac{\partial (\epsilon_p \rho_p \mathbf{v}_{slp})}{\partial t}+	\nabla \cdot \bigl( \epsilon_p \rho_p \, \mathbf{v}_{slp} \mathbf{v}_{slp} \bigr)\\
			&= \rho_p \mathbf{v}_{slp} \frac{\partial \epsilon_p}{\partial t} + \epsilon_p \rho_p \frac{\partial \mathbf{v}_{slp}}{\partial t}+\rho_p \mathbf{v}_{slp} \, \nabla \cdot (\epsilon_p\, \mathbf{v}_{slp}) + \epsilon_p \rho_p (\mathbf{v}_{slp} \cdot \nabla) \mathbf{v}_{slp}\\
			&= \rho_p \mathbf{v}_{slp}\underbrace{\left[ \frac{\partial \epsilon_p}{\partial t} + \nabla \cdot (\epsilon_p \, \mathbf{v}_{slp}) \right]}_{\substack{\\ \text{continuity equation of} \\ \text{the particle phase}}} + \epsilon_p \rho_p \left[\frac{\partial \mathbf{v}_{slp}}{\partial t} + (\mathbf{v}_{slp} \cdot \nabla) \mathbf{v}_{slp} \right].
		\end{split}
	\end{equation}
	
	For the stress term, we write:
	\begin{equation}
		\frac{1}{\epsilon_p \rho_p}\nabla \cdot \tau_p
		= \frac{\nu_{t}}{\epsilon_p} \nabla \cdot [\epsilon_p (\nabla \mathbf{v}_{slp} + \nabla \mathbf{v}_{slp}^\top)]
		= \nu_{t} \Bigl[\nabla^2 \mathbf{v}_{slp}
		+ \frac{1}{\epsilon_p}(\nabla \epsilon_p \cdot \nabla) \mathbf{v}_{slp}
		+ \frac{1}{\epsilon_p}(\nabla \epsilon_p \cdot \nabla)^\top \mathbf{v}_{slp}\Bigr]
	\end{equation}
	where $\nu_{t}:=\frac{\mu_{t,p}}{\rho_p}$ is the average turbulent eddy-viscosity.
	
	Substituting the stress term into Eq.~(\ref{eq:full_mom}) and invoking the first moment equation to represent particle-phase continuity, where \( c_M = \rho_{\text{dep}} \, \epsilon_p = \rho_p \, \epsilon_p \), \(\nabla \cdot \mathbf{v}_{\text{slp}} \approx 0\), and no additional particle injection is assumed (i.e., \( S_{c_M} = 0 \)), while also noting that the advection term has been replaced with \(\mathbf{v}_{\text{slp}}\), we obtain:
	\begin{equation} \label{eq:final}
		\begin{split}
			&\frac{\partial \mathbf{v}_{slp}}{\partial t} + (\mathbf{v}_{slp} \cdot \nabla) \mathbf{v}_{slp}
			= \nu_{t} \Bigl[\nabla^2 \mathbf{v}_{slp}
			+ \tfrac{1}{\epsilon_p}(\nabla \epsilon_p \cdot \nabla) \mathbf{v}_{slp}
			+ \tfrac{1}{\epsilon_p}(\nabla \epsilon_p \cdot \nabla)^\top \mathbf{v}_{slp}\Bigr]-\frac{\nabla p}{\rho_p}
			+ \mathbf{g}
			+ \frac{f_d}{\rho_p}-\\
			&\frac{\mathbf{v}_{slp}}{\epsilon_p}\big[\nabla \cdot\Bigl(\tilde{\mathcal{D}}\, \nabla \epsilon_p\Bigr) +\, f_v\bigl(\bar{m}(\mathbf{x},t),\mathbf{x},t\bigr) \, c_N(\mathbf{x},t)\big].
		\end{split}
	\end{equation}
	
	In general, the above equation is coupled with the fluid phase through the pressure and force terms, and with the first and second moments of the particle distribution. Notably, as \( f_d \) approaches the drag force on a single settling particle in an unbounded domain, the pressure gradient tends toward the hydrostatic condition, \( \nabla p = \rho_f \mathbf{g} \).
	
	The above equation can be written as:
	\begin{equation}\label{eq20}
		\begin{split}
			&\frac{\partial \mathbf{v}_{slp}}{\partial t} + (\mathbf{v}_{slp} \cdot \nabla) \mathbf{v}_{slp}
			= \nu_{t} \nabla^2 \mathbf{v}_{slp} +\bm{C}_{f}+\bm{C}_{c_M,c_N,f_v}.
		\end{split}
	\end{equation}	
	where
	$\bm{C}_{f}$ represents the coupling between $\mathbf{v}_{slp}$ and the fluid phase, and is expressed as 
	$\bm{C}_{f}=-\frac{\nabla p}{\rho_p} + \mathbf{g} + \frac{f_d}{\rho_p}$; 
	and $\bm{C}_{c_M,c_N,f_v}$ captures the coupling of $\mathbf{v}_{slp}$ to the mass and number concentration fields, as well as to the growth term, and is given by 
	$\bm{C}_{c_M,c_N,f_v}=-\frac{\mathbf{v}_{slp}}{\epsilon_p}\big[\nabla \cdot\Bigl(\tilde{\mathcal{D}}\, \nabla \epsilon_p\Bigr) +\, f_v\bigl(\bar{m}(\mathbf{x},t),\mathbf{x},t\bigr) \, c_N(\mathbf{x},t)\big]+\nu_{t}\Bigl[ 
	\tfrac{1}{\epsilon_p}(\nabla \epsilon_p \cdot \nabla) \mathbf{v}_{slp} 
	+ \tfrac{1}{\epsilon_p}(\nabla \epsilon_p \cdot \nabla)^\top \mathbf{v}_{slp}\Bigr]$.
	
	Accurate solution of the above equation presents formidable challenges in (i) specifying boundary and initial conditions, (ii) coupling the evolving fluid phase with particle number and mass concentrations, and (iii) incorporating microphysical growth rates. In addition, in practice, the eddy viscosity is not only space- and time-dependent, but also anisotropic-varying across different spatial directions. 

    Therefore, the general particle-phase momentum equation derived above provides the starting point for formulating the effective bulk settling velocity. However, its direct solution requires additional closures for the fluid–particle interactions and associated concentration fields. To obtain a computationally tractable representation suitable for the reduced-order contrail framework, we introduce a compressed low-order formulation that captures the leading-order influence of turbulent mixing on vertical particle settling, as detailed below.

\subsection{A Compressed Low-Order Model}
The development of the bulk settling-velocity model is motivated by the limitations of conventional approaches based on the terminal velocity of isolated particles. Specifically, traditional approaches to particle settling rely on the terminal velocity derived for individual particles in an unbounded domain using Stokes' law. However, in an Eulerian framework that describes the bulk concentration of particles (such as ice particles in contrails), the settling velocity must reflect the collective behaviour of particles that are also subject to turbulent mixing. In high-altitude regions, significant turbulent mixing implies that the ice particles initially 'hover' within the eddy-viscous layer while undergoing growth (provided that the growth conditions are met). This motivates a model in which the effective settling velocity is determined not only by gravitational and drag forces but also by turbulent (or self-) diffusion. To differentiate notations, we use \(v_s\) for the particle effective/bulk settling velocity and \(v_{\mathrm{ter}}\) for the terminal settling velocity, typically derived from Stokes' law (or alternative empirical formulas) for a single settling particle in an unbounded domain.

In turbulent flows, small particles can experience a phenomenon known as loitering. Due to their ability to be readily carried by rapidly fluctuating eddies (sometimes referred to as the eddy-locking, analogous to behaviour seen in nanofluids \cite{nano}; \cite{nano1}, these particles sample a broad range of flow directions and, as a consequence, effectively experience enhanced nonlinear drag and may follow a longer trajectory to settle across a given vertical distance. This results in an average fall speed that is reduced relative to predictions based on quiescent conditions. A second mechanism, commonly referred to as preferential sweeping, occurs when particles possess sufficient inertia to partially decouple from the turbulent eddies. They are centrifuged out of vortical cores and thus become concentrated in downward-moving regions of the flow, causing them to sample stronger downward velocity fluctuations and hence to acquire an enhanced mean settling velocity. The relative influence of these two mechanisms depends on the degree to which the particles are able to track turbulent fluctuations. Very small crystals tend to follow the turbulence closely, resulting in persistent agitation and reduced descent rates due to the loitering effect. Preferential sweeping becomes significant when particles grow large enough, or turbulence weakens sufficiently, to allow them to slip out of the smallest vortices. 

In the specific case of contrail cirrus, where ice crystals remain very small under considerable atmospheric turbulence, the loitering mechanism is expected to dominate the early stages of evolution. Over time, as the particles grow by nearly an order of magnitude, they are expected to begin sweeping toward their terminal velocity (or even exceed it). 

Specifically, this behaviour can be quantified using the Stokes number, \(St_\eta=\tau_p/\tau_\eta\), where \(\tau_p=\rho_i d_p^2/(18\mu_a)\) and \(\tau_\eta=(\nu_a/\epsilon)^{1/2}\), and the Galileo number, \(Ga=\sqrt{(\rho_i/\rho_a-1)gd_p^3}/\nu_a\). For representative cruise-altitude conditions (\(T=218\) K, \(p=250\) hPa, \(\rho_a=0.4\ \mathrm{kg\,m^{-3}}\), \(\mu_a=1.4\times10^{-5}\ \mathrm{Pa\,s}\), \(\rho_i=917\ \mathrm{kg\,m^{-3}}\), and \(\epsilon=10^{-5}\ \mathrm{m^2\,s^{-3}}\)), particles with radii between 1 and 17 µm yield \(St_\eta\sim10^{-5}-10^{-3}\) and \(Ga\sim10^{-2}-10^0\). {These relatively small values indicate weak particle inertia and strong coupling to turbulent eddies, suggesting that loitering dominates over preferential sweeping during the initial and intermediate stages of contrail evolution. As crystals grow and turbulence decays, preferential sweeping may become increasingly relevant (interested readers are referred to} \cite{33}; \cite{34}; \cite{35}; \cite{36}; \cite{setnew}).

The reduced-order model developed here captures these effects; specifically, the initial loitering behaviour decays asymptotically into preferential sweeping, thereby recovering the terminal velocity.

In turbulent flows, particularly under quasi-isotropic conditions, it is reasonable to assume that the turbulent diffusivity and the eddy viscosity are related by a turbulent Schmidt number, \(Sc_t\), such that \({K}\approx\nu_t/Sc_t\). For many atmospheric and geophysical flows, the turbulent Schmidt number is of order unity. In our formulation, we adopt an effective turbulent viscosity \(\nu_{t,\mathrm{ef}}\) that represents the three-dimensional averaged mixing. This is justified by the fact that the overall turbulent mixing experienced by the particles is inherently three-dimensional. Thus, we write \(\nu_{t,\mathrm{ef}}\approx\langle{K}(x,y,z)\rangle\), accounting for the total mixing of the particle-laden flow. The directional diffusivities vary with ambient turbulence and stratification; for far-wake, cruise-altitude contrails we adopt \({K}_x={K}_y\in[10,40]\,\mathrm{m^2\,s^{-1}}\), and \({K}_z\in[0.05,0.50]\,\mathrm{m^2\,s^{-1}}\), giving the arithmetic mean \(\nu_{t,\mathrm{ef}}\in[6.68,26.83]\,\mathrm{m^2\,s^{-1}}\).

The full Euler–Euler particle-phase momentum equation, including the coupling terms associated with the carrier-fluid phase and the particle concentration and growth fields, is given by Eq. (B5). Fully resolving this equation requires additional closures for the fluid phase, as well as for \(f_d\), which represents the average drag force experienced by individual particles, distinct from the drag force on a single settling particle in an unbounded domain. Its solution also depends critically on the specification of appropriate boundary and initial conditions.

The above equation can be expressed solely in terms of the particle bulk settling velocity \(v_s\) in the \(z\)-direction by collapsing the turbulent mixing in the vertical dimension. In other words, the 'quasi-hovering' (loitering) behaviour of horizontally distributed particles can be interpreted as an effective turbulent mixing in the vertical direction. For this reason, this type of one-dimensional modelling is referred to as a compressed low-order model.

This framework motivates defining an effective vertical eddy viscosity, \(\nu_{t,\mathrm{ef}}\approx\langle\nu_t(x,y,z)\rangle\), which encapsulates the total turbulent mixing effect.

Here, we assume that the particle bulk velocity is initially zero at a reference plane, \(z_{\mathrm{ref}}\), where the concentration is highest, and remains negligible for an extended period due to the strong early-time turbulent mixing effect. Far below \(z_{\mathrm{ref}}\), the local concentration decreases, and particle growth amplifies gravitational settling. This leads to an increase in both the Galileo and Stokes numbers. Consequently, the bulk net force \(f_d\) asymptotically converges to the drag force experienced by an individual particle in an unbounded domain, and the pressure gradient approaches hydrostatic balance, \(\nabla p=\rho_f\mathbf{g}\).

Within this framework, the fluid coupling term \(C_f\) is encoded within the boundary conditions. For simplicity, the coupling term associated with the concentration field, \(C_{c_M,c_N,f_V}\), is neglected. Nonetheless, because the bulk settling model is coupled to the microphysical ADEs, the effects of particle growth are implicitly represented through the time dependence of the terminal velocity, \(v_{\mathrm{ter}}:=v_{\mathrm{ter}}(z,t)\). Therefore, with appropriately chosen boundary conditions, Eq. (B5) reduces to the following one-dimensional model:
\begin{equation}
    \begin{split}
        &\frac{\partial v_s}{\partial t}
+v_s\frac{\partial v_s}{\partial z}
=\nu_{t,\mathrm{ef}}\frac{\partial^2v_s}{\partial z^2},\\
& v_s(z_{\mathrm{ref}},t=0)=0,\qquad
v_s(z\ll z_{\mathrm{ref}},t=0)=v_{\mathrm{ter}},
\qquad
\lim_{t\to\infty}v_s=v_{\mathrm{ter}}.
    \end{split}
\end{equation}

Here, \(v_{\mathrm{ter}}\) denotes the conventional terminal velocity of individual particles in an unbounded domain. Notably, by prescribing \(v_{\mathrm{ter}}\) as the outer boundary condition, the model inherently captures the interplay between (i) turbulent mixing in the highly concentrated upper region (loitering), (ii) the transition of the bulk net force \(f_d\) into the drag force on individual particles, and (iii) the acceleration of this transition by microphysical growth processes.

In addition, by expressing the asymptotic settling velocity as \(\lim_{t\to\infty}v_s=\alpha v_{\mathrm{ter}}\), where \(\alpha\propto St\) or \(Ga\), the model more accurately captures the sweeping mechanism. By allowing \(\alpha\) to exceed unity in inertia-dominated regimes, this formulation can represent the increased settling velocity associated with the sweeping mechanism.

We highlight that, by deliberately encoding the body force within the boundary conditions, the model circumvents the uncertainties and complexities associated with explicitly coupling the body force to the fluid phase and concentration field.

The governing equation requires a closure model to define the threshold distance beyond which turbulent mixing, characterized by the eddy viscosity \(\nu_{t,\mathrm{ef}}\), effectively smooths velocity deviations over a vertical extent denoted by \(z_{\mathrm{relax}}\). To characterize \(z_{\mathrm{relax}}\), we can either consider a local or global diffusion–advection balance. In local diffusion–advection balance, \(z_{\mathrm{relax}}\) is estimated by equating vertical advection and turbulent mixing rates: $v_{\mathrm{ter}}\frac{\Delta v_s}{\delta}
\sim
\nu_{t,\mathrm{ef}}\frac{\Delta v_s}{\delta^2},
\delta\equiv z_{\mathrm{relax}}\sim
\frac{\nu_{t,\mathrm{ef}}}{v_{\mathrm{ter}}}.$ However, in a global diffusion–advection balance, particles sample turbulent structures of size \(L_e\) as they settle; for long-term contrails, \(L_e\) represents the characteristic size of dominant turbulent structures, typically \(10\text{–}10^3\,\mathrm{m}\). The advection time across one eddy is defined as \(\tau_{\mathrm{adv}}:=L_e/v_{\mathrm{ter}}\), during which turbulent diffusion spreads the velocity deviations over a distance: $\delta\equiv z_{\mathrm{relax}}
\sim
\sqrt{\nu_{t,\mathrm{ef}}\tau_{\mathrm{adv}}}
=
\sqrt{\frac{\nu_{t,\mathrm{ef}}L_e}{v_{\mathrm{ter}}}}.$ Therefore, we rewrite Eq. (B6), adapted to both the axis direction and the settling sign on defining \(\hat v_s:=|v_s|\), together with the initial condition, as:
\begin{equation}
    \frac{\partial\hat v_s}{\partial t}
+\hat v_s\frac{\partial\hat v_s}{\partial z}
=
\nu_{t,\mathrm{ef}}\frac{\partial^2\hat v_s}{\partial z^2},
\qquad
\hat v_s(z,0)=
\begin{cases}
v_{\mathrm{ter}},&z\leq z_{\mathrm{relax}},\\
0,&z>z_{\mathrm{relax}}.
\end{cases}
\end{equation}
The resulting nonlinear advection–diffusion equation admits a closed-form solution through the Cole–Hopf transformation. The corresponding analytical derivation is presented in the following subsection.

\subsection{Closed-Form Solution for the Bulk Settling Velocity $v_s(z,t)$}
Having established the reduced-order governing equation and its associated initial condition in Appendix B.1, we now derive its analytical solution. The derivation employs the classical Cole–Hopf transformation, which converts the nonlinear Burgers equation into a linear heat equation.

Employing the Cole--Hopf transformation,
\begin{equation}\label{eq:colehopf}
	\hat{v}_s(z,t) = -2\nu_\textbf{t,ef}\,\frac{\partial}{\partial z}\ln\varphi(z,t) = -2\nu_\textbf{t,ef}\frac{\frac{\partial \varphi}{\partial z}}{\varphi}.
\end{equation}
where \(\varphi\) satisfies the linear heat equation:
\begin{equation}\label{eq:heat}
	\frac{\partial \varphi}{\partial t} = \nu_\textbf{t,ef}\,\frac{\partial^2 \varphi}{\partial z^2}.
\end{equation}
with the initial condition $\varphi(z,0)=:\varphi_0(z)$ as:
\begin{equation}
	\varphi_0(z)=\exp\!\Big(-\frac{1}{2\nu_\textbf{t,ef}}\int_{z_{\textbf{relax}}}^z \hat{v}_s(\xi,0)\,d\xi\Big)=
	\begin{cases}
		\exp\!\big(-\dfrac{v_\textbf{ter}}{2\nu_\textbf{t,ef}}(z-z_{\textbf{relax}})\big), & z\le z_{\textbf{relax}},\\
		1, & z>z_{\textbf{relax}}.
	\end{cases}.
\end{equation}

The solution of the heat equation with initial data \(\varphi_0\) is given by convolution with the heat kernel:
\begin{equation}
	G(\xi,t)=\frac{1}{\sqrt{4\pi\nu_\textbf{t,ef} t}}\,\exp\!\Big(-\frac{\xi^2}{4\nu_\textbf{t,ef} t}\Big),
	\qquad t>0,
\end{equation}
so for \(t>0\)
\begin{equation}
	\varphi(z,t)=\int_{-\infty}^{\infty} G(z-y,t)\,\varphi_0(y)\,dy.
\end{equation}
Splitting the integral at \(y=z_{\textbf{relax}}\), and setting $\Delta_0:=z-z_{\textbf{relax}}$, we write:
\begin{equation}\label{eqvs1}
	\varphi(z,t)= \int_{-\infty}^{z_{\textbf{relax}}} G(z-y,t)\,e^{-\frac{v_\textbf{ter}}{2\nu_\textbf{t,ef}}(y-z_{\textbf{relax}})}\,dy
	\;+\; \int_{z_{\textbf{relax}}}^{\infty} G(z-y,t)\,dy =: I_1 + I_2.
\end{equation}
\(I_2\) can be computed as:
\begin{equation}
	I_2 = \frac{1}{\sqrt{4\pi\nu_\textbf{t,ef} t}}\int_{z_{\textbf{relax}}}^{\infty}
	\exp\!\Big(-\frac{(z-y)^2}{4\nu_\textbf{t,ef} t}\Big)\,dy
	= \frac{1}{2}\Big(1+\operatorname{erf}\!\Big(\frac{\Delta_0}{2\sqrt{\nu_\textbf{t,ef} t}}\Big)\Big).
\end{equation}
For \(I_1\), we define \(s:=y-z_{\mathrm{relax}}\) so \(s\in(-\infty,0]\). Thus:
\begin{equation}
	I_1 = \frac{1}{\sqrt{4\pi\nu_\textbf{t,ef} t}}
	\int_{-\infty}^0 \exp\!\Big(-\frac{(\Delta_0-s)^2}{4\nu_\textbf{t,ef} t}-\frac{v_\textbf{ter}}{2\nu_\textbf{t,ef}}s\Big)\,ds.
\end{equation}
which can be written as:
\begin{equation}
	\begin{split}
	    &I_1
		= \exp\!\Big(\frac{v_\textbf{ter}^2 t}{4\nu_\textbf{t,ef}}-\frac{v_\textbf{ter}\Delta_0}{2\nu_\textbf{t,ef}}\Big)\cdot
		\frac{1}{\sqrt{4\pi\nu_\textbf{t,ef} t}}\int_{-\infty}^0
		\exp\!\Big(-\frac{(s-(\Delta_0 - v_\textbf{ter} t))^2}{4\nu_\textbf{t,ef} t}\Big)\,ds=\\
		&\exp\!\Big(\frac{v_\textbf{ter}^2 t}{4\nu_\textbf{t,ef}}-\frac{v_\textbf{ter}\Delta_0}{2\nu_\textbf{t,ef}}\Big)\cdot
		\frac{1}{2}\Big(1+\operatorname{erf}\!\Big(\frac{v_\textbf{ter} t - \Delta_0}{2\sqrt{\nu_\textbf{t,ef} t}}\Big)\Big).
	\end{split}
\end{equation}
Therefore, $\varphi(z,t)$ is obtained as $\varphi(z,t)=I_1+I_2$, and recovering \(\hat{v}_s\) using \eqref{eq:colehopf}, gives:
\begin{equation}\label{eq:final-u}
     \hat{v}_s(z,t) \;=\; v_\textbf{ter}\; \frac{\big(1+\erf(p_0)\big)\,p_2}{\big(1+\erf(p_0)\big)\,p_2 + \big(1+\erf(p_1)\big)}
\end{equation}
where:
\begin{equation}
	p_0:=\frac{v_\textbf{ter} t - \Delta_0}{2\sqrt{\nu_\textbf{t,ef} t}},\qquad
	p_1:=\frac{\Delta_0}{2\sqrt{\nu_\textbf{t,ef} t}},\qquad
	p_2:=\exp\!\Big(\frac{v_\textbf{ter}^2 t}{4\nu_\textbf{t,ef}}-\frac{v_\textbf{ter}\Delta_0}{2\nu_\textbf{t,ef}}\Big).
\end{equation}
The above equation can be also written in a more compact form by using the identity \((1+\erf(\xi))\)= \(\operatorname{erfc}(-\xi)\).

\subsection{Terminal Velocity, $v_\textbf{ter}$}
We present the method used to compute the terminal velocity $v_\textbf{ter}$ of randomly oriented spheroidal ice particles in a quiescent fluid under gravity, using the drag model by \cite{Ganser}.  Recall that for spheroids $\mathcal{V} = \tfrac{4}{3}\pi a^3\phi$, where $a$ is the equatorial semi‐axis and $\phi=c/a$ the aspect ratio.  The Ganser model requires the following geometrical definitions: the volume‐equivalent diameter $d_v=\bigl(6\mathcal{V}/\pi\bigr)^{1/3}=a\,(8\phi)^{1/3}$, and sphericity $\Psi=\pi d_v^2/S$ with $S$ being the total surface area of the spheroid, computed as 1) {Oblate} ($\phi<1$, $e=\sqrt{1-\phi^2}$): $S=2\pi a^2\bigl[1+\tfrac{1-e^2}{e}\tanh^{-1}(e)\bigr]$, 2) {Prolate} ($\phi>1$, $e=\sqrt{1-\phi^{-2}}$): $S=2\pi a^2\bigl[1+\tfrac{\phi}{e}\sin^{-1}(e)\bigr]$, and 3) {Sphere} ($\phi=1$): $S=4\pi a^2$.  
%\begin{itemize}
%\item \textbf{Oblate} ($\phi<1$, $e=\sqrt{1-\phi^2}$): $S=2\pi a^2\bigl[1+\tfrac{1-e^2}{e}\tanh^{-1}(e)\bigr]$;
%\item \textbf{Prolate} ($\phi>1$, $e=\sqrt{1-\phi^{-2}}$): $S=2\pi a^2\bigl[1+\tfrac{\phi}{e}\sin^{-1}(e)\bigr]$;
%\item \textbf{Sphere} ($\phi=1$): $S=4\pi a^2$.
%\end{itemize}

In addition, Ganser defines $d_n$ as the diameter of a sphere whose projected area normal to the motion equals that of the spheroid: thus one writes $A_{\rm proj}=\pi d_n^2/4$, and since the orientation‐averaged projected area satisfies $\langle A_{\rm proj}\rangle=S/4$ by Cauchy’s theorem, one has $d_n=\sqrt{4\,\langle A_{\rm proj}\rangle/\pi}=d_s$, where $d_s$ satisfies $S=\pi d_s^2\,(\Rightarrow d_s=\sqrt{S/\pi})$.  
	
Stokes’ shape factor (unbounded fluid) is $K_1=\tfrac13(d_n/d_v)+\tfrac23(d_s/d_v)$, and the generalized Reynolds number is $Re^*=Re\,K_1\,K_2$ with $Re=\frac{\rho_f v_\textbf{ter} d_v}{\mu_\text{ef}}$, where $\mu_\text{ef}$ is the effective dynamic viscosity.
	
Newton’s shape factor is fitted as $K_2=10^{1.8148(-\log_{10}\Psi)^{0.5743}}$.  
	
With these, Ganser’s drag model is given by:
	\begin{equation}
		C_D=\frac{24}{Re^*}\bigl(1+0.1118\,(Re^*)^{0.6567}\bigr)K_2+\frac{0.4305\,K_2}{1+3305/Re^*}.
	\end{equation}
	To determine the terminal velocity, we write:
	\begin{equation}
		\tfrac12\rho_f v_{ter}^2C_D\langle A_{\rm proj}\rangle=(\rho_p-\rho_f)\mathcal{V}g.
	\end{equation}  
	Let $B=\frac{4(\rho_p-\rho_f)\rho_fg d_v^3}{3\mu_\text{ef}^2}$, where $\mu_\text{ef}=\frac{\mu}{C(r_\text{eff})}$ with $C(r_\text{eff})$ being the Cunningham correction factor accounting for slip effects at very small particle sizes (equivalent to non‐continuum effects at high Knudsen numbers).    
	
	The Cunningham correction factor $C(a^*)$ is defined as:
	\begin{equation}
		C(r_\text{eff})=1 + \frac{\gamma}{r_\text{eff}}\Bigl(1.257 + 0.400\,e^{-1.1\,\frac{r_\text{eff}}{\gamma}}\Bigr).
	\end{equation}
	where $\gamma$ is the mean free path and $r_\text{eff}$ is the particle’s effective radius defined as $r_\text{eff}=a\,\phi^{1/3}$.   
	
	Substituting $v_\textbf{ter}=\frac{Re\,\mu_\text{ef}}{\rho_f d_v}$, gives the following implicit equation:
	\begin{equation}
		Re^2\,C_D(Re^*)=B.
	\end{equation}
	which is solved numerically for $Re$. Therefore, the terminal velocity is retrieved as:
	\begin{equation}
		v_\textbf{ter}=\frac{Re\,\mu_{ef}}{\rho_f d_v}.
	\end{equation}
    Notably, because $\langle A_{\rm proj}\rangle$ is the orientation‐averaged projected area, Ganser’s drag model inherently accounts for random tumbling orientation.

\section{Preliminary Results for the Bulk Settling Velocity $v_s$}\label{apC}	
\begin{figure}[H]
    \centering
    \begin{minipage}[c]{0.52\textwidth}
        \centering
        \includegraphics[width=\textwidth]{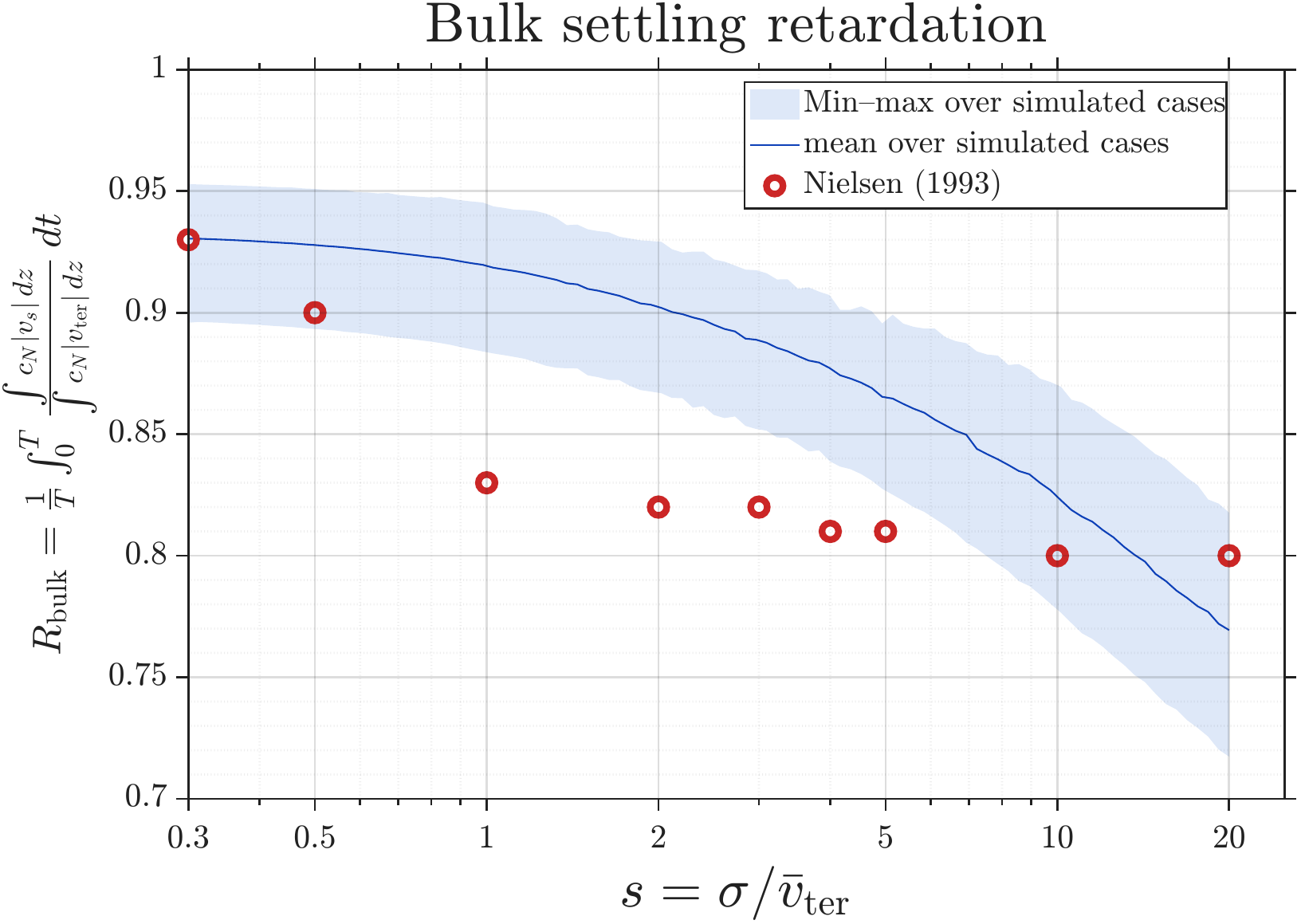}
        \subcaption{Preliminary comparison}
        \label{fig:settling-sensitivity}
    \end{minipage}%
    \hspace{0.01\textwidth}%
    \begin{minipage}[c]{0.25\textwidth}
        \centering
        \includegraphics[width=\textwidth]{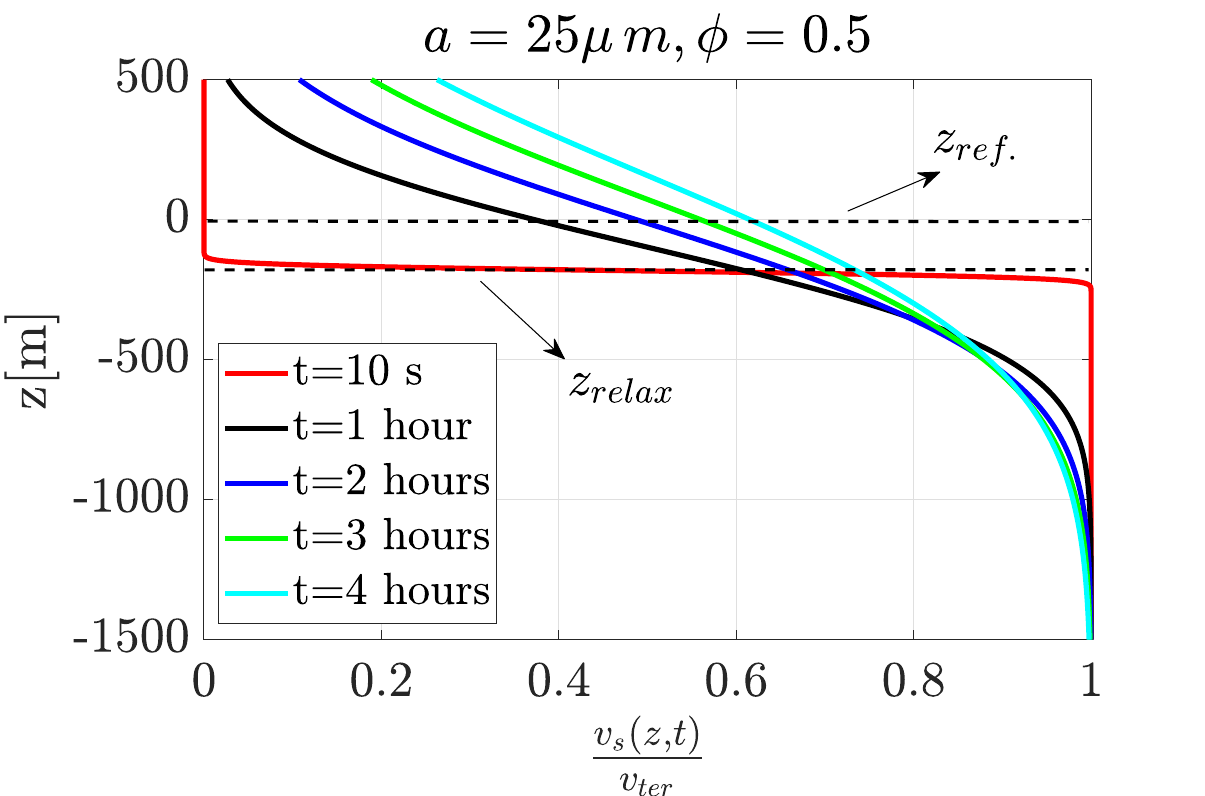}
        \subcaption{$\phi=0.5$}
        \label{fig:settling-phi05}

        \vspace{0.0cm}

        \includegraphics[width=\textwidth]{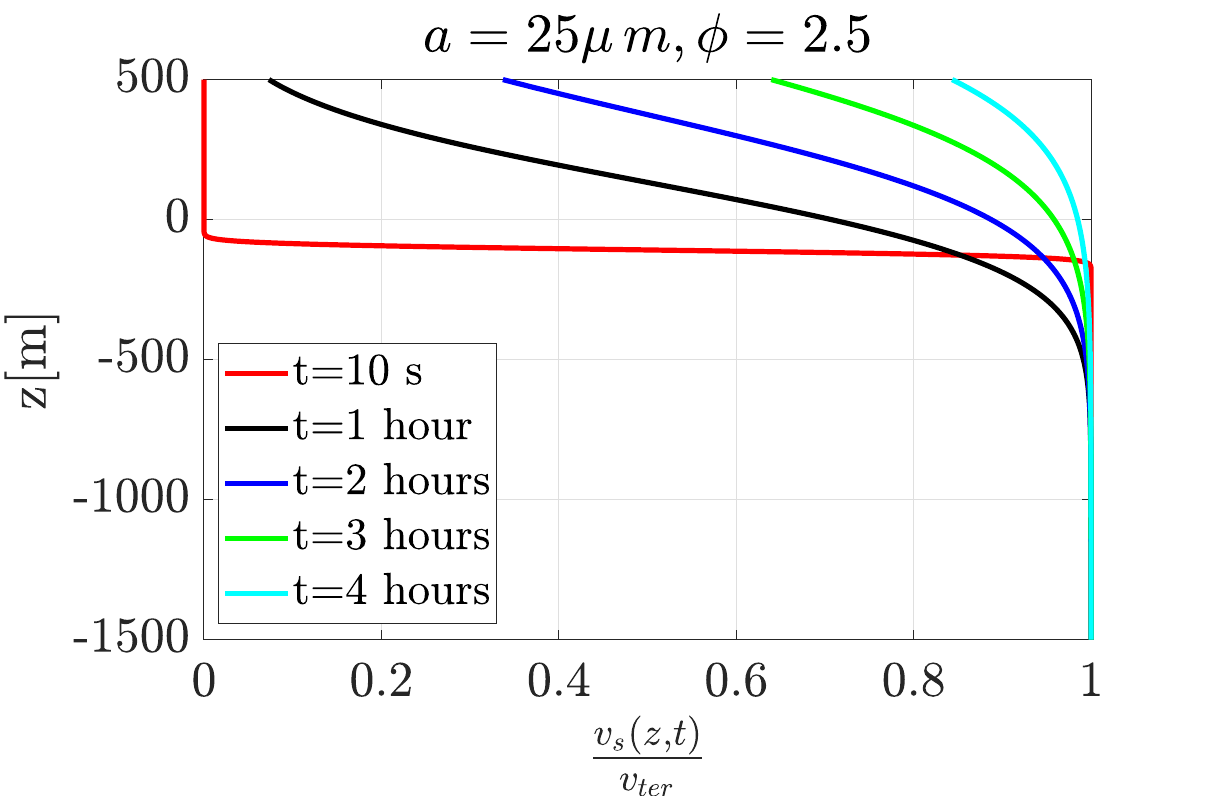}
        \subcaption{$\phi=2.5$}
        \label{fig:settling-phi25}
    \end{minipage}
    \caption{Right: Bulk settling retardation. Left: Bulk settling velocity ratio,
    $\frac{v_s(z,t)}{v_{\textbf{ter}}}$, at fixed equatorial radius
    $a=25\,\mu\mathrm{m}$ and varying shape index $\phi$.}
    \label{settling}
\end{figure}

To elucidate the bulk-settling closed-form formula, we evaluate the solution by fixing the equatorial radius at \( 25\,\mu\text{m} \) while varying the particle shape index \( \phi \). Note that here, shape index $\phi$ and equatorial radius $a$ are fixed in each run, resulting in fixed $v_{\textbf{ter}}$ which is not a realistic scenario and hence, Fig. \ref{settling}-b and c only show the behavior of the closed-form solution for the bulk settling velocity. 

In particular, the plots show the variation of the settling ratio, defined as the ratio of the bulk settling velocity of the particles (accounting for loitering and preferential sweeping effects) to their individual terminal settling velocity. In other words, this ratio, $v_s(z,t)/v_{\mathrm{ter}}$, quantifies the extent to which the actual bulk settling behavior deviates from the settling behavior of isolated particles. As shown in Fig. \ref{settling}-b and c, smaller ice crystals remain suspended for longer durations due to the dominance of turbulent mixing (\textit{loitering}) over gravitational settling (\textit{preferential sweeping}). In contrast, for larger particles, gravitational effects rapidly outweigh turbulent dispersion, resulting in the bulk settling velocity \( v_s \) converging more quickly to the individual terminal velocity \( v_{\textbf{ter}} \).

\textbf{{Sensitivity of bulk settling to turbulent mixing.}} We present a preliminary comparison with \cite{setcomp} for the bulk settling retardation. The settling response is evaluated at $T=218\,\mathrm{K}$ for four initial
peak ice-supersaturation values, $s_{i,\mathrm{peak}}=0.05$, $0.10$, $0.15$,
and $0.20$. In each case, the upper four atmospheric layers are initialized
at the selected value and the lower six layers at $-0.05$; all other initial
and microphysical conditions are similar to Sec. 5. The effective turbulent velocity scale and dimensionless turbulence parameter
are defined as $  \sigma=\frac{\nu_{\mathrm{t,ef}}}{L_\sigma},
  s=\frac{\sigma}{\overline v_{\mathrm{ter}}}
   =\frac{\nu_{\mathrm{t,ef}}}{L_\sigma\overline v_{\mathrm{ter}}},$ where $\nu_{\mathrm{t,ef}}$ is the effective eddy viscosity and
$\overline v_{\mathrm{ter}}$ is evaluated separately for each simulation to
account for particle growth and habit evolution:
\begin{equation}
  \overline v_{\mathrm{ter}}=
  \frac{\displaystyle\int_0^T\!\int_{\Omega_z}
    c_N(z,t)|v_{\mathrm{ter}}(z,t)|\,\mathrm dz\,\mathrm dt}
  {\displaystyle\int_0^T\!\int_{\Omega_z}
    c_N(z,t)\,\mathrm dz\,\mathrm dt}.
  \label{eq:appC_vter}
\end{equation}
The coherent-eddy prescription is imposed operationally as
$A_e=sL_e/L_\sigma=1$, giving $L_e=L_\sigma/s$. We use
$L_\sigma=100\,\mathrm{m}$, consistent with the $L_e=100\,\mathrm{m}$
reference scale used in Appendix C at $s=1$. Over $0.3\leq s\leq20$, this
implies $5\leq L_e\leq333\,\mathrm{m}$.  For each target $s_k$, $L_e$ is held fixed and $\nu_{\mathrm{t,ef}}$ is
obtained iteratively because $\overline v_{\mathrm{ter}}$ depends on the
simulation. With iteration index $n$, a run at $\nu_{\mathrm{t,ef}}^{(n)}$
gives $\overline v_{\mathrm{ter}}^{(n)}$ and
$s^{(n)}=\nu_{\mathrm{t,ef}}^{(n)}/(L_\sigma\overline v_{\mathrm{ter}}^{(n)})$;
the viscosity is updated by the multiplicative fixed-point rule $  \nu_{\mathrm{t,ef}}^{(n+1)}=
  \nu_{\mathrm{t,ef}}^{(n)}\frac{s_k}{s^{(n)}}.$
Iteration stops when $|s^{(n)}/s_k-1|\leq0.01$, with at most ten
complete simulations per target. The preceding converged terminal-speed
average provides the initial estimate for the next target. 

The instantaneous settling response for supersaturation case $j$ is the
ratio of particle-number-weighted deterministic bulk and individual-particle
terminal settling fluxes,
\begin{equation}
  R_j(t;s_k)=
  \frac{\displaystyle\int_{\Omega_z}c_{N,j}|v_{s,j}|\,\mathrm dz}
       {\displaystyle\int_{\Omega_z}c_{N,j}|v_{\mathrm{ter},j}|\,\mathrm dz},
  \qquad
  R_j(s_k)=\frac{1}{T_j}\int_0^{T_j}R_j(t;s_k)\,\mathrm dt.
  \label{eq:appC_R}
\end{equation}
Here $v_s$ is the deterministic Burgers bulk-settling velocity; stochastic
vertical air motion is excluded. The plotted mean is
$\overline R(s_k)=\frac14\sum_{j=1}^4R_j(s_k)$, and the shaded envelope is
the pointwise minimum--maximum across the four supersaturation cases, not a
confidence interval.

Nielsen (1993) reports experimental settling changes spanning a 40\% decrease
to an enhancement by a factor of four. The accessible original record does not
provide corresponding $s$-resolved coordinate pairs; accordingly, the plot
shows this reported range as a qualitative annotation, not as fabricated
pointwise data. The plotted ordinate $0\leq R\leq1$ covers the retardation
branch only; the reported enhancement lies outside this range. The Nielsen
summary is contextual rather than a pointwise validation because its
corresponding turbulence-intensity coordinates are unavailable.

\section{Inherent Growth Factor \(\Gamma^*\)}\label{apD}
{It should be emphasized that the current habit-dynamics framework does not represent bullet-rosette geometries, internal hollowing/branching morphologies, or ventilation effect. }

{To elaborate, incorporation of hollowing and branching together with habit-dependent ventilation is conceptually straightforward and can be implemented using well-documented microphysical relations \cite{newpaper1}; doing so does not invalidate the terminal-velocity formulations for generalized spheroidal particles (for example, the Ganser-type parameterizations employed in this study \cite{Ganser}) because those relationships remain applicable to the modified bulk shape and density descriptions. 
}

{By contrast, explicit treatment of bullet-rosette crystals poses practical difficulties. These habits typically develop after prolonged exposure to large ice supersaturations and therefore are likely less relevant to contrails. Specifically, in expanding contrails, the local relative humidity often relaxes to ice saturation level for a considerable period of time, and the subsequent recovery toward ambient humidity can be slow. Such a thermodynamic history does not generally provide the sustained, extreme supersaturation episodes necessary for bullet-rosette formation. Moreover, in-situ studies of contrail and contrail-cirrus microphysics usually report that plate- and column-like morphologies are observed more frequently than bullet-rosettes. Finally, parameterizations of the aerodynamic properties of complex rosette aggregates—especially broadly applicable terminal-velocity formulations—are scarce and poorly constrained, which further complicates their reliable inclusion in a habit-dynamics model.
}

{For these reasons — and in order to introduce a state-of-the-art baseline for habit dynamics in contrail microphysics — we omit bullet-rosettes, hollowing/branching, and ventilation effects in the present work and reserve them for future extension and validation.
}

%In addition, experimental literature for temperatures below \(-40\,^\circ\mathrm{C}\) remains too sparse to support their robust parameterization.

The available mathematical frameworks for habit dynamics are those by \cite{igf1} (based on the aspect-ratio hypothesis, defining $\frac{dc}{da}=\frac{\alpha_c}{\alpha_a}\frac{c}{a}:=\Gamma\, \phi$) and \cite{igf2} (based on the facet hypothesis, defining $\frac{dc}{da}=\frac{\alpha_c}{\alpha_a}:=\Gamma$), where $\alpha_a$ and $\alpha_c$ are the deposition coefficients of the basal and prism axes, respectively, $a$ is half the basal-plane maximum width (equatorial radius), $c$ is half the prism-plane height (transverse radius), and $\Gamma$ is known as the Inherent Growth Factor (IGF). It is straightforward to show that the above dynamics are equivalent to (\cite{igf6}):
\begin{equation}
	\frac{d\phi}{d\mathcal{V}}=\frac{\Gamma^*-1}{\Gamma^*+2}\frac{\phi}{\mathcal{V}},
\end{equation}
where $\Gamma^*=\Gamma$ in the Chen and Lamb model, and $\Gamma^*=\frac{\Gamma}{\phi}$ in the Nelson and Baker model.

{Here, we should also highlight that the mathematical derivation of the above differential equation is general and does not include assumptions regarding temperature ranges or mixed-phase cloud regimes (where tiny supercooled water droplets and ice crystals coexist) \cite{igf6}. Therefore, the framework is also valid for contrail crystals, provided that the IGF factor is properly estimated or tuned. Diagnostic approaches to habit classification have also been proposed in the literature (e.g., \cite{newpaper2}). However, a limitation of such diagnostic approaches is that temporal variations caused by local temperature and supersaturation changes may not be captured by static relations (\cite{newpaper1}). Another important point is that the well-known Schmidt--Appleman criterion sets the threshold for contrail formation, and the temperature required for contrail formation typically lies between approximately $-35$ and $-40\,^{\circ}\mathrm{C}$ (\cite{Ames}), which coincides with the ice-cloud regime. However, once contrail crystals are formed, they may be advected into a mixed-phase regime, altering the vapor-transfer pathways and further complicating habit microphysics. However, this possibility is not considered in the present work, based on the assumption that temperature variations over the course of the contrail lifetime may not be significant.
}

{Reviewing the literature on habit dynamics, predicting the IGF factor remains an active area of research, and in general, the available data are sparse for $T < -30\,^{\circ}\mathrm{C}$. However, extrapolations and fits to the limited available data have already been pursued in the literature (see, for example, \cite{igf6} and \cite{igf4}). To date, however, most IGF estimations rely on partially empirical or ad hoc closure assumptions.

In the present work, the IGF is formulated in a way to be consistent with the above state-of-the-art habit-dynamics models describing transitions between plate-like and columnar crystals, while remaining sufficiently simple to allow future extensions and refinement as controlled in-cloud observational data become available.}

To this end, we set $\Gamma=\Gamma(T)$ for $T > -30\,^{\circ}\mathrm{C}$ and directly fit a function $\Gamma(T)$ to the available data in \cite{igf1,igf3}. However, for $T < -30\,^{\circ}\mathrm{C}$ we use Nelson and Baker model which directly accounts for dislocation growth and step nucleation theories, characterized by supersaturation immediately above the surface (i.e., surface supersaturation $s_\text{surf}$), a temperature-dependent characteristic supersaturation describing the supersaturation-dependence of surface-kinetic mediated growth (i.e., $s_\text{char}$), and the parameter $N$ that describes the surface growth mode. An approximation for $\alpha$ that captures both dislocation growth and step nucleation theories was suggested by \cite{igf2}:
\begin{equation}\label{eqigf}
	\alpha(T,s_i)=\alpha_s\big(\frac{s_\text{surf}}{s_\text{char}}\big)^N\tanh\big(\frac{s_\text{char}}{s_\text{surf}}\big)^N
\end{equation}  
where $\alpha_s$ is the sticking probability/adsorption efficiency and is thought to be near unity \cite{igf4,igf5}. In addition, $N=1$ is consistent with the dislocation growth whereas $N\geq10$ is amenable to step nucleation.

{ Following \cite{igf4}, we compute $s_\text{surf}$ as:
}
\begin{equation}
	s_\text{surf}\approx s_\text{diff}^{1-\beta}\,s_\text{char}^\beta.
\end{equation}
with:
\begin{equation}
	\begin{split}
	  & s_\text{diff,a}=\frac{s_i}{1+L_a},\quad L_a=\frac{a\,c\,\bar{v}_v}{4D_vC_\Delta(c,a)},\quad s_\text{diff,c}=\frac{s_i}{1+L_c},\quad L_c=\frac{a^2\,\bar{v}_v}{4D_vC_\Delta(c,a)}.
	\end{split}
\end{equation}
where $D_v$ is the vapor diffusivity in air ($D_v = D_{v0}\Big(\frac{T}{T_{0}}\Big)^{1.94}\frac{p_0}{p},
D_{v0}=2.11\times10^{-5}\ \mathrm{m^2s^{-1}}$, $T_0=273.15$ K, $P_0=101325\,$ Pa), $\bar{v}_v$ is the mean speed of a vapor molecule ($\bar{v} \;=\; \sqrt{\frac{8k_B T}{\pi m_m}}
\;=\; \sqrt{\frac{8 R T}{\pi M}},$), where \(k_B\) is Boltzmann's constant, \(m_m\) the mass of one molecule, \(R\) the universal gas constant and \(M\) the molar mass. Moreover, $L_a$, and $L_c$ are unitless quantities that depend on the crystal geometry, and $C_\Delta$ is the capacitance \cite{growth1} evaluated one mean free path from the surface, i.e., $C_\Delta:=C(a+\Delta,c+\Delta)$ where $\Delta$ is approximately the mean free path. In addition, $\beta$ is a function of $N$ which can be directly obtained from \cite{igf4}. Moreover, $s_\text{diff}$, is the surface supersaturation over the respective axis when the deposition coefficient is unity.

{Notably, the Nelson and Baker formulation is expected to perform well at contrail-range temperatures and moderate supersaturation, but it falls short at low supersaturations \cite{igf4, igf6}. In this respect, it is suggested to account for residual errors by applying additional approximate parameterization techniques, more recently proposed in \cite{igf4}, to correct the surface supersaturation. However, the validity of such approximations has not yet been verified against experimental measurements (interested readers are referred to \cite{37,igf4,igf5,igf6,igf7}).
}

{From habit diagrams (e.g., \cite{newpaper4,igf8,igf9}), at temperatures colder than $-40,^{\circ}\mathrm{C}$ (contrail-range temperatures), there is a marked shift toward columnar behavior, except at low to moderate ice supersaturation (below $\approx 10\%$), where thick and irregular plates are observed. At moderate ice supersaturation (10\%–25\%), long solid columns and polycrystals with columnar and plate-like components are observed. Above approximately 25\% ice supersaturation, bullet rosettes, long columns, and column-containing polycrystals are observed, with the frequency of bullet rosettes and columns increasing as ice supersaturation increases.
}

{Therefore, for simplicity, consistency with the Nelson and Baker model, and correction of the low-supersaturation limit directly following the habit diagrams--while also explicitly addressing the frequently observed plate-like crystals in persistent contrails--we propose a small modification that corrects only the low-supersaturation shortcoming of the Nelson and Baker model by allowing plate-like crystals to reach a minimum of \(\Gamma = 0.1\), corresponding to nearly thick plates.
}

Here, we introduce an ad hoc modification to the habit model to account exclusively for the lower supersaturation range (mimicking the 2D‐nucleation–limited growth \cite{igf3, igf10}). The characteristic surface supersaturation $s_\text{char}=s_{\text{char}}(T)$ is specified using the Nelson--Baker/Harrington parameterization \cite{igf4} and fitting smooth functions to the available experimental range and extrapolating these fits for lower temperatures.

{We also point out that extrapolating the available data to cover a wider range, down to $-70,^{\circ}\mathrm{C}$, does not change the structure of the characteristic supersaturation, as long as the monotonic trend documented in the literature is preserved. However, the choice of extrapolation function primarily affects the upper-bound extent of the columnar crystals.
}

On choosing a critical supersaturation parameter $s_\text{critical}$, and a corresponding $\Gamma_\text{critical}$, the final form of the IGF for $T< -30\,^{\circ}\mathrm{C}$ is considered as:
\begin{equation}
\Gamma(T,s_i)=\Gamma_\text{critical}+\big(\frac{\alpha_c}{\alpha_a}-\Gamma_\text{critical}\big)e^{-(\frac{s_\text{critical}}{s_i})^n}.
\end{equation}
where $\alpha_a$, and $\alpha_c$ are obtained from Eq. (\ref{eqigf}).

Based on the foregoing discussion, we adopt the parameter ranges \(2\%\!<\!s_{\mathrm{critical}}\!<\!5\%\), \(\Gamma_{\mathrm{critical}}\in[0.05,0.2]\), and \(n\in[2,4]\). Numerical simulations show that, over these ranges, the model output exhibits at most moderate sensitivity to \(s_{\mathrm{critical}}\), negligible sensitivity to \(n\), and negligible-to-moderate sensitivity to \(\Gamma_{\mathrm{critical}}\). 
%
% In this study, we have set $\Gamma_\text{cr.}=0.1$, $s_\text{cr.} = 2\%$, and $n=2$.

{Further tuning of the model in the contrail regime should address a more general parameterization of low supersaturation and different types of monotonic extrapolation functions for the characteristic supersaturations. Such additional tuning or modification would mainly control the minimum and maximum bounds of \(\phi\).
}

\section{Vapor Deposition Volumetric Growth Rate \(f_\mathcal{V}\)}\label{apE}
The volumetric growth rate \(f_\mathcal{V}\) is computed from a standard diffusion-limited growth model with latent-heat (thermal-resistance) coupling:
\begin{equation}
	f_\mathcal{V} \;=\; \frac{4\pi\,C\,s_i}
	{\rho_{\text{dep}}\left(\dfrac{R_v T}{e_i D_v'} \;+\; \dfrac{L_s}{k_{\mathrm{air}}' T}\!\left(\dfrac{L_s}{R_v T}-1\right)\right)}.
\end{equation}
where $e_i = e_i(T)$ denotes the saturation vapor pressure over ice, and $L_s$ is the latent heat of sublimation. Empirical formulas from \cite{growth2} are used to evaluate $e_i$ and $L_s$. In addition, $R_v$ represents the specific gas constant for water vapor, with a value of approximately $461.5~\mathrm{J\,kg^{-1}\,K^{-1}}$. In the absence of hollowing and branching processes, the deposition density $\rho_{\text{dep}}$ is approximated by the ice density, i.e., $\rho_{\text{dep}} \approx \rho_{\text{ice}}$ $\approx\, 917$ kg m$^{-3}$.  Moreover, $D_v'$ denotes the kinetically limited vapor diffusivity, and $k_{\mathrm{air}}'$ the kinetically limited thermal conductivity, expressed as \cite{igf3}:
\begin{equation}
	D_v' = \frac{2}{3}\frac{D_v}{A+B_d} + \frac{1}{3}\frac{D_v}{A+C_d},\qquad
	k_{\mathrm{air}}' = \frac{2}{3}\frac{k_{\mathrm{air}}}{A+B_k} + \frac{1}{3}\frac{k_{\mathrm{air}}}{A+C_k}.
\end{equation}
where:
\begin{equation}
	\begin{split}
		&A=\frac{C}{C_\Delta},\quad B_d = \frac{4 D_v C}{\alpha_a v_v a^2 \phi},\quad
		C_d = \frac{4 D_v C}{\alpha_a \Gamma v_v a^2},\quad B_k = \frac{4 k_{\mathrm{air}} C}{0.7\,\rho_{\mathrm{air}} c_p v_v a^2 \phi},\quad
		C_k = \frac{4 k_{\mathrm{air}} C}{0.7\,\rho_{\mathrm{air}} c_p v_v a^2}.
	\end{split}
\end{equation}
In above, \(k_{\text{air}}\) is the thermal conductivity of air ($\approx\, 0.024$ W m$^{-1}$ K$^{-1}$), \(\rho_{\text{air}}\) is air density and \(c_p\) is the specific heat capacity of air ($\approx$ \(1005\ \mathrm{J\,kg^{-1}K^{-1}}\)). In addition, standard atmospheric relations are used to compute the air density, $\rho_{\text{air}}$, and pressure, $P$, from the temperature profile $T(z)$.

\section{Eulerian equation for supersaturation $s_i$}\label{apF}

Ice supersaturation is defined as
$s_i(\mathbf{x},t)=q_v/q_s(T,p)-1$, so that
$q_v=q_s(1+s_i)$, where $q_v$ is the water-vapour mixing ratio and
$q_s(T,p)=\epsilon e_{si}(T)/(p-e_{si}(T))$ is the saturation mixing
ratio over ice. The volumetric deposition rate is
$S_{\rm dep}=c_N\dot m=c_N\rho_{\rm dep}f_V$.
For compactness, write $q_s\equiv q_s(T,p)$.

We start from the conservative water-vapour balance
\begin{equation}
  \frac{\partial(\rho_{\rm air}q_v)}{\partial t}
  +\nabla\!\cdot(\rho_{\rm air}\mathbf{u}q_v)
  =\nabla\!\cdot\!\left(\rho_{\rm air}\mathbf{K}_v\nabla q_v\right)
   -S_{\rm dep},
  \tag{G1}\label{eq:G1-vapour}
\end{equation}
where $\mathbf{u}$ is the carrier-air velocity and $\mathbf{K}_v$ is the
effective molecular--turbulent diffusivity tensor for water vapour.
Assuming air-mass continuity,
$\partial_t\rho_{\rm air}+\nabla\cdot(\rho_{\rm air}\mathbf{u})=0$,
Eq.~\eqref{eq:G1-vapour} becomes:
\begin{equation}
  \frac{D_a q_v}{Dt}
  =-\frac{S_{\rm dep}}{\rho_{\rm air}}
   +\frac{1}{\rho_{\rm air}}
    \nabla\!\cdot\!\left(\rho_{\rm air}\mathbf{K}_v\nabla q_v\right),
  \qquad
  \frac{D_a}{Dt}=\frac{\partial}{\partial t}+\mathbf{u}\cdot\nabla.
  \tag{G2}\label{eq:G2-vapour-material}
\end{equation}
On the other hand, differentiating $q_v=q_s(1+s_i)$ along the air motion gives
\begin{equation}
  \frac{D_a q_v}{Dt}
  =q_s\frac{D_a s_i}{Dt}
   +(1+s_i)\frac{D_a q_s}{Dt}.
  \tag{G3}\label{eq:G3-chain-rule}
\end{equation}
Combining Eqs.~\eqref{eq:G2-vapour-material} and
\eqref{eq:G3-chain-rule}, and substituting $q_v=q_s(1+s_i)$ in the
diffusive flux, yields the full-dimensional supersaturation equation:
\begin{equation}
  \frac{D_a s_i}{Dt}
  =-\frac{S_{\rm dep}}{\rho_{\rm air}q_s}
   -\frac{1+s_i}{q_s}\frac{D_a q_s}{Dt}
   +\frac{1}{\rho_{\rm air}q_s}
    \nabla\!\cdot\!\left[
      \rho_{\rm air}\mathbf{K}_v
      \nabla\!\left(q_s(1+s_i)\right)
    \right].
  \tag{G4}\label{eq:G4-si-full}
\end{equation}

Notably, plume--environment exchange is represented by the
spatially resolved vapour-transport flux, provided that the surrounding
air and its ambient humidity are included through the domain and its
initial/boundary conditions. After reduction to the $(z,t)$ system, the
cross-sectional part of this exchange is no longer explicitly resolved.
Its reduced-order approximation, based on the wind-shear-dependent
cross-sectional geometry and its area growth, is therefore introduced
in Sec.~4 rather than in the full-dimensional supersaturation balance.

\section{Cross-sectional wind-shear closure}
\label{apG}

\subsection{Motivation and characteristic displacement}

The reduced MultiCon formulation retains the vertical coordinate $z$ as
its explicitly resolved spatial dimension, whereas the crosswind coordinate
$y$ is not independently discretized. Nevertheless, the cross-sectional
geometry is needed to estimate plume area and the associated entrainment of
ambient air. We therefore reconstruct the leading-order effect of normal
wind shear from particle trajectories while retaining the Eulerian vertical
number profile.

Let $x$ denote the direction along the contrail, $y$ the crosswind direction,
and $z$ the vertical direction. The present single-plume closure retains the
wind shear normal to the plume axis,
\begin{equation}
  u_y(z) = S_y\,(z-z_r),
  \qquad
  S_y = \frac{\partial u_y}{\partial z},
  \tag{H1}\label{eq:H1}
\end{equation}
where $z_r$ is a reference height; a uniform crosswind translation is
omitted because it shifts the plume without changing its cross-sectional
geometry. The streamwise shear $\partial u_x/\partial z$
is not included: it primarily redistributes plume material along the
unresolved $x$ direction and may alter the overlap of neighbouring plume
segments. Such along-track interactions are outside the scope of the present
single-plume cross-sectional closure. In contrast, $S_y$ displaces different
altitude levels across the plume and is directly relevant to its crosswind
geometry and lateral entrainment.

The shear-induced displacement is estimated from the characteristic system
\begin{equation}
  \frac{dZ_q}{dt}=w_p(Z_q,t),
  \qquad
  \frac{dY_q}{dt}=S_y\,[Z_q(t)-z_r],
  \qquad
  Z_q(0)=z_q,\quad Y_q(0)=0,
  \tag{H2}\label{eq:H2}
\end{equation}
where $q$ labels characteristics initialized over the initial vertical
profile, and $w_p$ is the signed vertical particle velocity, including the
settling and modelled vertical-air-motion contributions. The reference height
is taken as the initial number-weighted centroid,
\begin{equation}
  z_r =
  \frac{\displaystyle\int z\,g(z,0)\,dz}
       {\displaystyle\int g(z,0)\,dz}.
  \tag{H3}\label{eq:H3}
\end{equation}
The characteristic displacements are mapped onto the current vertical grid
to obtain $y_s(z,t)$. Thus, $y_s$ describes how the lateral position varies
with altitude; it does not introduce a separately resolved $y$-dependent
concentration field.

\subsection{Gaussian-equivalent cross-sectional geometry}

The current non-negative vertical number profile supplies the weights for
the geometrical moments. For compactness, define the number-weighted average
\begin{equation}
  \langle f\rangle_g
  =
  \frac{\displaystyle\int f(z)\,g_+(z,t)\,dz}
       {\displaystyle\int g_+(z,t)\,dz},
  \qquad g_+(z,t)=\max\{g(z,t),0\},
  \tag{H4}\label{eq:H4}
\end{equation}
and let $\bar z=\langle z\rangle_g$, $\bar y_s=\langle y_s\rangle_g$,
and $\widetilde y_s=y_s-\bar y_s$. The cross-sectional covariance components
are then estimated as
\begin{align}
  \sigma_{zz}
  &=\left\langle (z-\bar z)^2\right\rangle_g,
  \tag{H5a}\label{eq:H5a}\\
  \sigma_{yy}
  &=\sigma_{y,0}^{2}+2K_y t
    +\left\langle \widetilde y_s^{\,2}\right\rangle_g,
  \tag{H5b}\label{eq:H5b}\\
  \sigma_{yz}
  &=\left\langle (z-\bar z)\widetilde y_s\right\rangle_g.
  \tag{H5c}\label{eq:H5c}
\end{align}
Here, $\sigma_{y,0}$ is the initial crosswind standard deviation and $K_y$
is the effective crosswind diffusivity. The first two terms in $\sigma_{yy}$
represent the initial and diffusive crosswind spread, while the characteristic
displacement contributes the shear-induced spread and the crosswind--vertical
covariance. These moments define
\begin{equation}
  \boldsymbol{\Sigma}=
  \begin{pmatrix}
    \sigma_{yy} & \sigma_{yz}\\
    \sigma_{yz} & \sigma_{zz}
  \end{pmatrix}.
  \tag{H6}\label{eq:H6}
\end{equation}
We use the ellipse associated with the covariance contour
\begin{equation}
  \boldsymbol{r}^{\mathsf T}\boldsymbol{\Sigma}^{-1}\boldsymbol{r}=2,
  \qquad
  \boldsymbol{r}=\begin{pmatrix}y-\bar y_s\\z-\bar z\end{pmatrix},
  \tag{H7}\label{eq:H7}
\end{equation}
as a compact, Gaussian-equivalent representation of the cross-sectional
geometry. Its area and projected crosswind and vertical dimensions are
\begin{equation}
  A=2\pi\sqrt{\det\boldsymbol{\Sigma}},
  \qquad
  B=\sqrt{8\,\sigma_{yy}},
  \qquad
  D=\sqrt{8\,\sigma_{zz}}.
  \tag{H8}\label{eq:H8}
\end{equation}
The off-diagonal covariance $\sigma_{yz}$ accounts for the tilt of the
ellipse and therefore affects $A$, even though $B$ and $D$ are its projected
width and depth. The ellipse is a moment-based geometrical estimate; it is
not an assumption that the resolved vertical profile $g(z,t)$ is Gaussian.

% The area $A(t)$ is the principal geometrical quantity carried into the
% supersaturation closure: its positive temporal growth is used to estimate
% the entrainment timescale $\tau_{\rm entrain}$ (Appendix~J), which appears
% in the ambient-vapour relaxation term
% $(s_{i,\rm env}-s_i)/\tau_{\rm entrain}$ (Appendix~\ref{apG}). Thus, $A$ affects
% $s_i$ through entrainment, rather than as a direct source term. The width
% $B$ also enters the crosswind-averaged number-concentration estimate used
% for vapour deposition and depletion.

\begin{figure}[!htbp]
  \centering
  \begin{tikzpicture}[
      x=0.85cm,y=0.85cm,>=Stealth,
      every node/.style={font=\small},
      axis/.style={->,thick},
      guide/.style={gray!65,densely dashed},
      dim/.style={<->,thick}
    ]
    % MultiCon's resolved, generally non-Gaussian vertical number profile
    \draw[axis] (0.65,1.15) -- (0.65,5.55)
      node[above] {$z$};
    \path[draw=black!75,thick,fill=gray!18]
      (0.65,1.35)
      .. controls (1.00,1.62) and (1.48,1.95) .. (1.18,2.42)
      .. controls (0.98,2.78) and (2.30,3.05) .. (1.92,3.58)
      .. controls (1.62,4.00) and (1.02,4.28) .. (1.35,4.68)
      .. controls (1.58,5.02) and (1.05,5.30) .. (0.65,5.42)
      -- cycle;
    \node[align=center] at (1.55,6.05)
      {Resolved vertical profile};
    \node at (1.55,0.88) {$g(z,t)$};
    \node[font=\scriptsize] at (1.55,0.52) {not assumed Gaussian};

    % Mapping from the vertical solution to the cross-sectional moments
    \draw[->,thick] (2.55,3.45) -- (4.00,3.45);
    \node[font=\scriptsize,fill=white,inner sep=1pt]
      at (3.28,3.88) {characteristics};

    % Crosswind/vertical axes of the estimated cross-section
    \draw[axis] (4.55,0.95) -- (12.25,0.95)
      node[below right] {$y$ (crosswind)};
    \draw[axis] (4.55,0.95) -- (4.55,5.70)
      node[above] {$z$};

    % Projected B and D extents. The ellipse is constructed so that its
    % axis-aligned projections exactly match this dashed rectangle.
    \draw[guide] (5.70,1.70) rectangle (10.50,4.90);

    % Moment-equivalent ellipse with the same projected B and D extents.
    \begin{scope}[shift={(8.10,3.30)},rotate=18]
      \draw[draw=black!75,fill=blue!10,thick]
        (0,0) ellipse (2.4774 and 1.4773);
    \end{scope}

    % Cross-sectional area definition
    \node[fill=white,inner sep=2pt,align=center]
      at (8.05,3.30) {$A=2\pi\sqrt{\det\boldsymbol{\Sigma}}$};

    % Projected crosswind width B
    \draw[guide] (5.70,4.90) -- (5.70,5.20);
    \draw[guide] (10.50,4.90) -- (10.50,5.20);
    \draw[dim] (5.70,5.20) -- (10.50,5.20);
    \node[fill=white,inner sep=1.5pt] at (8.10,5.48)
      {$B=\sqrt{8\sigma_{yy}}$};

    % Projected vertical depth D
    \draw[guide] (10.50,1.70) -- (10.88,1.70);
    \draw[guide] (10.50,4.90) -- (10.88,4.90);
    \draw[dim] (10.88,1.70) -- (10.88,4.90);
    \node[align=left,anchor=west,fill=white,inner sep=1.5pt]
      at (11.12,3.30) {$D=\sqrt{8\sigma_{zz}}$};

    \node[font=\scriptsize] at (8.10,1.38)
      {Gaussian-equivalent cross-section};
  \end{tikzpicture}
  \caption{Schematic of the reduced cross-sectional wind-shear closure.
  MultiCon retains the generally non-Gaussian vertical number profile
  $g(z,t)$. Characteristic trajectories provide the height-dependent
  crosswind displacement, from which the covariance-based
  moment-equivalent ellipse is estimated. The ellipse area $A$ and its
  projected crosswind width $B$ and vertical depth $D$ are defined in
  Eq.~\eqref{eq:H8}. }
  \label{fig:H-windshear-geometry}
\end{figure}

The separate cross-sectional post-processing follows the same variance
decomposition. At each altitude it assigns a lateral Gaussian centred on
$y_s(z,t)$, with intrinsic variance
$\sigma_{y,G}^{2}=\sigma_{yy}-\langle\widetilde y_s^{\,2}\rangle_g
=\sigma_{y,0}^{2}+2K_y t$, weighted by the resolved IWC profile. This is a
diagnostic realization of the spread-plus-shear geometry underlying
$\boldsymbol{\Sigma}$ and $A$, not an additional prognostic solve. Because it
is IWC-weighted (whereas the moments are number-weighted) and may include
display-only vertical smoothing, its contours need not coincide exactly with
the moment-equivalent ellipse; it does not modify $A$, $\tau_{\rm entrain}$,
or the evolution of $s_i$.

% \begin{table}[H]
% 	\centering
% 	\caption{CoCiP specifications used for comparison}
% 	\label{tab:cocip-actual-specs}
% 	\begin{tabular}{lll}
% 		\toprule
% 		\textbf{Parameter} & \textbf{Value (actual)} & \textbf{Units} \\
% 		\midrule
% 		nvpm\_ei\_n & $1\times10^{14}$ & particles kg$^{-1}$ fuel \\
% 		nvpm\_ei\_n enhancement factor & $1.0$ & (dimensionless) \\
% 		fuel\_flow & $1.0$ & kg s$^{-1}$ \\
% 		true airspeed & $240.0$ & m s$^{-1}$ \\
% 		aircraft type & B738 & - \\
% 		engine UID & \texttt{CFM56-7B} & - \\
% 		number of engines & $2$ & - \\
% 		aircraft mass & $60000.0$ & kg \\
% 		wingspan & $34.32$ & m \\
% 		engine efficiency & $0.295$ & (dimensionless) \\
% 		flight altitude & $11000$ & m \\
% 		vertical shear (shear variable) & $-1.0\times10^{-3}$ & s$^{-1}$ \\
% 		wind field & $0$ (all levels) & m s$^{-1}$ \\
% 		temperature (MET field) & $212.0$ & K \\
% 		lon/lat center & $(0.1^\circ,\ 50.0^\circ)$ & degrees \\
% 		RHi & $\begin{cases}1.15 & 10800\le z<11200\\0.85 & \textbf{elsewhere}\end{cases}$ & - \\
% 		\bottomrule
% 	\end{tabular}
% \end{table}

\end{document}